\documentclass[11pt]{article}

\usepackage{lineno}

\pdfoutput=1
\usepackage[english]{babel}

\usepackage[letterpaper,top=1in,bottom=1in,left=1in,right=1in,marginparwidth=1in]{geometry}

\usepackage[utf8]{inputenc}
\usepackage{amssymb,amsmath,amsthm}
\usepackage{sansmathfonts}
\usepackage[sfdefault]{biolinum}
\usepackage[T1]{fontenc}
\usepackage[dvipsnames]{xcolor}

\RequirePackage{hyperref}
\hypersetup{colorlinks=true, urlcolor=Blue, citecolor=Green, linkcolor=BrickRed, breaklinks, unicode}

\usepackage[capitalise, noabbrev]{cleveref}
\usepackage{amsthm}
\usepackage{url}
\usepackage{standalone}
\usepackage[noadjust]{cite}
\usepackage{mathtools}
\usepackage{graphicx}
\usepackage{array}
\usepackage{xspace}
\usepackage{float}
\usepackage{amsfonts}
\usepackage{bm}
\usepackage{tikz}
\usepackage{todonotes}
\usepackage{wrapfig}
\usepackage{thmtools, thm-restate}
\usepackage[boxruled, nofillcomment, linesnumbered, lined]{algorithm2e}
\usepackage{subcaption}
\usepackage{tikz}
\usepackage{enumitem}

\theoremstyle{plain}
\newtheorem{theorem}{Theorem}
\numberwithin{theorem}{section}
\newtheorem{lemma}[theorem]{Lemma}
\newtheorem{corollary}[theorem]{Corollary}
\newtheorem*{claim*}{Claim}

\theoremstyle{definition}
\newtheorem{definition}[theorem]{Definition}
\newtheorem{question}[theorem]{Question}

\theoremstyle{remark}
\newtheorem{fact}[theorem]{Fact}

\newcommand{\dist}{\mathsf{dist}}

\newcommand{\rev}{\mathsf{rev}}

\newcommand{\Label}{\mathsf{Label}}
\newcommand{\Poly}{\text{poly}}
\newcommand{\Depth}{\mathsf{Depth}}
\newcommand{\Congest}{\mathsf{CONGEST}}

\newcommand{\eX}{e_{\!_X}}

\newcommand{\SX}{S_{\!_X}}
\newcommand{\SG}{S_{\!_G}}
\newcommand{\eH}{e_{\!_H}}
\newcommand{\SH}{S_{\!_H}}

\newcommand{\FX}{F_{\!_X}}

\date{}

\title{Distributed Maximum Flow in Planar Graphs\thanks{A preliminary version of this paper appeared in PODC 2025.}}

\author{%
 Yaseen Abd-Elhaleem\thanks{{\em Department of Computer Science, University of Haifa. Supported by The Israel Science Foundation (grant No. 2829/25 and grant No. 810/21).
}}\\
 \texttt{\href{mailto:yaseenuniacc@gmail.com}{yaseenuniacc@gmail.com}}\\
 \and Michal Dory\thanks{\em Department of Computer Science, University of Haifa. Supported by The Israel Science Foundation (grant No. 2829/25).}\\
 \texttt{\href{mailto:mdory@ds.haifa.ac.il}{mdory@ds.haifa.ac.il}}\\
 \and Merav Parter\thanks{\em Faculty of Mathematics and Computer Science, Weizmann Institute of Science. Supported by the European Research Council (ERC) under the European Union’s Horizon 2020 research and innovation programme (grant agreement No. 949083).}\\
 \texttt{\href{mailto:merav.parter@weizmann.ac.il}{merav.parter@weizmann.ac.il}}\\
 \and Oren Weimann\thanks{\em Department of Computer Science, University of Haifa. Supported by The Israel Science Foundation (grant No. 810/21).}\\
 \texttt{\href{mailto:oren@cs.haifa.ac.il}{oren@cs.haifa.ac.il}}
}

\begin{document}

\maketitle

\thispagestyle{empty}

\begin{abstract}
The dual of a planar graph $G$ is a planar graph $G^*$ that has a vertex for each face of $G$ and an edge for each pair of adjacent faces of $G$. The profound relationship between a planar graph and its dual has been the algorithmic basis for solving numerous (centralized) classical problems on planar graphs involving distances, flows, and cuts. In the distributed setting however, the only use of planar duality is for finding a recursive decomposition of $G$ [DISC 2017, STOC 2019].

In this paper, we initiate the study of distributed algorithms on dual planar graphs. 
Namely, we extend the distributed algorithmic toolkit (such as recursive decomposition and minor-aggregation) to work on the dual graph $G^*$. These tools can then facilitate various algorithms on $G$ by solving a suitable dual problem on $G^*$.

Given a directed planar graph $G$ with
positive and negative edge-lengths and undirected unweighted (hop) diameter $D$, our key result is an $\tilde O(D^2)$-round algorithm\footnote{The $\tilde{O}(\cdot)$ notation is used to omit $\Poly \log n$ factors.} for Single Source Shortest Paths on $G^*$. This algorithm implies an $\tilde O(D^2)$-round algorithm for Maximum $st$-Flow in $G$, and an $\tilde{O}(D^2)$-round algorithm for directed Global Minimum Cut in $G$. 
Prior to our work, no $\tilde{O}(\text{poly}(D))$-round algorithms were known for these problems.
When $G$ is undirected, we further obtain a near optimal $\tilde O(D)$-round algorithm for computing the weighted girth of $G$, and a $D\cdot n^{o(1)}$-rounds $(1-o(1))$-approximation algorithm for Maximum $st$-Flow in $G$ when $s$ and $t$ lie on the same face.
 
The main challenges in our work are that $G^*$ is not the communication graph (e.g., a vertex of $G$ is mapped to multiple vertices of $G^*$), and that the diameter of $G^*$ can be much larger than $D$ (i.e., possibly by a linear factor). We overcome these challenges by carefully defining and maintaining subgraphs of the dual graph $G^*$ while applying the recursive decomposition on the primal graph $G$. The main technical difficulty is that, along the recursive decomposition, a face of $G$ gets shattered into (disconnected) components, yet we still need to treat it as a dual node.    

We believe that the toolkit developed in this paper for exploiting planar duality will be used in future distributed algorithms for various other classical problems on planar graphs (as happened in the centralized setting). 
\end{abstract}

\newpage

{\tableofcontents}
\thispagestyle{empty}
\newpage

\newpage \setcounter{page}{1}

\section{Introduction}

Distributed algorithms for network optimization problems have a long and rich history. These problems are commonly studied under the $\Congest$ model \cite{peleg-book} where the network is abstracted as an $n$-vertex graph $G = (V, E)$ with hop-diameter $D$; communications occur in synchronous rounds. In each round, $O(\log n)$ bits can be sent along every edge. A sequence of breakthrough results provided $\tilde{O}(D+\sqrt{n})$-round algorithms for fundamental graph problems, such as minimum spanning tree (MST) \cite{GarayKP98}, approximate shortest-paths \cite{Nanongkai14}, minimum cuts \cite{DEMN21}, and approximate flow \cite{ghaffari2015near}. For general graphs, $\tilde{O}(D+\sqrt{n})$ rounds for solving the above mentioned problems is known to be near optimal, existentially \cite{SarmaHKKNPPW11}. 

A major and concentrated effort has been invested in designing improved solutions for special graph families that escape the topology of the worst-case lower bound graphs of \cite{SarmaHKKNPPW11}. The lower bound graph is sparse, and of arboricity two, so it belongs to many graph families. Arguably, one of the most interesting non-trivial families that escapes it, is the family of planar graphs. Thus, a significant focus has been given to the family of planar graphs, due to their frequent appearance in practice and because of their rich structural properties. In their seminal work, Ghaffari and Haeupler \cite{GH16a,GH16b} initiated the line of distributed planar graph algorithms based on the notion of {\em low-congestion shortcuts}. The latter serves the communication backbone for obtaining $\tilde{O}(D)$-round algorithms for MST \cite{GH16b}, minimum cut \cite{GH16b,GZ22} and approximate shortest paths \cite{GHSYZ22,GHLRZ22} in planar graphs. 

An additional key tool in working with planar graphs, starting with the seminal work of Lipton and Tarjan \cite{LT79}, is that of a planar \emph{path separator}: a path whose removal from the graph leaves connected components that are a constant factor smaller. Ghaffari and Parter \cite{GP17} presented a $\tilde{O}(D)$-round randomized algorithm for computing a cycle separator of size $O(D)$ which consists of a separator path plus one additional edge (that is possibly a {\em virtual} edge that is not in $G$). Recently, a deterministic construction of planar separators was given by Jauregui, Montealegre and Rapaport \cite{jauregui2025deterministic}. By now, planar separators are a key ingredient in a collection of $\tilde{O}(\Poly(D))$-round solutions for problems such as DFS \cite{GP17,jauregui2025deterministic}, distance computation \cite{LP19}, and reachability \cite{ParterReachability20a}. 
An important aspect of the planar separator algorithm of \cite{GP17} is that it employs a computation on the dual graph, by communicating over the primal graph.

\medskip
\noindent
{\bf Primal maximum flow via dual SSSP.}
Our goal in this paper is to expand the algorithmic toolkit for performing computations on the dual graph. This allows us to exploit the profound algorithmic duality in planar graphs, in which solving a problem $A$ in the dual graph provides a solution for problem $B$ in the primal graph. Within this context, our focus is on the \emph{Maximum $st$-Flow} problem (in directed planar graphs with edge capacities).
The Maximum $st$-Flow problem is arguably one of the most classical problems in theoretical computer science, extensively studied since the 50's, and still admitting breakthrough results in the sequential setting, such as the recent almost linear time algorithm by Chen, Kyng, Liu, Peng, Gutenberg and Sachdeva~\cite{ChenKLPGS22}. 
 Despite persistent attempts over the years, our understanding of the distributed complexity of this problem is still quite lacking. 
For general {\em undirected} $n$-vertex graphs, there is a $(1+o(1))$-approximation algorithm that runs in $(\sqrt{n}+D)n^{o(1)}$ rounds, by Ghaffari, Karrenbauer, Kuhn, Lenzen and Patt-Shamir~\cite{ghaffari2015near}. For directed $n$-vertex \emph{planar} graphs, a
$D \cdot n^{1/2+o(1)}$-round exact algorithm has been given by de Vos~\cite{de2023minimum}. 
No better tradeoffs are known for undirected planar graphs. In lack of any $\tilde{O}(\Poly(D))$-round maximum $st$-flow algorithm for directed planar graphs (not even when allowing approximation) we ask: 
\begin{question}\label{q:flow-D}
Is it possible to compute the maximum $st$-flow in directed planar graphs within $\tilde{O}(\Poly(D))$ rounds?
\end{question}

In directed planar graphs with integral edge-capacities, it is known from the 80's \cite{Venkatesan} that the maximum $st$-flow can be found by solving $\log \lambda$ instances of {\em Single Source Shortest Paths} (SSSP) with positive and negative edge-lengths on the {\em dual} graph $G^*$, where $\lambda$ is the maximum $st$-flow value. 
We answer \cref{q:flow-D} in the affirmative by designing a $\tilde{O}(D^2)$-round SSSP algorithm on the dual graph $G^*$. Our algorithm works w.h.p.\footnote{W.h.p. stands for a probability of $1-1/{n^c}$ for an arbitrary fixed constant $c>0$.} in the most general setting (i.e. when $G^*$ is directed and has positive and negative integral edge-lengths), and matches the fastest known exact SSSP algorithm in the primal graph. 
We show: 

\begin{restatable}[Exact maximum $st$-flow in directed planar graphs]{theorem}{thmstflow}
\label{thmstflow}
There is a randomized distributed algorithm that, given an $n$-vertex directed planar network $G$ with hop-diameter $D$ and integral edge-capacities, and two vertices $s, t$, computes the maximum $st$-flow value and assignment w.h.p. in $\tilde{O}(D^2)$ rounds.
\end{restatable}

No prior $\tilde{O}(\Poly(D))$ algorithm has been known for this problem, not even when allowing a constant approximation. 
We further improve the running time to $D\cdot n^{o(1)}$ rounds while introducing a $(1-o(1))$ approximation, provided that $G$ is undirected and that $s$ and $t$ both lie on the same face with respect to the given planar embedding:

\begin{restatable}[Approximate maximum $st$-flow in undirected $st$-planar graphs]{theorem}{thmstflowapprox}
\label{thmstflowapprox}
There is a randomized distributed algorithm that given an $n$-vertex undirected 
planar network $G$ with hop-diameter $D$ and integral edge-capacities, and two vertices $s,t$ lying on the same face, computes a $(1-o(1))$-approximation of the maximum $st$-flow value and a corresponding assignment 
in $D\cdot n^{o(1)}$ rounds w.h.p..
\end{restatable}

This latter algorithm is based on an approximate SSSP algorithm  that runs in $D \cdot n^{o(1)}$ rounds in planar graphs~\cite{GHSYZ22}. 
 Our implementation of the algorithm on the dual graph matches its round complexity on the primal graph. The obtained almost-optimal round complexity improves significantly over the current algorithm for general graphs that runs in $(\sqrt{n}+D)n^{o(1)}$ rounds~\cite{ghaffari2015near}. 

\medskip
\noindent
{\bf Minimum $st$-cut.}
 By the well-known Max Flow Min Cut theorem of~\cite{ff56}, our flow algorithms immediately give the value (or approximate value) of the minimum $st$-cut. We show that they can be extended to compute a corresponding bisection and the cut edges without any overhead in the round complexity. Moreover, since our exact flow algorithm works with directed planar graphs, it admits a solution to the {\em directed} minimum $st$-cut problem. 
 To the best of our knowledge, prior to our work, $\tilde{O}(\Poly(D))$-round  CONGEST algorithms for the minimum $st$-cut problem were known only for general graphs with constant cut values by~\cite{Parter19}. This is in contrast to the global (undirected) minimum cut problem that can be solved in $\tilde{O}(D)$ rounds in planar graphs~\cite{GH16b,GZ22}.

\medskip
\noindent
{\bf Directed global minimum cut.}
The directed global minimum weight cut problem is given a weighted directed graph and 
asks to find a bisection $(S,V\setminus S)$ of the vertex set, such that, the set of edges leaving $S$ to $V\setminus S$ is of minimal total weight.
To our knowledge, this problem was not studied in the $\Congest$ model and is open even for approximations, as opposed to the (directed or undirected) minimum $st$-cut  problem~\cite{de2023minimum, Parter19} and the undirected global minimum cut problem~\cite{GZ22, DEMN21, GH16b}. 
In light of efficient centralized results for planar graphs~\cite{planarDiCutMozes18, planarDiCutWullfNilsen09} and motivated by the importance of the problem, we ask:

\begin{question}\label{q:directed_global_cut}
Is it possible to compute the directed global minimum cut in planar graphs within $\tilde{O}(\Poly(D))$ rounds?
\end{question}

We answer \cref{q:directed_global_cut} in the affirmative by showing an extension to our SSSP algorithm in the dual, allowing us to compute the minimum weight directed cycle in the dual. By duality, the same set of edges corresponds to the desired cut and determines a corresponding bisection in the primal, all within the same $\tilde{O}(D^2)$ bound on the round complexity:

\begin{restatable}[Planar directed global minimum cut]{theorem}{theoremDirectedGlobalCut}
   \label{th: directed_global_min_cut}
   Given a directed planar graph $G$ with non-negative integral edge weights, there is an $\tilde{O}(D^2)$ round randomized algorithm that finds a global directed minimum cut
   w.h.p..
\end{restatable}

\medskip
\noindent
{\bf Primal weighted girth via dual cuts.} A distance parameter of considerable interest is the network \emph{girth}. For  unweighted graphs, the girth is the length of the shortest cycle in the graph. For  weighted graphs, the girth is the cycle of minimal total edge weight. Distributed girth computation has been studied over the years mainly for general $n$-vertex unweighted graphs. Frischknecht, Holzer and Wattenhofer~\cite{FrischknechtHW12} provided an $\Omega(\sqrt{n})$-round lower bound for computing a $(2-\epsilon)$ approximation of the unweighted girth. The state-of-the-art upper bound for the unweighted girth problem is a $(2-\epsilon)$ approximation in $\tilde{O}(n^{2/3}+D)$ rounds, obtained by combining the works of Peleg, Roditty and Tal \cite{PelegRT12}, and Holzer and Wattenhofer \cite{HolzerW12}. The weighted girth problem has been shown to admit a near-optimal lower bound of $\tilde{\Omega}(n)$ rounds in general graphs~\cite{HuaQYSJ21,ManoharanGirth}. Turning to planar graphs,
Parter \cite{ParterReachability20a} devised a $\tilde{O}(D^2)$ round algorithm for computing the weighted girth in directed planar graphs via SSSP computations. For undirected and unweighted planar graphs, the (unweighted) girth can be computed in $\tilde{O}(D)$ rounds by replacing the $\tilde{O}(D^2)$-round SSSP algorithm by a $O(D)$-round BFS algorithm. In light of this gap, we ask:

\begin{question}\label{q:girth-D}
Is it possible to compute the weighted girth of an undirected weighted planar graph within (near-optimal) $\tilde{O}(D)$ rounds? 
\end{question}

We answer this question in the affirmative by taking a different, non distance-related, approach than that taken in prior work. Our $\tilde{O}(D)$ round algorithm exploits the useful duality between cuts and cycles. By formulating the dual framework of the {\em minor-aggregation} model, we show how to simulate the primal exact minimum cut algorithm of Ghaffari and Zuzic~\cite{GZ22} on the dual graph. This dual simulation matches the primal round complexity. The solution to the dual cut problem immediately yields a solution to the primal weighted girth problem. 
We show:

\begin{restatable}[Planar weighted girth]{theorem}{theoremWeightedGirth}\label{thm:weighted-girth}
There is a randomized distributed algorithm that given an $n$-vertex undirected weighted planar  network $G$ with hop-diameter $D$, computes the weighted girth and finds the edges of a shortest cycle w.h.p. in $\tilde{O}(D)$ rounds. \end{restatable}

\section{Technical Overview}

The dual of a planar graph $G$ is a planar graph $G^*$ that has a node\footnote{Throughout, we refer to faces of the primal graph $G$ as \emph{nodes} (rather than vertices) of the dual graph $G^*$.} for each face of $G$. For every edge $e$ in $G$ there is an edge $e^*$ in $G^*$ that connects the nodes corresponding to the two faces of $G$ that contain $e$. 
Our results are based on two main primal tools that we extend to work on the dual graph: {\em Minor Aggregation} and {\em Bounded Diameter Decomposition}. We highlight the key ideas of these techniques and the challenges encountered in their dual implementation. For all the algorithms that we implement in the dual graph, we match the primal round complexity. 

\subsection{Minor-Aggregation in the Dual}
An important recent development in the field of distributed computing was a new model of computation, called the \emph{minor-aggregation} model introduced by Zuzic \textcircled{r}\footnote{\textcircled{r} is used to denote that the authors' ordering is randomized.} Goranci \textcircled{r} Ye \textcircled{r} Haeupler \textcircled{r} Sun \cite{GHSYZ22}, then extended by Ghaffari and Zuzic \cite{GZ22} to support working with {\em virtual nodes} added to the input graph. Recent state-of-art algorithms for various classical problems are being formulated in the minor-aggregation model (e.g., the exact min-cut algorithm of \cite{GZ22}, and the undirected shortest paths approximation algorithms of \cite{GHLRZ22,GHSYZ22}). Motivated by the algorithmic power of this model, we provide an implementation of the minor aggregation model on the dual graph. 
As noted by \cite{GHSYZ22}, minor aggregations can be implemented by solving the (simpler) part-wise aggregation task, where one needs to compute an aggregate function in a collection of vertex-disjoint connected parts of the graph. The distributed planar separator algorithm of \cite{GP17} implicitly implements a part-wise aggregation algorithm on the dual graph. Our contribution is in providing an explicit and generalized implementation of the dual part-wise aggregation algorithm and using it to implement the minor-aggregation model on the dual graph. 
Then, using this implementation of the minor-aggregation model along with planar duality yields a collection of algorithmic results for the primal graph. 
Next, we give a brief overview of the proof of the simulation theorem, where the full details appear in \cref{sec: dual_MA_model}.

\medskip
\noindent
{\bf The minor-aggregation model.}
In the minor-aggregation model, there is a given graph $G$, where the vertices and edges of $G$ are computational units, and an algorithm works in synchronous rounds, where in each round we can either contract some of the edges, or compute an aggregate function on certain sets of disjoint vertices or edges. For the full details of the model see \cref{def: basic_model}.

\medskip
\noindent
{\bf Part-wise aggregation.}
As shown in \cite{GHSYZ22}, the minor-aggregation model is simulated in $\Congest$ by solving the {\em part-wise aggregation (PA)} problem. 
Briefly, the problem considers the setting where a partition $\{G_i\}_{i=1}^N$ of $G$ is given s.t. each part $G_i$ is a connected subgraph of $G$, in addition each vertex $v\in V$ initially has some input $x_v$. The objective is then, for each subgraph $G_i$ to compute an {\em aggregate operator} over all inputs $x_v$ of vertices $v\in G_i$.
An aggregate operator is a function that allows to replace $\tilde{O}(1)$-bit strings by one $\tilde{O}(1)$-bit string. This is usually a simple commutative function such as taking a minimum or a sum.

\medskip
\noindent
{\bf Low-congestion shortcuts.}
In their influential work, Ghaffari and Haeupler~\cite{GH16a} define low-congestion shortcuts as follows. Let $\{G_i\}_{i=1}^{N}$ be a partition of $G$ into vertex-disjoint connected subgraphs, a $(\alpha, \beta)$-shortcut, is a set of subgraphs $\{H_i\}_{i=1}^{N}$ of $G$ such that the diameter of each $G_i\cup H_i$ is at most $\beta$ and each edge of $G$ participates in at most $\alpha$ subgraphs $H_i$.

Using this notion, the part-wise aggregation task is solved easily by constructing low-congestion shortcuts for the given partition $\{G_i\}_{i=1}^N$ and aggregating information over the graphs $G_i\cup H_i$ in $\tilde{O}(\alpha+\beta)$ rounds, which is $\tilde{O}(D)$ for planar graphs~\cite{GH16a}.

\medskip
\noindent
{\bf Face-disjoint graph.}
The {\em face disjoint graph} $\hat{G}$ was presented in \cite{GP17} as a way for simulating aggregations on (subtrees of) the dual graph $G^*$ in the distributed setting. This tool allows us to overcome several challenges and succeeds to perform computations on $G^*$ by communicating on $G$. The obvious challenge, among others, that is solved using $\hat{G}$, is that computational entities in the input (primal) graph $G$ are vertices and not faces. 
Moreover, a vertex on $G$ belongs to many (possibly a linear number of) faces, however, we wish to compute aggregate functions on sets of faces of $G$ using low-congestion shortcuts. Hence, to do so efficiently we want faces to be vertex-disjoint, which is the main property that $\hat{G}$ provides, in addition, it is a planar graph with diameter $O(D)$, and we can simulate $\Congest$ algorithms efficiently on it.
Then, we can define each face as its own part in the low-congestion shortcuts partition, therefore, we can compute low-congestion shortcuts and computes aggregates on faces of $G$ in $\tilde{O}(D)$ rounds.

\begin{figure}[htb]
    \centering
    \includegraphics[width=0.5\linewidth]{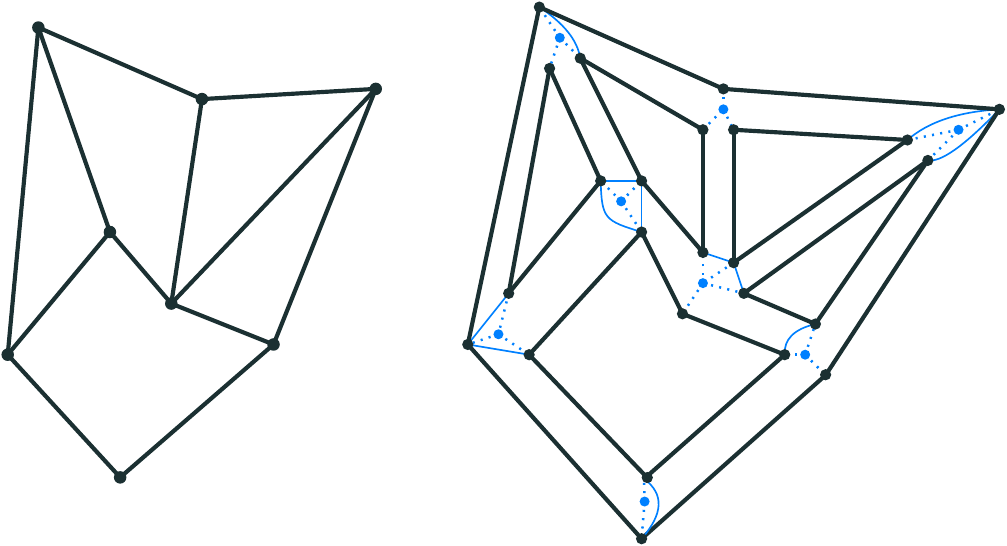}
    \caption{On the left, the primal graph $G$. On the right $\hat{G}$. Notice that each face of $G$ maps to a face (black cycle) of $\hat{G}$.
    \label{fig: face_disjoint_graph_overview}}
\end{figure}

To achieve that, $\hat{G}$ is (roughly) defined as the result of duplicating all edges of $G$ so that faces of $G$ map to distinct faces of $\hat{G}$ that are both vertex and edge disjoint. Then, in our definition of $\hat{G}$, those faces in $\hat{G}$ are connected with edges if their corresponding faces in $G$ share an edge (i.e., the dual nodes of those faces are connected by an edge in $G^*$). See \cref{fig: face_disjoint_graph_overview}. 

\medskip
\noindent
{\bf Simulation in the dual graph.}
Our goal is to use the face disjoint graph $\hat{G}$ to simulate a minor-aggregation algorithm on the dual graph, where the vertices of a face simulate the corresponding dual node, and each dual edge is simulated by the endpoints of the corresponding primal edge. By the above discussion, our main goal is to show how to solve the PA problem on the dual graph (i.e., the input this time is a partition $\{G^*_i\}_{i=1}^N$ of the {\em dual} graph into connected subgraphs).

We remark that Ghaffari and Parter \cite{GP17} showed how to solve specific aggregate functions on $G^*$ using the graph  $\hat{G}$ (aggregations on each face of $G$ and sub-tree sums on $G^*$).
For our purposes, we need something more general, so we show how to perform general part-wise aggregations on $G^*$. In particular, we need to perform aggregations that take into consideration the outgoing edges of each part $G^*_i$, something that was not done in \cite{GP17}. This specific task requires a small modification to $\hat{G}$ compared to the one defined in \cite{GP17} (see \cref{sec: face_disjoint_graph} and \cref{appendix: preliminaries_hat{G}} for more details). 

To solve the PA problem on $G^*$ we exploit the structure of $\hat{G}$. We prove that a PA problem in the dual graph $G^*$ can be translated to a corresponding PA problem in $\hat{G}$. In the graph $\hat{G}$ there is a disjoint cycle representing each face of $G$, where two such cycles are connected if there is a dual edge between the corresponding faces. Hence, if we take a partition $\{G^*_i\}_{i=1}^N$ of the dual graph into connected subgraphs, we can convert it to a partition of the vertices of $\hat{G}$, where each dual node $f \in G^*_i$ is replaced by the vertices of the cycle corresponding to $f$ in $\hat{G}$. Since cycles corresponding to different faces are vertex-disjoint in $\hat{G}$ we indeed get a partition.

In addition, by the properties of $\hat{G}$, it is a planar graph with diameter $O(D)$, and we can simulate $\Congest$ algorithms efficiently on $\hat{G}$. Based on this we can solve the PA problem in $\tilde{O}(D)$ rounds in $\hat{G}$, which leads to solving the PA problem on the dual graph $G^*$ in $\tilde{O}(D)$ rounds. For full details see \cref{sec: dual_MA_model}. We also prove that we can simulate an extended version of the minor-aggregation model defined in~\cite{GZ22} that allows adding $\tilde{O}(1)$ virtual nodes to the network, which is useful for our applications.

\medskip
\noindent
{\bf Dealing with parallel edges in \boldmath$G^*$.}
The the dual graph $G^*$ might be a multi-graph. E.g. two faces that share two edges in $G$ correspond to two nodes that are connected with two parallel edges in $G^*$. For some of our applications it is useful to think about $G^*$ as a simple graph. For example, when computing shortest paths we would like to keep only the edge of minimum weight connecting two dual nodes, and when computing a minimum cut we want to replace multiple parallel edges by one edge with the sum of weights. Note that if $G^*$ was the network of communication it was trivial to replace multiple parallel edges by one edge locally, but in our case each edge on a face $f$ is known by a different vertex of $f$. While we can compute an aggregate operator of all the edges adjacent to $f$ efficiently, if a face $f$ has many different neighboring faces $g$, we need to compute an aggregate operator for each such neighboring face of $f$, which is too expensive. To overcome this challenge, we use the low {\em arboricity} \footnote{The arboricity of a graph is the minimal number of forests into which its edges can be partitioned} of planar graphs to compute a {\em low out-degree orientation} of the edges of the dual graph. This orientation guarantees that from each face $f$ there are outgoing edges only to $O(1)$ other faces $g$, hence we can allow to compute an aggregate function on the edges between each such pair efficiently (each face takes responsibility to compute $O(1)$ aggregate operations). The algorithm for computing the orientation follow an algorithm of~\cite{BM10_NWdecomposition} that can be viewed as solving a series of $\tilde{O}(1)$ part-wise aggregation problems.

\medskip
\noindent
{\bf Applications.}
By simulating minor-aggregation algorithms on the dual graph $G^*$ we compute the exact minimum weighted cut in the dual graph, which by duality provides a solution to the {\em weighted girth} problem in the primal graph $G$. This approach allows us to find the weight of the minimum weight cycle, we also show that we can extend it to find the edges of the cycle, for details see \cref{sec: dual_MA_model}. 
We also use this model to simulate the recent approximate SSSP by \cite{GHSYZ22} on the dual graph, leading to our approximate max $st$-flow algorithm on the primal graph when $s$ and $t$ are on the same face in the given planar embedding (see \cref{section_flow}). Since currently there are fast SSSP minor-aggregation algorithms only for undirected graphs, this approach leads to an approximate max $st$-flow algorithm in undirected planar graphs. 
To solve the more general version of the max $st$-flow problem, we need additional tools described next.

\subsection{SSSP in the Dual via Distance Labels}
\medskip
\noindent
{\bf SSSP via distance labels.} 
In order to compute single source shortest paths (SSSP) in the dual graph we compute {\em distance labels}. That is, we assign each face of $G$ (node of $G^*$) a short $\tilde{O}(D)$-bit string known by all the vertices of the face s.t. the distance between any two nodes of $G^*$ can be deduced from their labels alone. This actually allows computing all pairs shortest paths (APSP) by broadcasting the label of any face to the entire graph (thus learning a shortest paths tree in $G^*$ from that face).

\begin{restatable}[Dual distance labeling]{theorem}{theoremDualLabeling}
    \label{th: dual_distance_labeling}
     There is a randomized distributed $\tilde{O}(D^2)$-round algorithm  that computes $\tilde{O}(D)$-bit distance labels for $G^*$, or reports that $G^*$ contains a negative cycle w.h.p.. Upon termination, each vertex of $G$ that lies on a face $f$, knows the label of the node $f$ in $G^*$.
\end{restatable}

\noindent
The above theorem with aggregations on $G^*$ allows learning an SSSP tree in the dual, i.e.,

\begin{restatable}[Dual single source shortest paths]{lemma}{lemmaDualSSSPTree}
\label{lemma_dual_SSSP}
     There is a randomized $\tilde{O}(D^2)$ round algorithm that w.h.p. computes a shortest paths tree from any given source $s\in G^*$, or reports that $G^*$ contains a negative cycle. Upon termination, each vertex of $G$ knows for each incident edge whether its dual is in the shortest paths tree or not.
\end{restatable}

Our labeling scheme follows the approach of~\cite{GPPR01} who gave a labeling scheme suitable for graph families that admit a small {\em separator}. We show the general idea on the primal graph and then our adaptation to the dual graph. 

As known from the 70's~\cite{LT79,Mil84}, the family of planar graphs admits cycle separators (a cycle whose removal disconnects the graph) of small size $O(\sqrt{n})$ or $O(D)$.
Since $\Omega(D)$ is a lower bound for all the problems we discuss in the distributed setting, we will focus on separators of size $\tilde{O}(D)$.
The centralized divide-and-conquer approach repetitively removes the separator vertices $\SG$ from the graph $G$ and recurses on the two remaining subgraphs that are a constant factor smaller, constituting a hierarchical decomposition of the graph with $O(\log n)$ levels.
The labeling scheme is defined recursively, where the label of a vertex $v$ in $G$ stores distances between $v$ and $\SG$ and recursively the label of $v$ in the subgraph that contains $v$ (either the interior or the exterior of $\SG$). 
Due to $\SG$ being a cycle, any shortest $u$-to-$v$ path $P$ for any two vertices $u,v$ in $G$ either  crosses $\SG$ (and the $u$-to-$v$ distance is known by the distances between $u,v$ and $\SG$ stored in their labels), or $P$ is enclosed in the interior or the exterior of $\SG$ (and the $u$-to-$v$ distance is decoded from the recursive labels of $u$ and $v$ in the subgraph that contains both of them). Moreover, since the separator is of size $\tilde{O}(D)$ and since there are $O(\log n)$ levels in the decomposition, the labels are of size $\tilde{O}(D)$.

\medskip
\noindent
{\bf Bounded diameter decomposition (BDD).} The BDD
is a distributed hierarchical decomposition for planar graphs,
which plays an analogous role to the centralized recursive separator decomposition. It was devised by Li and Parter~\cite{LP19} by carefully using and extending a distributed planar separator algorithm of Ghaffari and Parter~\cite{GP17}. The BDD has some useful properties that are specific to the distributed setting. In particular, all the separators and subgraphs obtained in each recursive level have small diameter of $\tilde{O}(D)$ and are nearly disjoint, which allows to broadcast information in all of them efficiently in $\Congest$. An immediate application of the BDD is a distributed distance labeling algorithm for primal planar graphs that follows the intuition above. In addition, BDDs have found other applications in $\Congest$ algorithms on the primal graph $G$ (e.g. diameter approximation, routing schemes and reachability \cite{DGHSW23,LP19,ParterReachability20a}).

 \medskip
\noindent
{\bf Challenges in constructing distance labels for the dual graph.} 
Ideally, we would like to apply the same idea on the dual graph. However, there are several challenges. First, if we apply it directly, the size of the separator and the running time of the algorithm will depend on the diameter of the dual graph, that can be much larger than the diameter $D$ of the primal graph (possibly by a linear factor). So it is unclear how to obtain a running time that depends on $D$.
Second, even though there are efficient distributed algorithms for constructing the BDD in the primal graph, it is unclear how to simulate an algorithm on the dual network, as this is not our communication network.

\begin{figure}[htb]
    \centering
    \includegraphics[width=1\linewidth]{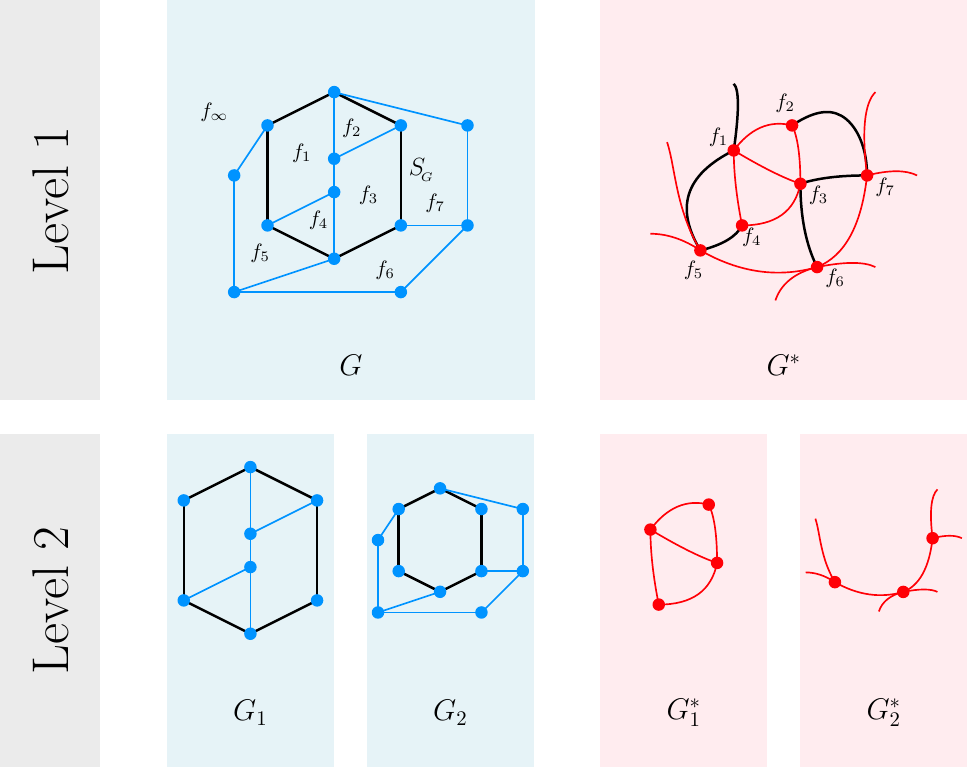}
   \caption{
    Two levels of the recursive separator decomposition of $G$ and their correspondent levels in the desired dual decomposition of $G^*$. The black edges demonstrate the separating cycle $\SG$ in $G$, the same edges map to a cut in $G^*$. For simplicity, the dual node corresponding to the face $f_\infty$ of $G$ is not illustrated in $G^*$ (its incident edges are the ones missing an endpoint). 
    \label{fig: simplified_decomposition}}
\end{figure}

 \medskip
\noindent
{\bf Our approach: recurse on primal, solve on dual.} 
To overcome this, we suggest a different approach. We use the primal BDD, and infer from it a decomposition of the dual graph.
To get an intuition, note that by the cycle-cut duality, a cycle in $G$ constitutes a cut in $G^*$. This means that the separator of $G$ can be used as a separator of $G^*$, in the sense that removing the cut edges (or their endpoints) disconnects $G^*$. I.e, a hierarchical decomposition of $G$  can (conceptually) be thought of as a hierarchical decomposition of $G^*$. See \cref{fig: simplified_decomposition}. 

Then, we would hope that using~\cite{LP19}'s algorithm for such a hierarchical decomposition for planar graphs would allow us to apply the same labeling scheme on $G^*$, where we consider the dual endpoints of $\SG$ edges as our separator in $G^*$.
However, taking a dual lens on the primal decomposition introduces several challenges that arise when one needs to define the dual subgraphs from the given primal subgraphs. This primal to dual translation is rather non-trivial due to critical gaps that arise when one needs to maintain information w.r.t. faces of $G$, rather than vertices of $G$. To add insult to injury, the nice structural property of the (primal) separator being a cycle of $G$ is simply not true any more. This is because the separator of~\cite{GP17} that the BDD of~\cite{LP19} uses is composed of two paths plus an additional edge that exists only when the graph is two-vertex connected. Otherwise, that edge is not in $G$ and cannot be used for communication. This {\em virtual} edge, if added, would split a face into two parts, introducing several challenges, including:

\begin{figure}[htb]
    \centering
    \includegraphics[width=.7\linewidth]{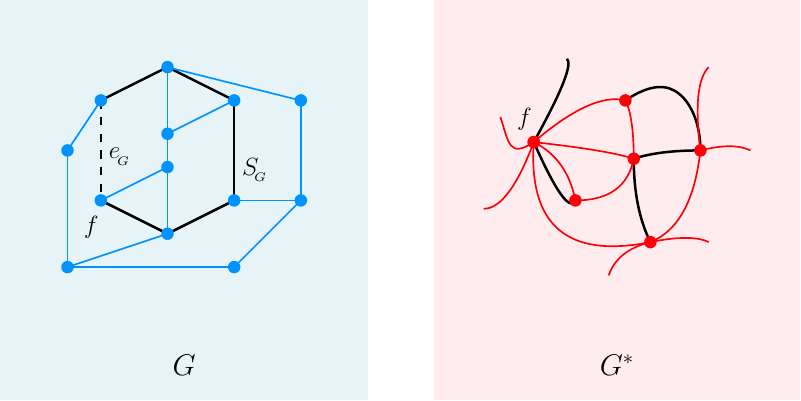}
   \caption{
   The edge $e_{\!_G}$ in the interior of the face $f$ is the virtual edge added to close a cycle, i.e., $P=\SG \setminus\{e_{\!_G}\}$. Note, since $e_{\!_G}$ is not an edge of $G$, it does not exist in $G^*$, and $f$ corresponds to a single node in $G^*$ instead of two. Thus, removing the separator vertices in $G$ disconnects $G$, however, removing its real (non-virtual) edges in $G^*$ does not disconnect it. 
    \label{fig: simplified_decomposition_problem_cut}}
\end{figure}

\medskip
\noindent
\textbf{Challenge I: dual subgraphs.}  There are three complications in this regard: (1) Taking the natural approach and defining the dual decomposition to contain the dual subgraph of each subgraph in the primal decomposition, does not work. This is because in the child subgraphs there are faces that do not exist in their parent subgraph (e.g. in \cref{fig: simplified_decomposition}, the "external" face of black edges in $G_1$).
More severely, the cycle $\SG$ may not even be a cycle of $G$, rather, a path $P$ accompanied with a virtual edge. Thus, $P$ is not necessarily a cut in $G^*$. 
(2) The second complication arises from the distributed hierarchical decomposition of~\cite{LP19} which may decompose each subgraph into as many as $\tilde{O}(D)$ subgraphs rather than two. This is problematic in the case where a given face (dual node) is split between these subgraphs.
(3) Finally, the useful property that a parent subgraph in the primal decomposition is given by the union of its child subgraphs is simply not true in the dual, since removing $P$ does not necessarily disconnect $G^*$ and may leave it connected by the face that contains the virtual edge (see \cref{fig: simplified_decomposition_problem_cut}). 
In order to disconnect $G^*$, that face must be split, which solves one problem but introduces another as explained next.

\medskip
\noindent
\textbf{Challenge II: face-parts.} In the primal graph, a vertex is an atomic unit, which keeps its identity throughout the computation. The situation in the dual graph is considerably more involved. Consider a primal graph $G$ with a large face $f$ containing $\Theta(n)$ edges. Throughout the recursive decomposition, the vertices of the face $f$ are split among multiple faces, denoted hereafter as \emph{face-parts}, and eventually $f$ is shattered among possibly a linear number of subgraphs. This means that a node in a dual subgraph no longer corresponds to a face $f$ of the primal subgraph, but rather to a subset of edges of $f$.
This creates a challenge in the divide-and-conquer computation, where one needs to assemble fragments of information from multiple subgraphs.

\medskip
\noindent
\textbf{New Structural Properties of the Primal BDD.}
We mitigate these technical difficulties by
characterizing the way that faces are partitioned during the primal BDD algorithm of~\cite{LP19}. While we run the primal BDD almost as is, our arguments analyze the primal procedure from a dual lens. 
Denote each subgraph in the BDD as a {\em bag}.
We use the primal BDD to define a suitable decomposition of $G^*$ that is more convenient to work with when performing computation on the dual graph. Interestingly, for various delicate reasons, in our dual BDD, the dual bag $X^*$ of a primal bag $X$ in the BDD is not necessarily the dual graph of $X$.  Finally, we show that each vertex in the primal graph can  
acquire the local distributed knowledge of this dual decomposition. 

Our dual perspective on the primal BDD allows us to provide a suitable labeling scheme for $G^*$. Next we give a more concrete description of our dual-based analysis of the primal BDD. The full details are provided in \cref{section_BDD}.

\medskip
\noindent
\textbf{Few face-parts.}
We show that in each bag $X$ of the BDD there is at most one face of $G$ that can be partitioned between different child bags of $X$ and was not partitioned in previous levels. This face is exactly the {\em critical} face $f$ that contains the virtual edge of the bag. We call the different parts of $f$ that appear in different child bags \emph{face-parts}. Since the decomposition has $O(\log{n})$ levels, overall we have at most $O(\log{n})$ face-parts in each bag. Note that we do not count face-parts that were obtained by splitting the critical face in ancestor bags (i.e., by splitting existing face-parts).  
For full details, the reader is referred to \cref{section_few_face_parts}.

\begin{figure}[htb]
\centering
\begin{subfigure}{0.32\textwidth}
\includegraphics[width=1\linewidth, height=1\linewidth]{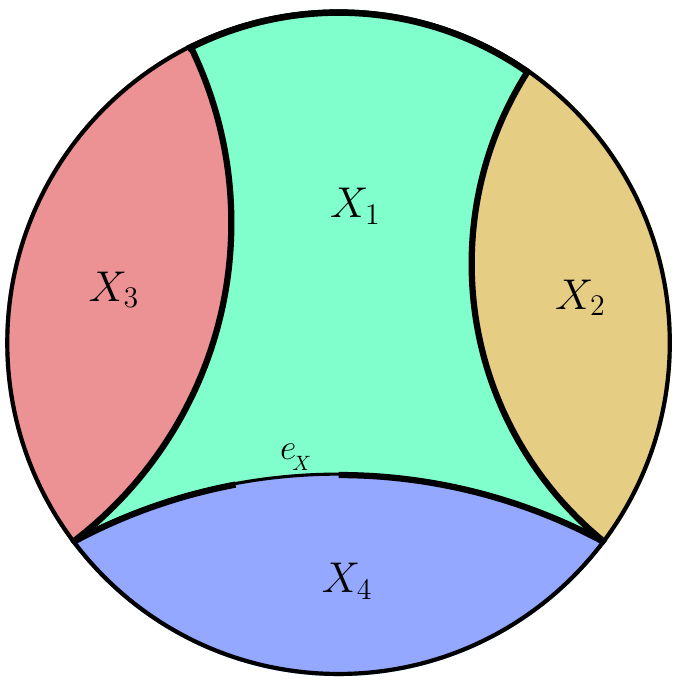} 
\end{subfigure}
\begin{subfigure}{0.32\textwidth}
\includegraphics[width=1\linewidth, height=1\linewidth]{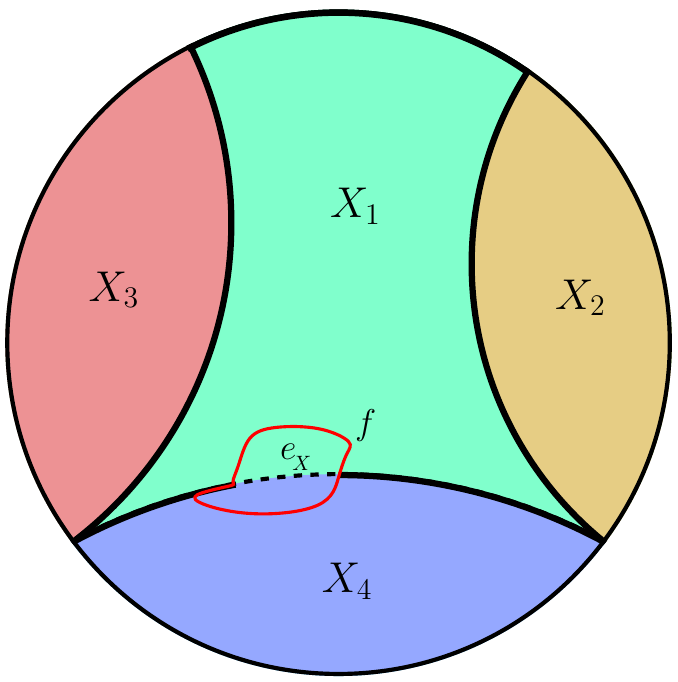}
\end{subfigure}
\begin{subfigure}{0.32\textwidth}
\includegraphics[width=1\linewidth, height=1\linewidth]{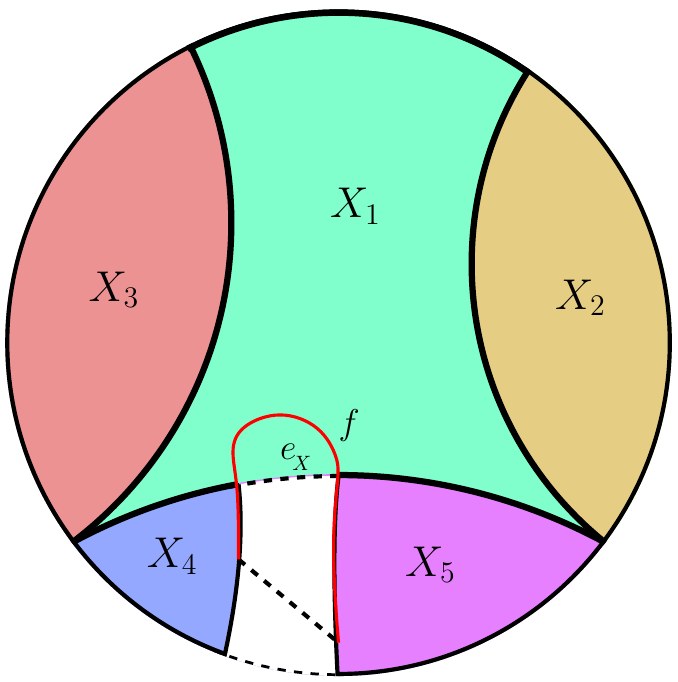}
\end{subfigure}
   \caption{
   Three examples of a bag $X$ and its child bags $X_i$ (of distinct colors). The bold black cycle is the separator $\SX$. The virtual edge $\eX$ is dashed. 
   The green child bag $X_1$ is the interior of $\SX$. 
    When $\eX \in E(G)$ (left image), we show that no face is partitioned between the child bags $X_i$. 
    When $\eX \notin E(G)$ (middle and right images), edges of the critical face $f$ are red.
    Notice, the face $f$ (middle) is partitioned into two child bags ($X_1$ and $X_4$), and the face-part $f$ (right) is further partitioned into three child bags ($X_1,X_4,X_5$); I.e., in the right image, $f$ was a face in a previous levels that got partitioned into face-parts, in particular, each dashed edge $\eH\neq \eX$ that appears in the empty white subgraph is an edge of a separator $\SH$ from an ancestor bag $H$ of $X$, such that $\eH$ had partitioned $f$ into parts. 
    }
    \label{fig: BDD_overview}
\end{figure}

\begin{lemma}[Few face-parts, Informal]
\label{lemma_informal_few_face_parts}
        Any bag $X$ of the BDD contains at most  $O(\log n)$  face-parts. 
\end{lemma}

To prove \cref{lemma_informal_few_face_parts}, we consider the separator $\SX$ of $X$ computed in the construction of the BDD by~\cite{LP19}. $\SX$ constitutes a separating cycle in $X$ containing at most one virtual edge $\eX$ that might not be in $E(G)$. The interior of $\SX$  defines one child bag $X_1$ of $X$, and its exterior may define as many as $\tilde{O}(D)$ child bags $X_2,X_3,\ldots$ (see \cref{fig: BDD_overview}). We prove that the face containing the endpoints of $\eX$ is the only face that might get partitioned in the bag $X$. First, we show that if $\eX \in E(G)$ then no face is partitioned, and every face is entirely contained in one of the child bags.
If $\eX \notin E(G)$, then we show that the critical face is the only face of $G$ that can be partitioned in $X$. We remark that face-parts can also be further partitioned, as shown in \cref{fig: BDD_overview}. 

\medskip
\noindent
\textbf{Dual bags and separators.}
Since we are interested in faces of $G$, we trace them down the decomposition and take face-parts into consideration when defining the dual bags. 
That is, both the faces and the face-parts that belong to a bag $X$ have corresponding dual nodes in $X^*$. We connect two nodes of $X^*$ by a dual edge only if their corresponding faces share a primal (non-virtual) edge in $X$. If $X=G$, then there are no face-parts, and $X^*$ is the standard dual graph $G^*$. For full details, the reader is referred to \cref{section_BDD}.

\begin{figure}[htb]
\centering

\begin{subfigure}{0.37\textwidth}
\includegraphics[width=1\linewidth, height=1\linewidth]{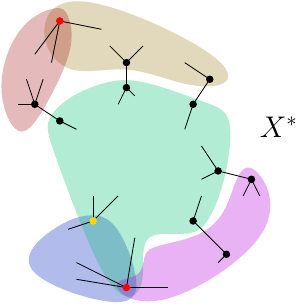} 
\end{subfigure}
\hspace{1.5 cm}
\begin{subfigure}{0.37\textwidth}
\includegraphics[width=1\linewidth, height=1\linewidth]{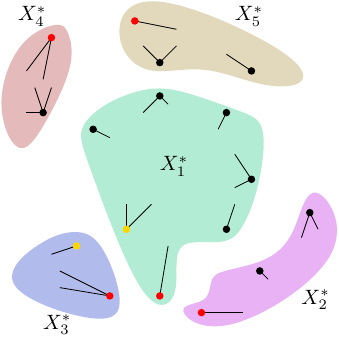}
\end{subfigure}
\caption{A bag $X^*$ on the left and its child bags on the right. Each shadow color demonstrates a distinct child bag of $X^*$. Each edge that is shadowed by a color, is entirely contained in the child bag corresponding to that color. The nodes shown in $X^*$ are the nodes of $\FX$. Some of these nodes are partitioned into several nodes in different child bags of $X^*$.
The black nodes are nodes of $X^*$ that are incident to a (dual) edge of $\SX$, the yellow node corresponds to the critical face of $X$, and the red nodes correspond to face-parts of $X$.
Notice how for the face-parts and the critical face of $X$ there are incident edges, each is entirely contained in a distinct child bag of $X^*$. 
Edges that are not contained in any child bag of $X^*$ are (dual) edges of $\SX$. 
} 
\label{fig: Dual_BDD} 
\end{figure}

As mentioned earlier, we would like the separator of a bag $X$ to constitute a separator for $X^*$. We show that this is almost the case. In particular, 
we define a set $\FX$ of size $\tilde{O}(D)$ that constitutes a separator in $X^*$. The set $\FX$ consists of (1) nodes incident to dual edges of $\SX$ in $X^*$, and (2) nodes corresponding to faces or face-parts that are partitioned between child bags of $X^*$. See \cref{fig: Dual_BDD}.  

 Following the high-level idea of the labeling scheme, the set $\FX$  is crucial.
The main structural property it provides is the following.
\begin{lemma}[Dual Separator - Informal]
\label{lemma_FX_informal}
    Any path in $X^*$ that is not entirely contained in a single child bag of $X^*$ must intersect $\FX$.
\end{lemma}
The above lemma is proved in \cref{section_dual_bags}. Intuitively, when $X^*$ is split into its child bags, some of its nodes are "shattered" (those that correspond to its critical face or its $O(\log n)$ face-parts), and some of its edges are removed (dual $\SX$ edges). We show how to assemble $X^*$ from its child bags by considering faces and face-parts of child bags that correspond to nodes of $\FX$.

\medskip
\noindent
{\bf Distributed knowledge.} 
A node $f$ in $X^*$ is simulated by the vertices of the corresponding face or face-part $f$ of $X$, and each dual edge adjacent to $f$ will be known by one of its endpoints. 
In particular, we assign each face and face-part in a bag $X$ a unique $\tilde{O}(1)$-bit identifier and learn for each vertex $v\in X$, the set of faces and face-parts containing it and its adjacent edges.
This is done by keeping track of $G$'s edges along the decomposition. At first we learn $G^*$ and then extend this recursively to its child bags. The implementation relies on the fact that we have only a few face-parts in each bag (\cref{lemma_informal_few_face_parts}) preventing high congestion when we broadcast information about  face-parts to learn the dual nodes. For details see \cref{section_distributed_knowledge}.

\medskip
\noindent
\textbf{The labeling scheme.} 
Our next goal is to use the decomposition in order to compute distances in the dual graph. 
 More concretely, we (distributively) compute, for every node in every bag $X^*$, a short label of size $\tilde{O}(D)$, such that given only the labels of any two nodes in $X^*$ we can deduce their distance. 
The labeling scheme is a refinement of the intuition given earlier. For full details, the reader is referred to \cref{section_distance_labels}.

Recall, the set $\FX$ plays the role of a node separator of $X^*$ (\cref{lemma_FX_informal}). 
A node $g\notin \FX$ of $X^*$ corresponds to a real face of $G$ (i.e. not a face-part) contained in $X$. The distance label $\Label_{X^*}(g)$ of $g\notin \FX$  is defined recursively.
If $X^*$ is a leaf in the decomposition then $\Label_{X^*}(g)$ stores ID($g$) and the distances between $g$ and all other nodes $h\in X^*$. Otherwise, the label consists of the ID of $X$\footnote{In the BDD of~\cite{LP19}, each bag is assigned a $O(\log n$)-bit identifier.}, the distances in $X^*$ between $g$ and all nodes of $\FX$, and (recursively) the label of $g$ in the child bag of $X^*$ that entirely contains $g$.
In case $g\in \FX$, we simply store the distances between $g$ and all other nodes in $\FX$.
The labels are of size $\tilde{O}(D)$-bits since $|\FX|=\tilde{O}(D)$ and since the BDD is of a logarithmic height.

The correctness follows from \cref{lemma_FX_informal}.
I.e, a $g$-to-$h$ shortest path $P$ in $X^*$ is either: (1) Entirely contained in $X^*_i$, and $\dist_{X^*}(g,h)$ can be deduced from $\Label_{X^*_i}(g)$ and $\Label_{X^*_i}(h)$, or (2) $P$ intersects $\FX$ with an edge, then $\dist_{X^*}(g,h) = \min_{f\in \FX} \{\dist_{X^*}(g,f)+\dist_{X^*}(f,h)\}$. 
If one of $g$ or $h$ is a node of $\FX$ then distances are retrieved instantly, as each node in $X^*$ stores its distances to $\FX$.

\medskip
\noindent
\textbf{The algorithm.}
We compute the distance labels bottom-up on the decomposition. In the leaf bags, due to their small size ($\tilde{O}(D)$-bits) we allow ourselves to gather the entire bag in each of its vertices. Then, vertices locally compute the labels of faces and face-parts that contain them. If a negative cycle is detected, it gets reported via a global broadcast on $G$ and the algorithm terminates.

\begin{figure}[htb]
    \centering
    \includegraphics[width=0.5\linewidth]{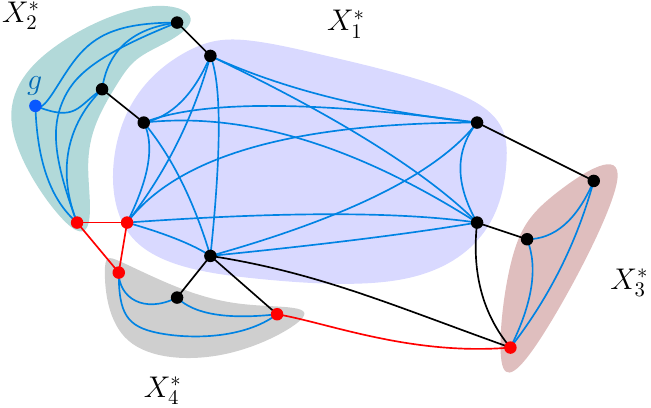}
    \caption{
    The graph $DDG(g)$. The blue node is $g$, other nodes are nodes of $\FX$, in particular, red nodes correspond to face-parts and black nodes to endpoints of $\SX$.
    The blue edges represent distances between nodes that belong to the same child bag of $X^*$, red edges connect face-parts of the same face in $X^*$, and black edges are dual edges to $\SX$ edges. 
     For simplicity, the figure is undirected.
    }
    \label{fig: overview_DDG}
\end{figure}

In non-leaf bags, we broadcast $\tilde{O}(D^2)$ information that includes labels of nodes in $\FX$ in the child bags ($\tilde{O}(D)$ labels, each of size $\tilde{O}(D)$), and the edges of the separator. Since all the bags in the same level have $\tilde{O}(D)$ diameter and are nearly disjoint, we can broadcast all the information in $\tilde{O}(D^2)$ rounds in all bags.  We prove that based on this information nodes can  locally deduce their distance label in $X^*$. 

In particular, each vertex $v\in X$ that belongs to a face or a face-part $g$ of $X$ constructs locally a {\em Dense Distance Graph} denoted $DDG(g)$.  $DDG(g)$ is a small (non-planar) graph (with $\tilde{O}(D)$ vertices and $\tilde{O}(D^2)$ edges) that preserves the distances in $X^*$ between pairs of nodes in $\FX \cup \{g\}$.
See \cref{fig: overview_DDG}. 
Via (local) APSP computations on the DDGs, $v$ learns the distances between $g$ and $\FX$ nodes in $X^*$ (i.e., constructs $\Label_{X^*}(g)$). Again, vertices check for negative cycles locally (in the DDG).
To prove correctness, we show that each shortest path $P$ in $X^*$ can be decomposed into a set of subpaths and edges whose endpoints are in $\FX \cup \{g\}$. Each such subpath is entirely contained in a child bag of $X^*$ (so its weight can be retrieved from the already computed labels) and each such edge is a (dual) edge of $\SX$ (whose identity and weight were broadcast). For full details, the reader is referred to \cref{section_dual_labeling_algorithm}.

Finally, in order to compute distances from any given node $s\in G^*$, it suffices to broadcast its label to the entire graph. Thus, vertices can learn locally for each face of $G$ that contains them the distance from $s$. To learn an SSSP tree from $s$ in $G^*$, for each node $g$ in $G^*$ we mark its incident edge $e=(f,g)$ that minimizes $\dist(s,f)+w(e)$. 
For this task, we use our Minor-Aggregation implementation on $G^*$.  Concluding our main result, a labeling scheme that gives dual SSSP and hence primal maximum $st$-flow.

 \subsection{Flow Assignments and Cuts}

\medskip
\noindent
{\bf Maximum $st$-flow.}
It was shown in the 90's by Miller and Naor~\cite{MN95}, building on a result of Venkatesan~\cite{Venkatesan}, that an exact $st$-flow algorithm on a directed planar graph $G$ with positive edge-capacities can be obtained by $\log \lambda$ applications of a SSSP algorithm on the dual graph $G^*$ with positive and negative edge-weighs (each application with possibly different edge weights). 
We use this result as a black-box. At the end, we obtain the flow value and the flow assignment to the edges of $G$. This works in $\tilde{O}(D^2)$ rounds since $\lambda$ (the value of the maximum $st$-flow) is polynomial in $n$, and since our dual SSSP algorithm terminates within $\tilde{O}(D^2)$ rounds. 

\medskip
\noindent
{\bf Approximate maximum $st$-flow.}
For the case of undirected planar graphs where $s$ and $t$ lie on the same face with respect to the given embedding, we further obtain an improved round complexity at the cost of solving the problem approximately. Namely, it was shown in the 80's by Hassin~\cite{Hassin} that in this setting it suffices to use an SSSP algorithm on the dual graph after augmenting it with a single edge, dual to the edge $(s,t)$, that is limited to positive edge-weights and to undirected graphs. Since $s$ and $t$ are on the same face it is possible to add such an edge while preserving planarity. Note that adding this edge splits a node in the dual graph. We show that we can still simulate efficently minor-aggregation algorithms on the augmented dual graph. To compute SSSP on that graph we use an SSSP algorithm by~\cite{GHSYZ22} that works in the minor-aggregation model. In particular, it computes a $(1+o(1))$-approximate SSSP tree within $D\cdot n^{o(1)}$ rounds, which allows us to find a $(1-o(1))$-approximate flow value in the same time bound.

\medskip
\noindent
{\bf Finding the flow assignment.}
As mentioned above, our exact maximum $st$-flow algorithm also gives a flow assignment.
However, in the approximate case,
our algorithm easily gives the approximate flow value, but it is not guaranteed to give a corresponding flow assignment due to the fact that the reduction we use by~\cite{Hassin} is meant to be used with an exact algorithm. 
In particular, to have an assignment, the outputted SSSP tree needs to (approximately) satisfy the triangle inequality. However, most distance approximation algorithms do not satisfy that quality.
To overcome this, we apply a method of~\cite{RozhonHMGZ23}, that given any approximate SSSP algorithm
produces an SSSP tree that (approximately) satisfies the triangle inequality. To obtain an assignment, by~\cite{RozhonHMGZ23}, we have to apply the approximate SSSP algorithm $\tilde{O}(1)$ times each on different virtual graphs related to the dual graph.
This is not immediate, and raises several challenges. Most importantly, their algorithm was not implemented in the minor-aggregation model so it is not immediate to simulate it on the dual graph.
We exploit the features of the minor-aggregation model excessively in order to resolve this complication.

\medskip
\noindent
{\bf Minimum $st$-cut.}
By the celebrated min-max theorem of Ford and Fulkerson~\cite{ff56}, the exact (or approximate) value of the max $st$-flow is equivalent to that of the min $st$-cut. 
We show that our algorithms can also be extended to compute a corresponding bisection and mark the cut edges of the min $st$-cut. 
In the exact case, we use classic textbook methods (e.g, residual graphs), and since this flow algorithm works for directed planar graphs, we obtain a {\em directed} minimum $st$-cut. In the approximate case, we have to work harder and exploit the minor-aggregation model on the dual graph one last time, in which, we apply a special argument on the cycle-cut duality that was given in the 80's by Reif~\cite{Reif83}.

\medskip
\noindent
{\bf Directed global minimum cut.} We exploit our techniques to also find the directed global minimum cut.
The general idea is simple, we use the cycle-cut duality: finding the (directed) minimum cut in $G$ amounts to finding the minimum weight (directed) simple cycle $C$ in $G^*$. Thus,
given an extended BDD of $G$ and having distance labels already computed in $\tilde{O}(D^2)$ rounds for all dual bags in the BDD, we have that any shortest cycle of $G^*$ either intersects a dual separator node or is entirely contained in one of $G^*$'s child bags. Thus, in order to find $C$, we take the minimum over all: (i) directed cycles that intersect separator nodes of $G^*$ (deduced by pairwise decoding labels of the separator nodes), and (ii) minimum directed cycles in $G^*$'s child bags (found recursively).
However, we note that the duality does not hold in general directed planar graphs, which complicates the above approach. I.e., in order for the duality to hold,  we have to work with a graph where for each edge, its {\em reversal dart} (an edge in the opposite direction) is present, so we add those darts. Then, we should ensure that the cycles we consider are simple darts cycles (i.e. a cycle where no edge and its reversal dart appear together), which we do by investigating two options related to the dual separator. 
Finding the cut bisection is done simply by detecting connected components (in the undirected sense) after the removal of the cut edges and their reversal darts.

\subsection{Organization, Summary and Open Questions}
The rest of the paper is organized as follows. \cref{section: preliminaries} discusses preliminaries. \cref{sec: dual_MA_model} gives the proofs for our minor-aggregation simulation on the dual graph and its applications for the weighted girth.
Our labeling scheme for the dual graph appears in \cref{sec: dual SSSP}, and our applications for max $st$-flow, min $st$-cut appear in \cref{section_flow_cut},
while the directed global min-cut appears in \cref{section_directed_global_mincut}. 
See also \cref{fig: roadmap}. 

\begin{figure}[htb]
	\centering
	\begin{tikzpicture}
	[
	every node/.style={
		align = center,
		fill          =  black!5,
		inner sep     = 4pt,
		minimum width = 18pt,
		rounded corners=.1cm}
	]
	
	\node (MA) at (-4,9) {Dual Minor-Aggregation \\ 
	(\cref{sec: dual_MA_model})};

        \node (girth) at (-6,7) {Undirected Weighted Girth \\ 
	(\cref{thm:weighted-girth})};

        \node (approxSSSP) at (-2.5,5.5) {Approximate Dual SSSP};

        \node (BDD) at (4,9) {Extended BDD \\ 
	(\cref{section_BDD})};

        \node (labeling) at (4,7) {Dual Distance Labeling \\ 
        (\cref{section_distance_labels})};

        \node (SSSP) at (2,5.3) {Exact Dual SSSP\\ 
        (\cref{section_dual_labeling_algorithm})};

        \node (flow) at (-2,4) {Primal Max $st$-Flow\\ 
        (\cref{section_flow})};

        \node (globalcut) at (4,3) {Primal Directed Global Min-Cut\\ 
        (\cref{section_directed_global_mincut})};

        \node (cut) at (-2,2.3) {Primal Min $st$-Cut\\ 
        (\cref{section_st_cut})};

	 \draw[->, line width=1.5] (MA) -- (girth);
      \draw[->, line width=1.5] (MA) -- (approxSSSP);
      \draw[->, line width=1.5] (approxSSSP) -- (flow);

      \draw[->, line width=1.5] (BDD) -- (labeling);
      \draw[->, line width=1.5] (labeling) -- (SSSP);
      \draw[->, line width=1.5] (labeling) -- (globalcut);
      \draw[->, line width=1.5] (SSSP) -- (flow);
      \draw[->, line width=1.5] (flow) -- (cut);

	
	
	\end{tikzpicture}

	\caption{A roadmap of the paper.}
	\label{fig: roadmap}
\end{figure}

Next, we summarize the techniques by their advantages and limitations, then, we conclude with an interesting open question. 

\paragraph{Minor-aggregation.}
In general, the minor aggregation technique is useful when dealing with undirected graphs and especially when we want to simulate some related graph to the graph of input, which includes a small number (usually $\tilde{O}(1)$) virtual nodes. I.e., nodes that all vertices of $G$ simulate. 
Albeit the dual graph $G^*$ is entirely virtual, we extend previous work~\cite{GP17} and show that the round complexity of simulating a minor aggregation round on it matches the near-optimal bound~\cite{GHLRZ22} known for $G$, which allows then to add up to $\tilde{O}(1)$ virtual nodes to the dual graph that are not necessarily related to the faces of $G$.
Then, applying black-box algorithms on $G^*$ after manipulating its topology (adding/ replacing nodes with virtual nodes), constitutes a distributed implementation of classic centralized reductions, some of which are immediate (e.g. min-cut in $G^*$ gives the girth of $G$) and some of which require more work (e.g. shortest paths with positive edge-lengths in $G^*$ implies maximum $st$-flow in $G$ when $s,t$ lie on the same face).

This closes the problem of minor-aggregation on the dual and expands the algorithmic design toolkit to include all the very useful efficient minor-aggregation procedures (e.g. subtree sums, heavy-light decompositions~\cite{GZ22}, and more) and even complicated algorithms (e.g. min-cut~\cite{GZ22}, approximate undirected shortest path~\cite{GHSYZ22}) for the dual.

\paragraph{Recursive decompositions.}
Our recursive decomposition based on the BDD of~\cite{LP19} provides a framework to divide-and-conquer in $G^*$ making this approach feasible albeit the difficulties encountered.
In this scenario, usually, one solves the problem in a brute-force manner on the leaf-bags then broadcasts some information to the parent bags in order to assemble a solution for the parent bag, and so on, bubbling up to the root of the decomposition $G^*$. The most obvious example is the SSSP algorithm (\cref{sec: dual SSSP}) we give to compute maximum $st$-flow in $G$. An additional example is given in \cref{section_directed_global_mincut}, where we show how to use the distance labels computed in $G^*$ in a non-trivial manner to compute its minimal weight (directed) cycle, which results with the global directed minimum cut of $G$.

Having a recursive separator decomposition for $G^*$, opens the question also to other kinds of graph decompositions known to the primal graph with efficient distributed constructions (e.g. shortest-path separator recursive decomposition~\cite{LP19}).
Although the dual decomposition itself can be constructed and distributively learned in near-optimal time, the barrier for a better maximum flow algorithm lies within the dual SSSP algorithm, which is discussed next.

\paragraph{Open questions.}
Since the round complexity of the algorithm that we devise for directed single source shortest paths with positive and negative edge-lengths for $G^*$ matches the state-of-the-art for $G$, then,
perhaps the most obvious question is to obtain better SSSP algorithms for the primal graph $G$ (where we do not have the challenges of simulating an entirely virtual graph as $G^*$).
I.e., in $G$, one should probably first improve the $\tilde{O}(D^2)$ bound of the (easier) directed SSSP problem (even with non-negative weights) before trying to achieve such an improvement for (the harder) maximum $st$-flow problem.

It seems that this is the bottleneck to overcome in order to obtain better distributed planar algorithms for various optimization problems, since most such problems in $G$ reduce to computing distances either in $G$ or in $G^*$.

It is intuitive to believe that an improvement upon the bound of $\tilde{O}(D^2)$ is possible, because every step in the SSSP algorithm is near-optimal except when we simply broadcast the labels of all separator vertices ($\tilde{\Theta}(D^2)$ bits), if one can compress those distances to $\tilde{o}(D^2)$ bits or manage to compute SSSP in a better round complexity via a different approach, then likely better algorithms for many planar optimization problems would follow.
An indication that this is possible was given by Parter~\cite{ParterReachability20a}, where she shows near-optimal algorithms for planar {\em reachability} problems using the BDD, by compressing the reachability labels of all separator vertices into $\tilde{O}(D)$ bits before the broadcast step.

\section{Preliminaries}
\label{section: preliminaries}

We denote by $G=(V,E)$ the directed (possibly weighted) simple planar network of communication, and by $D$ the network's undirected and unweighted (hop) diameter. 
Let $S$ be a set of vertices or edges, we denote by $G[S]$ the  subgraph of $G$ induced by $S$.

\medskip
\noindent
{\bf The \boldmath$\Congest$ model.}
We work in the standard distributed $\Congest$ model~\cite{peleg-book}. Initially, each vertex knows only its unique $O(\log n)$-bit identifier and the identifiers  of its neighbors. Communication occurs in synchronous rounds. 
In each round, each vertex can send (and receive) an $O(\log n)$-bit message on each of its incident edges (different edges can transmit different messages). When the edges of $G$ are weighted we assume that the weights are polynomially bounded integers.
Thus, the weight of an edge can be transmitted in $O(1)$ rounds. This is a standard assumption in the $\Congest$ model.

\medskip
\noindent
{\bf Distributed storage.}
When referring to a distributed algorithm that solves some problem on a network $G$, the input (and later the output) is stored distributively. That is, each vertex knows only a small (local) part of the information. E.g. when computing a shortest paths tree from a single source vertex $s$, we assume that all vertices know (as an input) the ID of $s$ and the weights and direction of their incident edges. When the algorithm halts, each vertex knows its distance from $s$ and which of its incident edges are in the shortest paths tree.

\medskip
\noindent
{\bf Planar embedding.}
Let $G = (V, E)$ be a directed planar graph.
The {\em geometric} planar embedding of $G$ is a drawing of $G$ on a plane so that edges intersect only in vertices.
A {\em combinatorial} planar embedding of $G$ provides for each $v\in G$, the local clockwise order of its incident edges, such that, the ordering of edges is consistent with some geometric planar embedding of $G$.
For a detailed discussion, the reader is referred to~\cite{KM}.
Throughout, we assume that a combinatorial embedding of $G$ is known locally for each vertex. This is done in $\tilde{O}(D)$ rounds using the planar embedding algorithm of Ghaffari and Haeupler~\cite{GH16a}.

\begin{figure}[htb]
 \centering
  \includegraphics[width=0.35\linewidth]{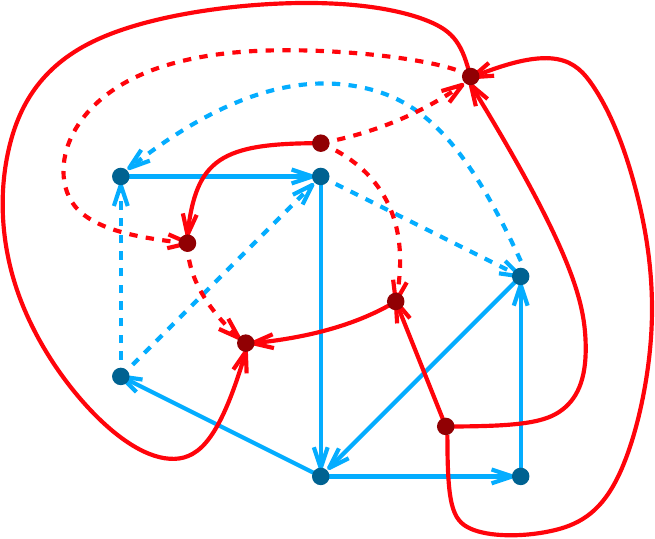}
  \caption{The graph $G$ in blue and its dual graph $G^*$ in red. A primal cycle and its dual cut are dashed.  \label{fig:dual_graph}}
 \end{figure}

\medskip
\noindent
{\bf The dual graph \boldmath$G^*$.}
\label{def: dual_graph}
The {\em dual} of the {\em primal} planar graph $G$ is a planar graph, denoted $G^*$. The nodes of $G^*$ correspond to the faces of $G$. 
The dual graph has an edge $e^*$ for each pair of faces in $G$ that share an edge $e\in G$ (and a self-loop when the same face appears on both sides of an edge). If $e$ is directed then the direction of $e^*$ is from the face on the left of $e$ to the face on the right of $e$ where left and right are defined with respect to the direction of $e$. Observe that if two faces of $G$ share several edges then $G^*$ has several parallel edges between these faces. I.e., $G^*$ might be a multi-graph even when the primal graph $G$ is a simple graph. See \cref{fig:dual_graph}.
Since the mapping between edges of $G$ and $G^*$ is a bijection, we sometimes abuse notation and refer to both $e$ and $e^*$ as $e$. 
We refer to the vertices of $G$ as {\em vertices} and to the vertices of $G^*$ as {\em nodes}. 

We use the well-known duality between primal cycles and dual cuts:
\begin{fact}
    [Undirected Cycle-Cut Duality~\cite{KM}]
    \label{fac: cycle_cut_duality}
   A set of edges $C$ is a simple cycle in an embedded connected planar graph $G$ if and only if $C$ is a simple cut in $G^*$ (a cut $C$ is said to be simple if when removed, the resulting graph has exactly two connected components).
\end{fact}

\medskip
\noindent
{\bf The face-disjoint graph \boldmath$\hat{G}$.} \label{sec: face_disjoint_graph}
In the distributed setting, the computational entities are vertices and not faces. Since a vertex can belong to many (possibly a linear number of) faces of $G$, this raises several challenges for simulating computations over the dual graph $G^*$: (1) We want to communicate on $G^*$ but the communication network is $G$.  
(2) $G^*$ might be of a large diameter (up to $\Omega(n)$ even when the diameter of $G$ is $D=O(1)$). 
 (3) We wish to compute aggregate functions on sets of faces of $G$. To do so efficiently we want faces to be vertex-disjoint.

\begin{figure}[htb]
    \centering
    \includegraphics[width=0.65\linewidth]{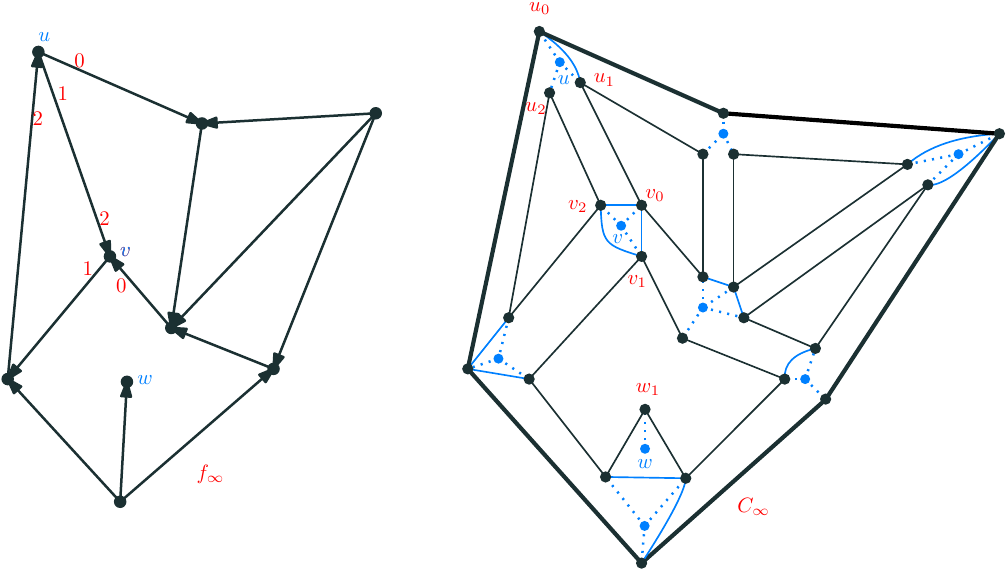}
    \caption{On the left, the primal graph $G$. The clockwise local numbering of $u$ and $v$'s edges is in red. 
    $f_\infty$ denotes the infinite face of $G$. 
    On the right, the face-disjoint graph $\hat{G}$. The face $f_\infty$ of $G$ maps to the face $C_\infty$ (thick black) of $\hat{G}[E_R]$.
    Vertex $v$ in $G$ has a star-center copy $v$ in $V_s\subset V(\hat{G})$, and $\deg(v)$ additional duplicates $v_0,v_1,v_2$ in $V(\hat{G})$ connected by star edges $E_S \subset E(\hat{G})$ (dashed blue).
    The edge $(u,v)$ of $G$ maps to three edges in $\hat{G}$: an edge $(v_0,v_2)$ in $E_C \subset E(\hat{G})$ (solid blue) and two edges $(u_2,v_2)$ and $(u_1,v_0)$ in $E_R\subset E(\hat{G})$ (black edges).
    The vertex $w\in G$ is of degree one, thus, it has a star-center copy and one additional copy in $\hat{G}$ connected by a dashed blue edge. Its only incident edge in $G$ has two duplicates, each connected to a distinct non star-center copy of its only neighbor. The non star-center copies of $w$'s neighbor are connected by a solid blue edge (corresponding to the edge incident to $w$), this edge corresponds to a self loop in $G^*$ as it connects the same face to itself (the face containing $w$).
    \label{fig: face_disjoint_graph}}
\end{figure}

To overcome these problems, the {\em face disjoint graph} $\hat{G}$ was presented in \cite{GP17} as a way for simulating aggregations on the dual graph $G^*$ in the distributed setting, which we slightly modify here. 
As it is mainly a communication tool, we think of $\hat{G}$ as an undirected unweighted graph, however, it will still allow working with weighted and directed input graphs $G$, as each endpoint of an edge in $\hat{G}$ that represents an edge of $G^*$, shall know the weight and direction of that edge.
Intuitively, $\hat{G}$ can be thought of as the result of duplicating all edges of $G$ so that faces of $G$ map to distinct faces of $\hat{G}$ that are both vertex and edge disjoint. See \cref{fig: face_disjoint_graph}.

 We next give the full definition of the graph $\hat{G}$.
Let $\deg(v)$ denote the undirected degree of a vertex $v$ in $G$. Then, a vertex $v$ in $G$ has $\deg(v)+1$ copies in $\hat{G}$: A copy $v$ referred to as a {\em star center} (we denote by $V_S=V(G)$ the set of all star centers) and a copy $v_i$ for each {\em local region} of $v$. A local region of $v$ is defined by two consecutive edges in $v$'s local embedding. Notice that $v$ has $\deg(v)$ local regions, and that each local region belongs to a face of $G$ (it is possible that several local regions of $v$ belong to the same face). 
Eventually, we will have that the faces of $G$  correspond to (vertex and edge disjoint) faces of $\hat{G}$. If $i$ local regions of $v$ belong to the same face $f$ of $G$, then $v$ will have $i$ copies in the face of $\hat{G}$ that corresponds to $f$. 
The vertices of $\hat{G}$ are therefore $V_S \bigcup (\cup_{v\in G}N(v))$ where $N(v)=\{v_1,\ldots, v_{\deg(v)}\}$. 

We now define the edges of $\hat{G}$ in three parts $E(\hat{G})=E_S\cup E_R \cup E_C$. The {\em star edges} $E_S$ are of the form $(v,v_i)$, the edges $E_R$ are of the form $(v_i,u_j)$ (for $u\neq v$) and the edges $E_C$ are of the form $(v_i,v_j)$ (for $i\neq j$). For every edge $e=(u,v)$ we need endpoints of $e$ to figure out which edges $(v_i,u_j)$ belong to $\hat{G}$ so that $\hat{G}$ is planar and faces of $G$ map to faces of $\hat{G}$.
While a vertex of $G$ does not know any global information about the faces of $G$, adjacent vertices know that there is some face that contains both of them.

\begin{description}
\item[\boldmath$E_S$:]
These edges connect every star center $v$ and its copies $v_i\in N(v)$.

\item[\boldmath$E_R$:]
For each edge $(u,v)\in G$, there are two edges in $E_R$, one copy for each face of $G$ that contains $(u,v)$. Let $(u,v)$ be the $i$'th and $j$'th edge in the clockwise local embedding of $u$ and $v$, respectively.
Then the edges added to $E_R$ are $(u_{i+1},v_j)$ and $(u_{i},v_{j+1})$. This is because the face containing the $i$'th and $(i+1)$'th edges of $u$ ($i$'th local region of $u$) is the face containing the $j$'th and $(j-1)$'th edges of $v$ ($(j-1)$'th local region of $v$). Symmetrically, the face containing the $i$'th and $(i-1)$'th edges of $u$ is the  face containing the $j$'th and $(j+1)$'th edges of $v$.

\item[\boldmath$E_C$:]
In order to simulate computations on $G^*$, we make sure that if two faces of $G$ share an edge $e$ (i.e. are adjacent nodes in $G^*$) then their corresponding faces in $\hat{G}$ are connected by a corresponding edge in $E_C$, such an edge connects two vertices $v_j,v_{j+1}$ of the same star, because each of $v_j, v_{j+1}$ corresponds to a face that contains $e$.
For an edge $(u,v)$ we can add an edge in $E_C$ between copies of $u$ or copies of $v$, the choice of which such edge to add is arbitrary: 
For every edge $(u,v)$ in $G$ we add the edge $(v_j,v_{j+1})$ to $E_C$ if $v$'s ID is larger than $u$'s, and $(u,v)$ is the $j$'th edge in the clockwise local embedding of $v$. 
\end{description}

In \cite{GP17}, $\hat{G}$ was defined without the edges $E_C$, as their purposes did not require it. However, the following properties (except for Property \ref{property: hat(G)_mapping_to_E(G^*)}) of the face-disjoint graph were proven in \cite{GP17} and are yet valid. For proofs of all properties see \cref{appendix: preliminaries_hat{G}}.

\begin{enumerate}
    \item    
    \label{property: hat(G)_construction_representation}
    $\hat{G}$ is planar and can be constructed in $O(1)$ rounds (after which every vertex in $G$ knows the information of all its copies in $\hat{G}$ and their adjacent edges). 
    \item
    \label{property: hat(G)_diameter}
    $\hat{G}$ has diameter at most $3D$.
    \item
    \label{property: hat(G)_simulation}
    Any $r$-round algorithm on $\hat{G}$ can be simulated by a  $2r$-round  algorithm on  $G$.
    \item
    \label{property: hat(G)_face_identification}
    There is an $\tilde{O}(D)$-round algorithm that identifies $G$'s faces by detecting their corresponding faces in $\hat{G}$. 
    When the algorithm terminates, every such face of $\hat{G}$ is assigned a face leader that knows the face's ID. Finally, the vertices of $G$ know the IDs of all faces that contain them, and for each of their incident edges the two IDs of the faces that contain them.
    Thus, each vertex of $G$ knows for each pair of consecutive edges adjacent to it (using its clockwise ordering of edges) the ID of the face that contains them.
    \item
    \label{property: hat(G)_mapping_to_E(G^*)}
    There is a 1-1 mapping between edges of $G^*$ and a subset of edges $E_C$ of $\hat{G}$. Both endpoints of an edge in $E_C$ know the weight and direction of its corresponding edge in $G^*$ (if any). 
\end{enumerate}

\section{Weighted Girth via Minor Aggregations}
\label{sec: dual_MA_model}

We provide an implementation of the minor-aggregation model in the dual graph, which immediately yields a collection of algorithmic results in the primal graph.

\subsection{Preliminaries}

\medskip
\noindent
{\bf Low-congestion shortcuts and part-wise aggregation.}
In recent years, several breakthroughs in distributed computing were based on the notion of {\em low-congestion shortcuts} introduced by Ghaffari and Haeupler \cite{GH16b} (first for planar graphs, then for other graph families \cite{HIZ21, GH21}). Consider a partition $\{S_1,\ldots,S_N\}$ of $V$ where every $G[S_i]$ is connected.
\begin{definition}
    [Low-congestion shortcuts]
    An $\alpha$-congestion $\beta$-dilation shortcut of $G$, denoted $(\alpha, \beta)$-shortcut, is a set of subgraphs $\{H_1,\ldots,H_N\}$ of $G$ such that the diameter of each $G[S_i]\cup H_i$ is at most $O(\beta)$ and each edge of $G$ participates in at most $O(\alpha)$ subgraphs $H_i$.
\end{definition} 

\begin{definition}
    [Shortcut quality]
    The shortcut quality $SQ(G)$ of a graph $G$ is the smallest $\alpha$ for which $G$ admits an $(\alpha, \alpha)$ low-congestion shortcut.
    \end{definition} 

\begin{definition}
    [Aggregation operators]
      Given a set of $b$-bit strings $\{x_1,\ldots,x_m\}$ and an aggregation operator $\oplus$ (AND /OR /SUM etc.), their aggregate $\bigoplus_i x_i$ is defined as the result of repeatedly replacing any two strings $x_i, x_j$ with the ($b$-bit) string  $x_i \oplus x_j$, until a single string remains.
\end{definition} 

\begin{definition}
    [Part-wise aggregation]
    Assume each $v\in S_i$ has an $\tilde{O}(1)$-bit string $x_v$. The PA problem asks that each vertex of $S_i$ knows the aggregate function $\bigoplus_{v\in S_i} x_v$.
\end{definition}

\noindent
The following lemma captures the relation between the PA problem and shortcuts.

\begin{lemma}
    [Part-wise aggregation via low-congestion shortcuts \cite{GH16b, GH21, HIZ21}]
    \label{lem:PA}
    If $G$ admits an $(\alpha, \beta)$-shortcut that can be constructed in $r$ rounds, then the PA problem can be solved on $G$ in $\tilde{O}(r+\alpha+\beta)$ rounds. In particular, the PA problem can be solved on $G$ in $\tilde{O}(r+SQ(G))$ rounds, where $r$ is the number of rounds needed for constructing low-congestion shortcuts of quality $SQ(G)$.
\end{lemma}

\noindent Since a planar graph $G$ of hop-diameter $D$ admits low-congestion shortcuts of quality $SQ(G)=\tilde{O}(D)$ \cite{GH16b, GH21, HIZ21} that can be constructed in $\tilde{O}(D)$ rounds we get:

\begin{corollary}
\label{cor: planar_PA}
The PA problem can be solved on a planar graph $G$ of hop-diameter $D$ in $\tilde{O}(D)$ rounds.  
\end{corollary}

\paragraph{The minor-aggregation model.}
We define the minor-aggregation model as first defined by~\cite{GHSYZ22}. We mention that~\cite{GZ22} has defined an extended version of the model that supports {\em virtual nodes}. However, the definitions related to the {\em extended} model are delayed until we prove that we can simulate it as well, which would simply follow from simulating the  {\em basic} (original) model.

\begin{definition}
[Basic minor-aggregation model \cite{GHSYZ22}]
\label{def: basic_model}
Given an undirected graph $G=(V,E)$, both vertices and edges are computational entities (i.e. have their own processor and their own private memory). Each vertex has a unique $\tilde{O}(1)$-bit ID. Communication occurs in synchronous rounds. All entities wake up at the same time, each vertex knows its ID, and each edge knows its endpoints' IDs.
Each round of communication consists of the following three steps (in that order):

\begin{enumerate}
    \item 
    {\em Contraction step.}
    Each edge $e$ chooses a value in $\{{0},{1}\}$. Contracting all edges that chose ${1}$ defines a minor $G'=(V',E')$ of $G$. 
    The nodes $V'$ of $G'$ are subsets of $V$ called  {\em super-nodes}.
    
    \item 
    {\em Consensus step.}
    Each node $v\in V$ chooses an $\tilde{O}(1)$-bit value $x_v$. For each super-node $s\in V'$, let $y_s=\bigoplus_{v\in s} x_v$, where $\oplus$ is a pre-defined aggregation operator. All $v\in s$ learn $y_s$.
    
    \item
    {\em Aggregation step.}
    Each edge $e=(a,b)\in E'$ learns $y_a$ and $y_b$, then chooses two $\tilde{O}(1)$-bit values, $z_{e,a}$ and $z_{e,b}$. For each super-node $s\in V'$, let $I(s)$ be the set of incident edges of $s$, then,  $z_s=\bigotimes_{e\in I(s)} z_{e,s}$, where $\otimes$ is a pre-defined aggregation operator. All $v\in s$ learn the same $z_s$. 
    
\end{enumerate}
\end{definition}

It was shown by \cite{GHSYZ22} that using low-congestion shortcuts, any minor-aggregation algorithm can be simulated in the standard $\Congest$ model. It is easy to check that the consensus and aggregation steps of the minor-aggregation model can indeed be phrased as PA problems, and it was shown in \cite{GHSYZ22} that the contraction step can be broken to a series of $O(\log{n})$ PA problems.
More specifically, it was proved that any efficient ($\tilde{O}(1)$-round) minor-aggregation algorithm implies an efficient $\Congest$ algorithm on $G$ if one can solve the PA problem efficiently on $G$. Our goal is to show a similar result for the dual graph $G^*$.

\begin{lemma}[Minor-aggregation model simulation in $\Congest$ via PA \cite{GHSYZ22,GZ22}]
\label{lem: MA_via_PA}
Any $\tau$-round minor-aggregation algorithm can be simulated in the $\Congest$ model on $G$ in $\tilde{O}(\tau\cdot N)$ rounds where $N$ is the $\Congest$ round complexity for  solving the PA problem on $G$.
\end{lemma}
Recall (\cref{lem:PA}) that the PA problem on $G$ can be solved in $N = \tilde{O}(r+SQ(G))$ $\Congest$ rounds where $r$ is the round complexity for constructing shortcuts of quality $SQ(G)$. Therefore, the above lemma implies that any $\tau$-round minor-aggregation algorithm can be simulated in the $\Congest$ model in $\tilde{O}(\tau\cdot (r+SQ(G)))$ rounds, which is $\tilde{O}(\tau\cdot D)$ for planar graphs (Corollary~\ref{cor: planar_PA}).

\subsection{Minor-Aggregation  for the Dual Graph}
\label{sec: ma_for_dual_simulation}
We prove that any minor-aggregation algorithm $A$ can be compiled into a $\Congest$ algorithm that runs in the planar network of communication $G$, and simulates $A$ on the dual graph $G^*$. 
An application of this, is a $\Congest$ algorithm for finding the minimum-weight cycle of an undirected weighted planar graph $G$. In this case, the algorithm $A$ is a min-cut algorithm, and so running $A$ on $G^*$ gives the minimum-weight cycle of $G$ by \cref{fac: cycle_cut_duality} (to be shown later in this section). Another application is an approximation for maximum $st$-flow when the graph is undirected and $s,t$ lie on the same face. In that case, the algorithm $A$ is an approximate single source shortest paths algorithm (see \cref{section_flow_cut} for details and proof). 

\subsubsection{PA on $G^*$ and Basic Model Simulation}
Since $G^*$ is not in hand, a straightforward $\Congest$ simulation on $G^*$ might suffer a linear blow-up in the round complexity. Moreover, we cannot directly follow the method of \cite{GHSYZ22} of compiling the minor-aggregation model to $\Congest$, since it is unclear how to do so on $G^*$. Instead, we combine their approach with part-wise aggregations on $G^*$ via $\hat{G}$.

Using aggregations on $\hat{G}$, aggregations on $G^*$ were previously done by Ghaffari and Parter \cite{GP17} to solve the specific problem of computing a primal {\em balanced separator}, thus, their aggregation tasks were problem-specific (aggregations on faces of $G$ and sub-tree sums on $G^*$).
For our purposes, we need something more general, so we extend their method to perform general part-wise aggregations on $G^*$. In particular, we need to perform aggregations that take into consideration the outgoing edges of each part $S_i$ (i.e., edges that have one endpoint in $S_i$ and the other endpoint in $S_j\neq S_i$), something that was not done in \cite{GP17}. This specific task required our small modification to $\hat{G}$ compared to the one defined by Ghaffari and Parter.

\begin{lemma}
    [Part-Wise Aggregations on $G^*$]
    \label{lem: PA_on_G^*}
    Let $P=\{S_1,S_2,\ldots,S_N\}$ be a partition of $V(G^*)$, such that, (1) all $G^*[S_i]$ are connected, (2) each vertex on a face $\hat{f}$ of $\hat{G}$ that correspond to a node $f\in S_i$, knows the (same) ID for $S_i$, (3) each input of a node $f$ is known to the leader vertex of $\hat{f}$ in $\hat{G}$ and (4) each input of an edge $e^*$ (if any) is known to the two endpoints in $\hat{G}$ of the edge $\hat{e}\in E_C$ that maps to $e^*$.
    Then, any aggregate operator over $S_i$ nodes or edges can be computed simultaneously in $\tilde{O}(D)$ $\Congest$ rounds, such that, the output value of each $S_i$ is known to all vertices of $G$ that lie on the faces whose corresponding dual node is in $S_i$.
\end{lemma}

\begin{proof}    
The computation is done via aggregations inside $\hat{G}$, in particular, a partition of $G^*$ induces a partition on $\hat{G}$. Notice that $\hat{G}$ is planar, is built in $O(1)$ rounds, of diameter $O(D)$ and allows $\Congest$ simulation in $G$ with a constant blow-up factor in the round complexity (Properties \ref{property: hat(G)_construction_representation}, \ref{property: hat(G)_diameter} and \ref{property: hat(G)_simulation} of $\hat{G}$).  Thus, $\hat{G}$ admits $\tilde{O}(D)$ quality low-congestion shortcuts in $\tilde{O}(D)$ rounds and the PA problem is solved in $\tilde{O}(D)$ rounds on it (\cref{lem:PA} and \cref{cor: planar_PA}).

A partition $P$ of $G^*$ induces a partition $\hat{P}$ on $\hat{G}$ as follows.
A node $f$ of $G^*$ maps to a face $\hat{f}$ of $\hat{G}$, the part $\hat{S}_i\in \hat{P}$ is defined to contain all vertices on faces $\hat{f}$ of $\hat{G}$ that map to nodes $f$ of $G^*$ in $S_i$. All $\hat{S}_i$ are connected and vertex-disjoint, as each two distinct $f,g\in G^*$ map to distinct vertex- and edge-disjoint faces $\hat{f},\hat{g}$ of $\hat{G}$, in addition, $\hat{f},\hat{g}$ are connected by an $E_C$ edge for each edge in $G^*$ connecting $f,g$.
Notice that star-center vertices ($V_S$) are not in any part $\hat{S_i}$, so we define each to be its own part. Hence, one can see that $\hat{P}=(\cup_{i\in [N]} \{\hat{S_i}\})\bigcup (\cup_{v\in V_S}\{v\})$ is a partition of $\hat{G}$ into connected vertex-disjoint subgraphs. Thus, to compute part-wise aggregation on $P$, we compute part-wise aggregations on $\hat{P}$, in which, vertices of $V_S$ do not participate as an input to the part-wise aggregation task and are used exclusively for communication, henceforth, we do not refer to these vertices.

Initially, all vertices in a part $\hat{S}_i$ know the same ID of $\hat{S}_i$ (equivalently, ID of $S_i$). Thus, $\tilde{O}(D)$ quality low-congestion shortcuts on $\hat{G}$ are constructed for the partition $\hat{P}$ in $\tilde{O}(D)$ rounds as the assumptions for constructing low-congestion shortcuts are met, and because $\hat{G}$ admits such shortcuts as mentioned above.

Recall that in the PA task, each vertex $v$ in a part of the partition has an $\tilde{O}(1)$-bit input,
and the aggregate operator is computed over all inputs of vertices which are in the same part.
When computing a PA task over nodes of $G^*$, the input of $f\in G^*$ is required to be known to some vertex $\hat{v} \in \hat{G}$ (w.l.o.g. the leader) on the face of $\hat{G}$ that maps to $f$ (Property \ref{property: hat(G)_face_identification} of $\hat{G}$). Then, aggregations on all $\hat{S}_i$ shall simulate aggregations on $S_i$, where only the input of the leader of $\hat{f}$ is considered as an input of the dual node $f$. That is, the aggregation task considers input from leaders only, and all other vertices of $\hat{G}$ participate in the algorithm just for the sake of communication (their input is an identity element).

If the PA task is over inside-part edges or over outgoing edges, then, each vertex $v$ of $\hat{G}$ that is incident to an edge $\hat{e}=\{u,v\}\in E_C$ (recall, edges of $E_C$ map 1-1 to $E(G^*)$ by Property \ref{property: hat(G)_mapping_to_E(G^*)} of $\hat{G}$)), knows if $\hat{e}$'s input should or should not be considered in the aggregation - as $v$ knows the ID of the part $\hat{S}_i$ that contains it, and know the ID of $\hat{S}_j$ that contains $u$ in one communication round.
Then, aggregations on all $\hat{S}_i$ simulate aggregations on $S_i$, where any $v\in \hat{G}$ that participates in the aggregation considers some of its edges as an input depending on the task and on whether they are inside-part edges or not. Vertices that have no incident input edge, participate in the algorithm just for the sake of communication (their input is an identity element). This PA task on $\hat{G}$ is computed within $\tilde{O}(D)$ rounds as mentioned at the beginning of the proof.

Finally, after the aggregation task is done, all vertices in $\hat{G}\setminus V_S$ know the output value of their part, which directly translates from $\hat{G}$ to $G$ (as in Property \ref{property: hat(G)_construction_representation} of $\hat{G}$).
Notice, even though for any specific node $f\in G^*$ there is no entity in $\hat{G}$ that knows all its "local" information  (i.e. a list of its incident edges), we still can compute part-wise aggregations over $G^*$.
\end{proof}

Now, having part-wise aggregations on $G^*$, we show a sketch of simulating of the minor-aggregation model in $\Congest$ via PA (as in \cite{GHSYZ22}), where the input graph is the same as the network of communication. Then describe the differences with the similar proof for simulation on $G^*$. In the simulation, a vertex of $G$ simulates itself, and the two endpoints of an edge $e$ simulate $e$. The simulation consists of three main steps: 
(1) A few ($\tilde{O}(1)$) part-wise aggregation tasks are executed to elect a leader for each super node, defining the nodes of the obtained minor, 
(2) A part-wise aggregation task is executed to simulate consensus, where a part is $G[s]$ ($s$ is some super-node),
and finally (3) The endpoints of each edge exchange their parts' consensus values and one more part-wise aggregation task is performed to compute the aggregation values. 
Lemmas \ref{lem: MA_via_PA} and \ref{lem: PA_on_G^*} pave the way for the following theorem.
\begin{restatable}[Basic minor-aggregation model for $G^*$]{theorem}{MaOnDual}
    
    \label{th: MA_on_dual}
    Any $\tau$-round minor aggregation algorithm $A$ can be simulated on $G^*$ within $\tilde{O}(\tau\cdot D)$  $\Congest$ rounds on $G$, such that, the output of each node $f\in G^*$ is known to all vertices of the corresponding face $f$ of $G$, and the output of an edge $e^*\in G^*$ is known to the endpoints of $e\in G$.
\end{restatable}

\begin{proof}
Let $s=\{f_1,f_2,\ldots\}$ be a super-node of $G^*$ ($f_i\in V(G^*)$), the vertices of $\hat{G}$ that map to $s$ are the vertices that lie on the faces $\hat{f}_i$ of $\hat{G}$, such that, $\hat{f}_i$ maps to the node $f_i$.
Then, the computational power of a super-node of $G^*$ is simulated by all vertices of $\hat{G}$ that map to it.
The computational power of an edge in (a minor of) $G^*$ shall be simulated by the endpoints of its corresponding $E_C$ edge in $\hat{G}$ (Property \ref{property: hat(G)_mapping_to_E(G^*)}). Often, we will use Lemma \ref{lem: PA_on_G^*} 
as a black-box for PA on $G^*$ in order to implement the simulation.

There are three steps in a minor-aggregate round: contraction, consensus and aggregation. In fact, there is an implicit initial step at the start of each round for electing a leader of each super-node. We follow the method of \cite{GHSYZ22} and show how all steps compile down to PA in $G^*$ and a few simple operations in $\hat{G}$ as follows.
\begin{enumerate}
    \item
    {\em Electing a leader (and contraction).}
    Recall, in the contraction step, each edge has a value of zero or one, and considered to be contracted if and only if its value is one. To implement that, we have mainly two computational tasks. First, merge {\em clusters} of nodes of $G^*$ over their outgoing edges of value one (clusters eventually define super-nodes of $G^*$). Second, assign each cluster an ID that is known to all vertices of $\hat{G}$ that simulate it.
    Initially, a cluster is a single node of $G^*$, then, clusters are grown to match the set of super-nodes of (the minor of) $G^*$. 
    
    Implementing that, (1) connected components of $\hat{G}[E_R]$ are detected (those are faces of $\hat{G}$ that map to nodes $v\in G^*$), where the ID of a node of $G^*$ is the minimal ID of a vertex in $\hat{G}$ that maps to it, this is done in $\tilde{O}(D)$ rounds by property \ref{property: hat(G)_face_identification} of $\hat{G}$.
    Now each vertex of $\hat{G}$ knows in which of these connected components it is (defines a partition on $G^*$ where each part is a single node) defining each node as a cluster. Notice that now Lemma \ref{lem: PA_on_G^*} (PA in $G^*$) is applicable, due to the fact that all vertices of $\hat{G}$ that correspond to a node $f\in G^*$ in a cluster, know the same ID for that cluster.
    Then, (2) we grow clusters. The merging process follows Boruvka's classic scheme, where the process breaks down into $O(\log n)$ star-shaped merges, the classic implementation is randomized and similar to that of \cite{GH16b, GP17}, however, one can implement this merging process deterministically with the derandomization of \cite{GZ22}. 
    Namely, the PA task on $G^*$ here is when each cluster defines a part in a partition of $V(G^*)$, and the aggregate operator is computed over the outgoing edges of each cluster that chose a value of 1 in the contraction step, deciding the merges; the operator computed is over labels of neighboring clusters incident to those edges, where each cluster is labeled as a {\em joiner} or a {\em receiver}, such that, joiners suggest merges and receivers accept them (i.e., receivers are the star centers in the star-shaped merges). Randomly, the labels are decided by a fair coin flip that the leader of the cluster tosses and broadcasts to the whole cluster. Deterministically, the classic algorithm for 3-coloring by Cole and Vishkin \cite{cole-vishkin} is used (E.g. see~\cite{GZ22} for more details).
    At most $\tilde{O}(1)$ merges are repeatedly done until clusters match super-nodes of $G^*$. 
    In each phase of merges, all nodes maintain the minimum ID of a node $v$ (of $G^*$) in the cluster. 
    The vertex of $\hat{G}$ that has that exact ID is considered to be the leader of vertices in $\hat{G}$ that map to the super-nodes represented by the cluster. Notice, there is such a vertex in $\hat{G}$ that has that exact ID since IDs of nodes of $G^*$ are IDs of vertices in $\hat{G}$ that map to them.
    In addition, at the end of each merging phase Lemma \ref{lem: PA_on_G^*} (PA on $G^*$) is applicable, thus, it can be applied repeatedly $\tilde{O}(1)$ times, once for each merging phase. By \cref{lem: PA_on_G^*}, computing $\tilde{O}(1)$ PA tasks on $G^*$ is done in $\tilde{O}(D)$ rounds on $G$.
    
    The above implements the contraction step, after which, all vertices of $\hat{G}$ that simulate the same super node of $G^*$ have a cluster-unique shared leader and ID.

    \item
    {\em  Consensus.}
    After assigning IDs to clusters, as said, each vertex of $\hat{G}$ knows the ID of the cluster that it maps to, which is used as the part ID in the part-wise aggregation tasks on $G^*$. 
    Regarding input, the input is (perhaps unique) per node of $G^*$, which elected leaders in $\hat{G}$ know (those leaders were elected in the first phase of cluster merging in the previous step). Hence, Lemma \ref{lem: PA_on_G^*} is applicable and is used to implement the consensus step in $\tilde{O}(D)$ rounds on $G$. 
    
    \item
    {\em Aggregation.}
    After performing the consensus step, each endpoint $u$ in $\hat{G}$ of an edge in $E_C$
    informs the other endpoint $v$ with the consensus value of the cluster that $u$ maps to. Then, an endpoint $v\in \hat{G}$ of an edge $e\in E_C$ that maps to the edge $e^*=(a_1,a_2)$ (where $a_1,a_2$ are distinct super nodes of $G^*$), does the following. If $v$ is in the cluster (that maps to) $a_i$, $v$ thinks of the edge $e$ as if it was assigned the value/weight of $z_{e,a_i}$, and an identity element otherwise. Then, the aggregate operator is computed over the outgoing edges of super-nodes.
    Again, the aggregation is done by Lemma \ref{lem: PA_on_G^*} in $\tilde{O}(D)$ rounds on $G$.
    Finally, the two endpoints of each edge know the output of the aggregate operator, thus, they can choose zero or one for that edge, and we are ready for the next round of minor-aggregation.
        
\end{enumerate}
Notice that after the simulation finishes, each vertex in a cluster knows the output value of its cluster's corresponding super-node. Hence, by definition of $\hat{G}$ (see Property \ref{property: hat(G)_construction_representation} of $\hat{G}$), for each $f^*\in G^*$ its output is known to all vertices on the corresponding face $f$ of $G$, and the value of an edge $e^*$ that maps to $e\in \hat{G}$ is known to $e$'s endpoints in $\hat{G}$ as specified in the statement of Lemma~\ref{lem: PA_on_G^*}.
\end{proof}

\subsubsection{Extended Model Simulation}
We define now the extended minor-aggregation model as presented in~\cite{GZ22} and show that it can be simulated in $G^*$. 
\begin{definition}
[Extended minor-aggregation model \cite{GZ22}]
\label{def: extende_model}
This model extends the minor-aggregation model (\cref{def: basic_model}) to work on virtual graphs that have a small number of virtual nodes, without strongly affecting the round complexity. In particular, (1) arbitrarily connected $\tilde{O}(1)$ virtual nodes can be added to the network, and (2) $\tilde{O}(1)$ nodes of the network, each can be replaced with an arbitrarily connected virtual node.
\end{definition}
A virtual node is a node that does not exist in the input network (has no computational power) and must be simulated by nodes that do actually exist. 
Adding a virtual node $v_{virt}$ means to decide to which other (virtual and real) nodes it is connected. Replacing a node $v$ of the network with a virtual node $v_{virt}$, means that $v$ does not participate in computations, instead, $v_{virt}$ (that might be connected to different neighbors than $v$) is simulated by other (than $v$) non-virtual nodes of the network.
An edge that is connected to a virtual node is a {\em virtual edge}. The obtained graph from the input network $G$ by adding virtual nodes is called a {\em virtual graph} and denoted $G_{virt}$. 
All non-virtual nodes of $G_{virt}$ are required to know the list of all (IDs of) virtual nodes in $G_{virt}$.
A virtual edge connecting a non-virtual $u\in G_{virt}$ and a virtual $v\in G_{virt}$ is only known to $u$.
A virtual edge between two virtual nodes in $G_{virt}$ is required to be known by all non-virtual nodes of $G_{virt}$.

In our context, a node of the graph is a face of $\hat{G}$, thus, all vertices on a face $\hat{f}$ of $\hat{G}$ that maps to a node $f$ of $G^*$ shall know the (IDs of) all virtual nodes, virtual edges connecting $f$ to virtual nodes of $G^*_{virt}$ and virtual edges connecting two virtual nodes.
Notice, this is feasible, as there are at most $\tilde{O}(1)$ virtual nodes, thus, there is a total of $\tilde{O}(1)$-bits that shall be known to vertices of $\hat{f}$.
We now want to simulate a minor-aggregation algorithm $A$ on $G^*_{virt}$ that has at most $O(\beta)$ virtual nodes. That is, we want to (1) represent (store) the obtained network $G^*_{virt}$ distributively as defined above, and (2) run any minor aggregation algorithm $A$ on $G^*_{virt}$ with $O(\beta)$ blow up factor in the minor-aggregate round complexity of $A$.

Given the additional $O(\beta)$ virtual nodes and virtual nodes that replace real nodes as input - that is, real nodes that are incident to a virtual node know its ID, and a real node that shall be replaced with a virtual node, knows itself as such. 
Ghaffari and Zuzic \cite{GZ22} showed the following.

\begin{lemma}
[Lemma 15 and Theorem 14 of \cite{GZ22}]
\label{lem: MA_stored_virtual_nodes}
There is an $O(\beta)$ minor-aggregation rounds deterministic algorithm $A_1$ that stores a virtual network $G_{virt}$.
If multiple edges connect some virtual node $v_{virt}$ with some neighbor, $G_{virt}$ will contain a single edge with a weight equal to the sum (or any aggregate operator) of such edges in $G$.
\end{lemma}

\begin{lemma}
     [Lemma 15 and Theorem 14 of \cite{GZ22}]
     \label{lem: MA_simulation_virtual_nodes}
     Let $G_{virt}$ be a virtual graph obtained from a connected graph $G$, and $A$ be an $O(\tau)$-round minor-aggregation algorithm. Then, there is a minor aggregation algorithm $A_2$ on $G$ that simulates $A$ on $G_{virt}$ within $O(\tau \beta)$ minor-aggregate rounds. Upon termination, each non-virtual node $v$ learns all information learned by $v$ and all virtual nodes. The output of a virtual edge is known to all nodes if both of its endpoints are virtual, otherwise, its output is only known to its non-virtual endpoint.
\end{lemma}

Our proof would make a black-box use of \cref{lem: MA_simulation_virtual_nodes}, however, we provide a sketch of the proof of~\cite{GZ22} just to make things a bit clearer (for the full proof see Theorem 14 of \cite{GZ22}).
(1 - Contraction) Contract non-virtual edges that shall be contracted, then, in $O(\beta)$ rounds, each super node shall learn the IDs of virtual-nodes it is connected to, via computing an aggregate over its incident virtual edges. Then, the super-nodes' IDs get computed,
(2 - Consensus) At first, all super-nodes that do not contain any virtual node perform their consensus step, then, we iterate over all virtual nodes and compute the consensus of the super-node they are contained in, such that, all nodes of the graph know the output value of each super node.
Finally (3 - Aggregation) A real edge simulates itself, a virtual edge that has one non-virtual endpoint node is simulated by that endpoint and a virtual edge that the two of its endpoints are virtual is simulated by all nodes. The $z$ value of each edge is computed and an aggregation is performed as in the consensus step.

Simulating the extended minor-aggregation model to run on $G^*$ is a direct corollary of \cref{lem: MA_stored_virtual_nodes}, \cref{lem: MA_simulation_virtual_nodes}, and \cref{th: MA_on_dual}.
Let $G_{virt}^*$ be a virtual graph that is obtained from $G^*$ by adding and replacing nodes with up to $\beta$ arbitrarily connected virtual nodes, then, 

\begin{theorem}
\label{th: extended_MA_on_dual}
     Any $\tau$-round minor-aggregation algorithm $A$
    can be simulated on $G^*_{virt}$ within $\tilde{O}(\tau \beta D)$ $\Congest$ rounds of communication on $G$, such that, the output of each non-virtual $f\in G^*_{virt}$ is known to all vertices on the corresponding face $f$ of $G$, and the output of each virtual $f_{virt}\in G^*_{virt}$ is known to all $v\in G$. The output of a virtual edge $e$ is known to all vertices of $G$ if both of its endpoints are virtual in $G^*_{virt}$, otherwise, $e$'s output is only known to vertices of $G$ that lie on the faces that correspond to $e$'s non-virtual endpoint.
\end{theorem}

 Notice that adding virtual nodes and edges to $G^*$ might violate planarity, but that is not a problem, as these virtual nodes and edges are simulated by (on top of) the same underlying network $\hat{G}$ that is used to simulate the basic model, without making any changes on it. In particular, this extended model, is simulated by the basic model unconditionally, i.e., Lemma~\ref{lem: MA_simulation_virtual_nodes} and~\ref{lem: MA_stored_virtual_nodes} constitute a universal algorithm that runs in the basic model for simulating the extended model, so we simply apply that algorithm on $G^*$ using \cref{th: MA_on_dual}.

\subsubsection{Dealing with Parallel Edges in $G^*$}
We now address the problem of $G^*$ being a (multi)  non-simple graph, i.e. has self-loops and parallel edges. 
 This might be a problem due to the fact that many algorithms assume their input graph is simple. We provide a procedure for deactivating parallel edges of $G^*$: 
 Consider two nodes $u,v$ of $G^*$ that are connected by multiple edges $e_1,e_2,\ldots,e_m$. We deactivate all these edges except for one (called the {\em active} edge). If edges are assigned a weight $w(\cdot)$, the active edge gets assigned a new weight $\otimes_iw(e_i)$, where $\otimes$ is a predefined aggregation operator.
 E.g., $\otimes$ is the summation operator if one is interested in computing a minimum cut, and the minimum operator if one is interested in computing shortest paths.
 
 One cannot straightforwardly use part-wise aggregations on $G^*$ in order to do the above, since a node of $G^*$ might be connected to many nodes, and the required task is an aggregation for each neighbor rather than one aggregation over all incident edges, this may lead to high congestion if all nodes do those aggregations naively and simultaneously.
We use the the fact that the underlying simple graph of $G^*$ \footnote{I.e. $G^*$ after the removal of self-loops and "collapsing" parallel edges to a single edge.} has a small {\em arboricity} of $\alpha=3$ (the arboricity of a graph is the minimal number of forests into which its edges can be partitioned) to obtain a minor-aggregation procedure that requires $\tilde{O}(\alpha)$ minor-aggregate rounds for orienting $G^*$ edges, so that in the underlying simple graph of $G^*$ each node has at most a constant out-degree, which we utilize for load balancing the aggregation tasks. 
In the original $G^*$ (the multi graph) this means that each node would have outgoing edges to a constant number of neighbors, rather than a constant number of outgoing edges. I.e., we allow many edges to be oriented outwards a node as long as they connect it to a constant number of neighbors.
We then apply this procedure on $G^*$ in a black-box manner using \cref{th: extended_MA_on_dual}, resulting with an $\tilde{O}(D)$ $\Congest$ algorithm on $G$ simulating the procedure on $G^*$. We henceforth assume that $G^*$ is a simple graph. Formally, we obtain the following lemma. 

 \begin{restatable}[Deactivating parallel edges and self-loops]{lemma}{deactivateParallelEdges}
    \label{lem: deactivating_parallel_edges}
      There is a minor-aggregation algorithm that runs in $\tilde{O}(\alpha)$-rounds and deactivates parallel edges and self-loops in an input graph of arboricity $\alpha$, assigning each active edge $(u,v)$ a weight equivalent to a predefined aggregate operator over the weights of all edges with the same endpoints.
\end{restatable}

Let $\alpha$ be the arboricity of the input graph, we implement an algorithm of \cite{BM10_NWdecomposition} in the minor-aggregation model that runs in $\tilde{O}(\alpha)$ minor-aggregation rounds and produces such an orientation. Since the arboricity of the underlying simple graph of $G^*$ is three, we get our constant out-degree orientation.

 \begin{proof}

    \medskip
    \noindent
    First, we omit self loops, each edge whose endpoints have the same ID, deactivates itself, this is done in only one minor-aggregate round.
    Next, we describe \cite{BM10_NWdecomposition}'s algorithm and then describe how to implement each step in the minor-aggregation model. At the start of the algorithm, all nodes are {\em white}. For each node, there is a time where it becomes {\em black}, the algorithm terminates when all nodes turn black. 
    The algorithm produces a partition  of the node set into $\ell =\lceil2 \log n\rceil$ parts, $H_1,\ldots,H_\ell$. Upon termination, each node knows the part $H_i$ that contains it. Moreover, the following orientation of edges, results with an orientation with at most $O(\alpha)$ outgoing edges for each nodes (counting parallel edges between two nodes as one, that is, in the underlying simple graph).
    The orientation is as follows. An edge $e=(u,v)$ where $u\in H_i$ and $v\in H_j$ s.t. $i<j$, gets oriented towards $v$. Otherwise, if both $u,v$ belong to the same part, $e$ is oriented towards the endpoint with a greater ID.
    
    The algorithm for computing such a partition is simple.
    Let $v$ be a (super-)node, then, for a phase $i\in [\ell]$:
    If $v$ is white with at most $3\alpha $ white neighbors, (1) make $v$ black, (2) add $v$ to $H_i$ and (3) notify $v$'s neighbors that $v$ has turned black and has joined $H_i$.
    This is implemented this in the minor aggregation model is as follows.
    \begin{enumerate}
        \item We want to know for each node $v$ if it has at most $3\alpha$ white neighbors. 
        We do that with at most $3\alpha$ consensus and aggregation steps. That is, a node $v$ computes an aggregate over IDs and labels (black/white) of its neighboring nodes, iteratively counting white nodes that were not already counted by $v$ in the current phase, then the ID of that node that just got counted is broadcast (inside the super node $v$). Thus, given the previously broadcast IDs one can count a new node that was not already counted and avoid counting nodes that were counted already. If $v$ counts more than $3\alpha$ values it stops counting.
        
        \item If the super-node $v$ has counted $3\alpha$ neighbors, via a consensus step, this information ($v$ shall turn black) is broadcasted to all nodes inside $v$. I.e., edges incident to $v$ learn this information, and $v$'s neighbors learn that $v$ is black and belongs to $H_i$.
        
    \end{enumerate}

    After the algorithm terminates, each edge knows to orient itself as was described earlier.
    Each node now has at most $\alpha$ outgoing edges (counting parallel edges as one). This procedure terminates in $\tilde{O}(\alpha)$ rounds. Note, the orientation of an edge can be decided locally by the two vertices in the communication graph which simulate that edge (in our case, these are vertices of $\hat{G}$).

    \medskip
    \noindent
    { \em Deactivate parallel edges.}
    Assume edges are assigned unique $\tilde{O}(1)$-bit IDs in $\tilde{O}(1)$ rounds. 
     Having self-loops deactivated, we now use the orientation to deactivate parallel edges. Each node deactivates its outgoing parallel edges in $O(\alpha)$ rounds. First, enumerate outgoing edges from $1$ to $3\alpha$, where all edges that connect $v$ to the same neighbor have the same number, this is done in $O(\alpha)$ rounds, in an iterative way similar to that of counting neighbors.
     Then, for all $i\in [3\alpha]$, for edges with number $i$: (1) An aggregation step over their weights is computed, (2) An edge is chosen to be the active edge by another aggregation step (say the minimum ID one) and other edges deactivate themselves.
    This process terminates in $O(\alpha)$ rounds.
\end{proof}
Note, each edge in $G^*$ can be identified by its primal endpoints IDs (as the primal graph is simple), we consider this to be the edge's ID and is known locally by the edge's endpoints in the communication graph $\hat{G}$.

\subsection{Weighted Girth}
In this subsection we show how to compute the {\em weighted girth} (i.e. the minimum-weight cycle) of an undirected weighted planar graph $G$.
We do this by exploiting the result of \cref{sec: ma_for_dual_simulation} in order to simulate a minor-aggregate weighted minimum-cut algorithm on $G^*$, which (by \cref{fac: cycle_cut_duality}) gives the weighted girth of $G$.

Recently, a {\em universally optimal}
\footnote{An algorithm that has an optimal round complexity (up to poly$\log n$ factor) on each specific network topology, conditioned on an efficient construction of low-congestion shortcuts for that topology. For an extended discussion, see \cite{GHSYZ22,GHLRZ22,GZ22}.}
algorithm for the exact weighted minimum cut problem was discovered by Ghaffari and Zuzic \cite{GZ22}, generalizing a breakthrough of Dory, Efron, Mukhopadhyay and Nanongkai \cite{DEMN21}. Their algorithm is formulated in the extended minor-aggregation model.

\begin{theorem}
    [Minor-aggregate exact min-cut, Theorems 12, 18 and 40 of \cite{GZ22}]
    \label{th: GZ_min_cut}
    Let $G$ be an $\exp(\tilde{O}(1))$ edge-weighted input graph. There is a randomized  $\tilde{O}(1)$-round minor aggregation algorithm that computes the weight of the minimum weight cut $C$ of $G$ w.h.p.. Upon termination, the value of $C$ is known to all vertices.
\end{theorem}
Notice, applying \cref{{th: GZ_min_cut}} above on $G^*$  already allows us to find the weighted girth of our planar network in $\tilde{O}(D)$ rounds. 
In \cite{GZ22}'s work, the focus was on finding the weight of the minimum cut, but it can be easily extended to find also the edges of the minimum cut.
Let $T$ be a spanning tree of $G$ and $C=(S,V(G)\setminus S)$ be a cut of $G$, then, $C$ is, respectively, {\em (one-) two-respecting} of $T$ if and only if there are exactly (one) two edges of $T$ that cross the cut. Given such a tree, where its edges that cross the cut are known, in $O(1)$ minor aggregation rounds, one can mark all edges of the cut.

\begin{restatable}[Mark cut edges]{lemma}{markcut}
    \label{lem: mark min-cut}
    Let $G$ be any network, $T$ a spanning tree of $G$ and $C$ any cut that one- (two-) respects $T$, defined by edges $e_1, e_2\in T$ (possibly $e_1=e_2$). Given $G$, assuming that edges of $T$ and the edges $e_1,e_2$ know themselves, there is an $O(1)$ deterministic minor-aggregation algorithm that marks all $C$ edges in $G$ (i.e., upon termination, each edge $e\in C$ knows itself as such).
\end{restatable}

\begin{proof}
Consider the tree $T$, we start by (a contraction step) contracting all $T$ edges except for $e_1,e_2$, which results in a minor $G'$ of $G$ with at most three super nodes that are connected by $e_1,e_2$ and possibly other (parallel) edges.
Next, each super-node shall compute its cost (defined as the number of edges from $e_1,e_2$ that are incident to it).
This is done by an aggregation step, each edge in $G'$ chooses a weight of zero, except for $e_1,e_2$, each chooses a weight of one, and an aggregation step is performed with the sum operator, resulting with the sum of outgoing edges for each super-node, which is its cost.
Now, the ID of the maximum-cost node is computed and broadcast to all the graph - by contracting all the graph into a single super-node $s$, initiating the input of each node $v$ to a tuple that contains the ID and the cost of the super-node $s'\in G'$ s.t. $v\in s'$, then, we compute the maximum cost by a consensus step (breaking ties arbitrarily).
Finally, we return to consider $G'$ as our graph (we undo the last contraction). Notice, edges of $G'$ know the ID of the maximum-cost node, so, they mark themselves if only if they are incident to it.

The round complexity is clear.
The correctness is as follows. $G'$ is a graph that is composed of two or three super-nodes, each super-node corresponds to a connected component of $T\setminus C$, each induces a connected subgraph of $G$, such that, one of which is incident to all edges $e_1,e_2$ in $G$, that subgraph defines one side of the cut (denoted $S$) and the other (at most two) subgraphs define the other side of the cut.
Since $S$ is incident to $e_1,e_2$ it has the highest cost, thus, all edges exiting $S$ in $G$ get marked, these are exactly the edges of $G$ that cross the cut, hence, all edges of $C$. 
\end{proof}

The algorithm of \cite{GZ22} (\cref{th: GZ_min_cut}), like many min-cut algorithms (both centralized and distributed, see \cite{Kar94, Kar00, Th01, GH16b, GG18, GMW20,MN20, DEMN21}) consists of three main steps: (1) pack spanning trees of $G$, (2) compute a minimum 1-respecting cut of each packed tree and (3) compute a minimum 2-respecting cut of each packed tree, the final output is the overall minimum.
As said, upon termination the value of $C$ is known to all nodes, in addition, it is implied in \cite{GZ22} that the edges of a spanning tree $T$ of $G$ that $C$ (1-) 2-respects know themselves, in particular, edges in $T\cap C$ know themselves as such. 
Using \cref{th: extended_MA_on_dual} on $G^*$ in order to apply the exact min-cut algorithm (\cref{th: GZ_min_cut}) and to apply the procedure for marking cut edges (\cref{lem: mark min-cut}), we get an undirected planar weighted girth algorithm: 

\theoremWeightedGirth*

\begin{proof}
    First we make $G^*$ simple by applying \cref{lem: deactivating_parallel_edges} for deactivating parallel edges, the (only) active edge left between any two nodes $u,v$, gets assigned a weight equal to the sum of weights of edges with the same endpoints in $G^*$.
    Then, we compute the minimum cut $C$ of the dual network $G^*$ by applying \cref{th: extended_MA_on_dual} to simulate \cite{GZ22}'s minor-aggregate exact min-cut algorithm (\cref{th: GZ_min_cut}) on $G^*$.
    Afterwhich, we can assume that we have the min-cut $C$ value, that $C$ is (1-) 2-respecting to some tree $T$, and that each edge in $T$ along with $T$'s edges $e_1,e_2$ that cross the cut know themselves ($e_1=e_2$ in case $C$ is 1-respecting).
    Now we use \cref{th: extended_MA_on_dual} for applying \cref{lem: mark min-cut} in order to mark the edges of $C$ in $G^*$. 
    By the cycle-cut duality (\cref{fac: cycle_cut_duality}), the weight of $C$ is the weight of $G$'s minimum cycle and the edges of $C$ constitute a minimum weight cycle in $G$.
    Since all procedures we use on $G^*$ are of $\tilde{O}(1)$ minor-aggregate round complexity, then, the $\Congest$ round complexity for simulating them over $G$ is $\tilde{O}(D)$ by \cref{th: extended_MA_on_dual}.
\end{proof}

    We clarify a low-level detail regarding marking the edges of $C$. 
    Recall, the minor-aggregation model simulation on $G^*$ is done via the face-disjoint graph $\hat{G}$.
    Since the edges of $C$  are edges of $G^*$, each edge $e^*$ of them maps to exactly one edge $\hat{e}$ of $\hat{G}$ (Property \ref{property: hat(G)_mapping_to_E(G^*)}), which itself maps to the edge $e\in G$ (the dual of $e^*$), that is, indeed, marking edges of $G^*$ marks the corresponding edges of $G$.
    Note however, edges $\hat{e}\in \hat{G}$ that map to marked edges might not form a cut nor a cycle in $\hat{G}$ but that is fine as it is only a communication tool and not our input graph for the problem.

\section{Dual Distance Labeling}
\label{sec: dual SSSP}

In this section we show how to compute single source shortest paths (SSSP) in the dual graph $G^*$ using a labeling scheme. I.e., we assign each dual node a label, such that, using the labels alone of any two dual nodes $s$ and $t$, one can deduce the $s$-to-$t$ distance in $G^*$.
Notice that such labeling actually allows computation of all pairs shortest paths (APSP), i.e, we can solve SSSP from any source node by broadcasting its label to the entire graph. In this section, our algorithms are stated to be randomized, however, the only randomized component is an algorithm of \cite{GP17} that we implicitly apply when using the BDD of \cite{LP19} in a black-box manner. A derandomization for that algorithm would instantly imply a derandomization of our results in this section.

\medskip
\noindent
{\bf Section organization.} First, in \cref{section_BDD} we extend the analysis of the BDD, divided into three subsections, in each we provide the proofs of some new properties of the BDD. In particular, \cref{section_few_face_parts} focuses on analyzing the way faces get split, \cref{section_dual_bags} focuses on the dual decomposition, and \cref{section_distributed_knowledge} shows how to learn that decomposition distributively. After which, we would be ready to formally define the labeling scheme in \cref{section_distance_labels}. In \cref{section_dual_labeling_algorithm} we show an algorithm that constructs the labels, and finally in \cref{section_dual_SSSP_tree}
 we show how to learn an SSSP tree.

\subsection {Extending the BDD} 
\label{section_BDD}

The Bounded Diameter Decomposition (BDD), introduced by Li and Parter \cite{LP19}, is a hierarchical decomposition of the planar graph $G$ using cycle separators. It is a rooted tree $\mathcal{T}$ of depth $O(\log n)$ whose nodes (called {\em bags}) correspond to connected subgraphs of $G$ with small diameter $\tilde{O}(D)$. The root of $\mathcal{T}$ corresponds to the entire graph $G$, and the leaves of $\mathcal{T}$ correspond to subgraphs of small size $\tilde{O}(D)$. 
In \cite{LP19}, each bag $X$ was defined as a subset of vertices. For our purposes, it is convenient to define $X$ as a subset of edges (a subgraph). Given that we want to work with the dual graph $G^*$, the bijection between the edges of $G^*$ and the edges of $G$ allows us to move smoothly between working on $G$ and $G^*$.

In this section, our goal is to extend the BDD of \cite{LP19} to the dual network $G^*$. We stress that the extension to the BDD is not obtained by a simple black-box application  of the BDD on $G^*$ (for the two reasons mentioned earlier: $G^*$ is not the network of communication, and $G^*$'s diameter might be much larger than the diameter of $G$). 
Instead, we use the same BDD $\mathcal T$ and carefully define the duals $X^*$ of bags $X$. 
We begin with the following lemma summarizing the properties of the BDD of \cite{LP19}. 

\begin{lemma}
        [Bounded diameter decomposition \cite{LP19}]
        \label{lem: BDD}
        Let $G$ be an embedded planar graph with hop diameter $D$. There is a distributed randomized $\tilde{O}(D)$-rounds algorithm that w.h.p. computes a bounded diameter decomposition $\mathcal{T}$ of $G$, satisfying:
        \begin{enumerate}
        \item \label{property: BDD_depth} $\mathcal{T}$ is of depth $O(\log n)$.

        \item The root of \ $\mathcal{T}$ corresponds to the graph $G$. 
        \item \label{property: BDD_leaf_size} The leaves of \ $\mathcal{T}$ correspond to graphs of size $O(D \log n)$.
        \item \label{property: BDD_separator} For a non-leaf bag $X$, let $\SX$ be the set of vertices of $X$ which are present in more than one child bag of $X$. Then, $|\SX|=O(D \log n)$.

        \item \label{property: BDD_diameter}  For every bag $X$, the diameter of $X$ is at most $O(D \log n)$.
        
        \item \label{property: BDD_children} For every  bag $X$ with child bags $X_1,X_2,\ldots$, we have $X=\cup_i X_i$.
        
        \item \label{property: BDD_parallel_work_in_level} Each edge $e$ of $G$ is in at most two distinct bags of the same level of $\mathcal{T}$.
              
        \item  \label{property: BDD_ID} Every bag has a unique $\tilde{O}(1)$-bit ID. Every vertex knows the IDs of all bags containing it. 
    \end{enumerate}

\end{lemma}

\noindent
By the construction of the BDD, the set $\SX$ plays the role of a cycle separator of $X$.

Since our focus is on the dual graph, we need to understand how faces of $G$ are affected by the BDD. 
Namely, decomposing a bag $X$ into child bags $X_1,X_2,\ldots$ might partition some faces of $X$ into multiple parts, where each such {\em face-part} is contained in a distinct child bag of $X$. We next discuss this in detail. Since each edge belongs to two faces of $G$, it is convenient to view an edge as two copies (called {\em darts}) in opposite directions. 
This is fairly common in the centralized literature on planar graphs~\cite{KM}.

\begin{figure}[htb]
  \centering 
  \includegraphics[width=0.35\linewidth]{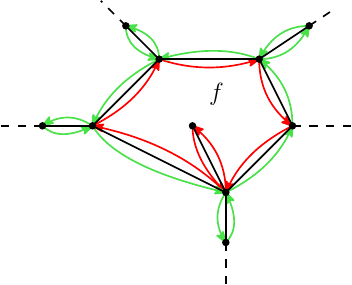}
  \caption{Edges are black, each edge has two darts (green or red). The red clockwise cycle of darts corresponds to the face $f$.}
  \label{fig:darts}
\end{figure}

\medskip
\noindent 
{\bf Darts.}
\label{def: darts}
Each edge $e\in E(X)$ is represented by two darts $d^-$ and $d^+$ (if the graph is directed, $d^+$ has the same direction as $e$, and $d^-$ is in the opposite direction).
The reverse dart is defined as $\rev(d^+)=d^-$ and $\rev(d^-)=d^+$. 
We think of the two darts as embedded one on top of the other (i.e., replacing the edges with darts does not create new faces). When we mention a path or a cycle $C$ of darts, we denote by $\rev(C)$ the path or cycle obtained by reversing all darts of $C$.
The faces of $G$ define a partition over the set of darts (i.e., each dart is in exactly one face) such that each face of $G$ is a clockwise cycle of darts~\cite{KM}. See Figure~\ref{fig:darts}.

\medskip
\noindent 
{\bf Face-parts.}
Ideally, we would like  each face of a bag $X$ to be entirely contained in one of its child bags. However, it is possible that a face $f$ is split into multiple child bags, each containing a subset of $f$'s darts (which we call a {\em face-part}). 
This issue was not discussed in \cite{LP19} as they work with the primal graph $G$.

We keep track of the faces of $G$ and their partition into face-parts. In $G$, each face is a cycle of directed darts. 
A face-part in a bag $X$ is identified by a collection of directed paths of darts, all belonging to the same face of $G$. Notice that a face-part might be disconnected. A bag $X$ therefore contains darts that form faces of $G$ that are entirely contained in $X$, and darts that form face-parts. If a dart $d$ belongs to $X$ but $\rev(d)$ does not belong to $X$, we say that $\rev(d)$ lies on a {\em hole} (rather than a face or a face-part) of $X$. This happens when $d$ is a dart of $\SH$ for some ancestor bag $H$ of $X$.

\begin{figure}[!htb]
    \centering
    \includegraphics[width=1\linewidth]{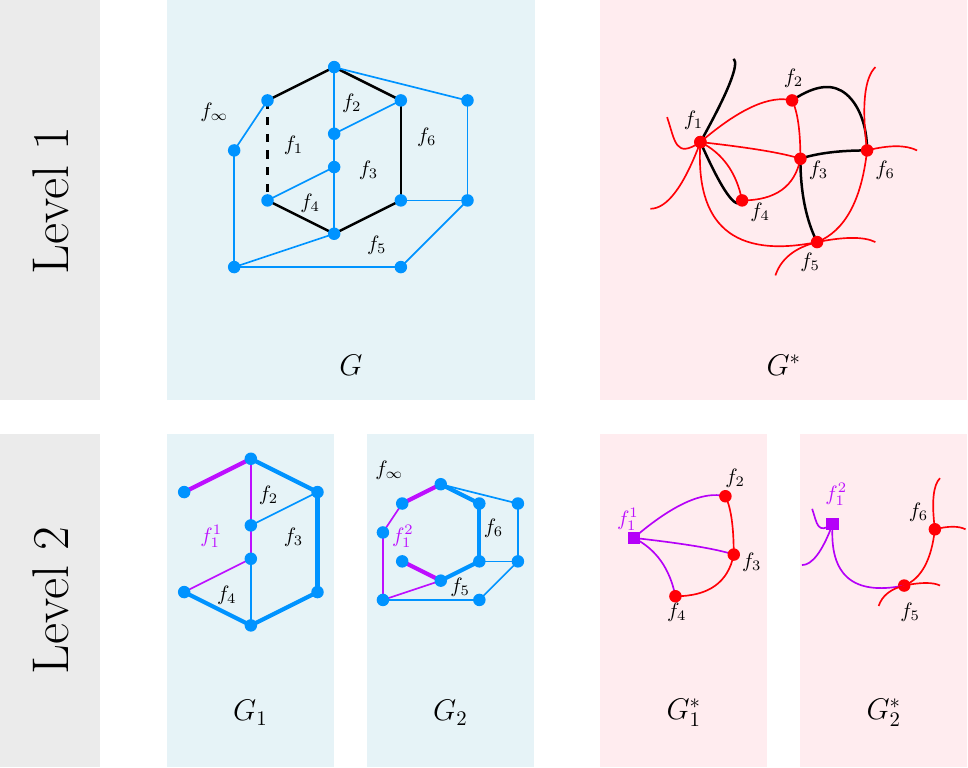}
    \caption{
    Two levels of the BDD (blue) and their corresponding levels in the dual decomposition (red). For simplicity, the figure is undirected, in addition, the dual node corresponding to the face $f_\infty$ of $G$ is not illustrated in $G^*$ (its incident edges are the ones missing an endpoint).\\
    First level: the graph $G$ and its dual $G^*$. Black edges are the separator $\SG$ edges, the dashed one is the virtual edge $e_{\!_G}$. Note, $e_{\!_G}$ splits the critical face, $f_1$.\\
    Second level: the child bags $G_1, G_2$ of $G$ and their corresponding dual bags $G^*_1,G^*_2$ (respectively). Purple edges in $G_1$ and $G_2$ are edges of the critical face $f_1$ of $G$. In $G_1$, those edges define the face-part $f_1^1$ and in $G_2$ they define the face-part $f_1^2$. Both face-parts are represented by dual nodes (squared purple).  Bold edges in $G_1,G_2$ are edges of $\SG$ and non of them has a dual edge in $G^*_1,G^*_2$ as they lie on a hole. Each other edge has a dual edge, in particular, edges of $f_1^1$ ($f_1^2$) connect them to their neighbors from $G^*$ in $G^*_1$ ($G^*_2$).}
   \label{fig: dual_decomposition}
\end{figure}

We think of obtaining face-parts as a recursive process that starts from the root of $\mathcal{T}$. In each level of $\mathcal{T}$, each bag $X$ has its faces of $G$ and face-parts of $G$. Both of these may get partitioned in the next level between child bags of $X$, resulting with new (smaller) face-parts each contained in one child bag of $X$.

\medskip
\noindent
{\bf Dual bags.}
We define the dual bag $X^*$ of a bag $X$ to be the dual graph of the graph $X$ when treating face-parts (and not only faces) of $G$ in $X$ as nodes of $X^*$. 
Formally, each face or face-part $f$ of $G$ in $X$ has a corresponding dual node $f^*$ in $X^*$.
The dual edges $e^*$ incident to $f^*$ in $X^*$ are the edges $e$ (collection of paths) of $X$ that define $f$ and have both their darts in $X$. Recall that each dart of $e$ is part of exactly one face or face-part, and the dual edge $e^*$ connects the corresponding dual nodes (if only one dart of $e$ is in $X$ then $e$ lies on a hole of $X$ and we do not consider holes as dual nodes).
The weight of $e^*$ is the same as that of $e$ and its direction is defined as in \cref{section: preliminaries}.
For an illustration see \cref{fig: dual_decomposition}.
When discussing dual bags, we refer to a node that corresponds to a face-part as a {\em node-part}.

Ideally, we would like the child bags $X^*_i$ to be subgraphs of $X^*$. If that were true, then $X^*$ would be the union of all $X^*_i$ plus the edges dual to $\SX$ (since they are the only edges in more than a bag of the same level). This is almost the case, except that $X^*$ may also have as many as $O(\log n)$ (we prove this next) faces and face-parts that are further partitioned into smaller face-parts in the child bags $X^*_i$.  
 I.e., the union of all $X^*_i$ has a different node set than that of $X^*$.
 Thus, in $X^*$ we have $O(\log n)$ nodes that might be divided to multiple nodes, each in some child bag $X^*_i$.

\begin{theorem}[BDD additional properties]
\label{theorem_additional_BDD_properties}
Let  $\mathcal{T}$ be a Bounded Diameter Decomposition of an embedded planar graph $G$ with hop-diameter $D$. Within additional $\tilde{O}(D)$ rounds, $\mathcal{T}$ satisfies the following additional properties.

\begin{enumerate}
\setcounter{enumi}{8}

    \item[] {\em \textbf{Few face-parts property:}}
        
    \item \label{BDD_number_face_parts} Each bag $X$ contains $O(\log n)$ face-parts. Moreover, the number of face-parts resulting from partitioning faces and face-parts of $X$ across its child bags is $O(D \log^2 n)$.

    \item[] {\em \textbf{Dual bags' properties:}}

    \item \label{BDD_dual_leaf_size} The dual bags of leaves of $\mathcal{T}$ are of size $O(D \log n)$.

    \item \label{BDD_Fx_cut} 
    For a non-leaf bag $X^*$, let $\FX$ be the set of nodes whose incident edges are not contained in a single child bag of $X^*$. Then, $\FX$ is a node-cut (separator) of $X^*$ of size $|\FX|=O(D \log n)$. Concretely, $\FX$ is the set of (a) nodes incident to dual edges of $\SX$, and (b) nodes corresponding to faces or face-parts that are partitioned between child bags of $X^*$.

    \item \label{BDD_dual_children} Let $X^*$ be a non-leaf bag with child bags $X^*_1,X^*_2,\ldots$; Then, $X^*$ is equivalent to $\cup_i X^*_i \cup X^*[\SX]$ \footnote{We abuse notation, even though $\SX$ is defined as a set of vertices of $X$, we denote by $X^*[\SX]$ the induced subgraph of $X^*$ on dual edges to (the edges) of $\SX$.} after connecting all node-parts $f_1,f_2,\ldots$ in $X^*_1,X^*_2,\ldots$ (resp.) corresponding to the same $f\in \FX$ with a clique and contracting it. \footnote{For an intuition, see \cref{fig: dual_decomposition}: 
merging $f_1^1$ and $f_1^2$ to one node would result with $G^*$ after adding the separator edges.}

    \item[] {\em \textbf{Distributed knowledge:}}
    
    \item \label{BDD_know_faces} Each node $f$ of $X^*$ has an $\tilde{O}(1)$-bit ID, s.t. each vertex $v$ of $X$ lying on $f$ knows its ID. Moreover, $v$  knows whether $f$ corresponds to a face, a face-part or the critical face of $G$.

    \item \label{BDD_know_edges} For each bag $X$, and vertices $v$ of $X$, $v$ knows the dual edges in $X^*$ (if exist) corresponding to each of $v$'s incident edges.
    
 \end{enumerate}

\end{theorem}

\subsubsection{The Few Face-parts Property}
\label{section_few_face_parts}

First, we prove the following lemma. Then, we derive a corollary concluding the proof of Property~\ref{BDD_number_face_parts}. 
\begin{restatable}[Few face-parts]{lemma}{lemmaNumberOfFaceParts}
\label{lemma: number of new face-parts}
    Any bag $X\in \mathcal{T}$ contains at most $O(\log n)$ face-parts. Moreover, there is at most one face of $G$ that is entirely contained in $X$ and is  partitioned in between child bags of  $X$.
\end{restatable}

\begin{proof}
Since the height of $\mathcal{T}$ is $O(\log n)$, it is sufficient to prove that in each bag $X$ of $\mathcal{T}$ there is at most one face of $G$ that is entirely contained in $X$ and is partitioned into face-parts in $X$'s child bags.\footnote{In \cite{LP19}, a single hole of $X$ (the union of some $\SH$ edges of ancestors $H$ of $X$, referred to in \cite{LP19} as the {\em boundary} of $X$) might be partitioned between $X$'s child bags. In our case, since we do not care about holes, we do not count this as a face (or a face-part) that gets partitioned.} This would imply that each face-part of $X$ can be associated with the unique leafmost ancestor of $X$ in which this face was whole (and thus there are only $O(\log n)$ face-parts in $X$). Note, in bags $X$ of deeper recursion levels, a face-part might consist of multiple disconnected paths. However, we still count it as one face-part of $X$ since all of those paths belong to the same face of $G$.

To show that indeed there is only one such face, we need to understand the cases in which faces are partitioned in the BDD of \cite{LP19}. 
The set $\SX$ in a bag $X\in \mathcal{T}$ is a simple  (unoriented) cycle consisting of two paths in a spanning tree $T$ of $X$ and an additional edge $\eX$. The edge $\eX$ might not be an edge of $E(G)$, in which case, we refer to it as the {\em virtual} edge of $X$. We think of $\eX$ as if it is temporarily added to $X$, in order to define its child bags, and then removed. 
The cycle $\SX$ is a {\em balanced} cycle separator, i.e. a cycle that contains a constant factor of $V(G)$ in each of its strict interior and exterior\footnote{Formally, the interior (resp. exterior) of $\SX$ is defined as the subgraph that is enclosed by the  cw (resp. ccw) cycle of its darts. Those definitions are standard in the literature of planar graphs. See~\cite{KM}.}.  

The edge $\eX$ is found in \cite{GP17} by carefully identifying two vertices $u,v$, then setting $\eX = (u,v)$ closing a cycle with $T$. $\eX$ is embedded without violating planarity. Given the two endpoints and the embedding of $\eX$, one can identify the {\em critical} face (face-part) $f$ that contains $\eX$ by simply considering the face (face-part) that contains the two edges in the local ordering of $v$ which $\eX$ is embedded in between.

\begin{figure}[htb]
    \centering
    \begin{subfigure}{0.42\textwidth}
    \includegraphics[width=1\linewidth, height=1\linewidth]{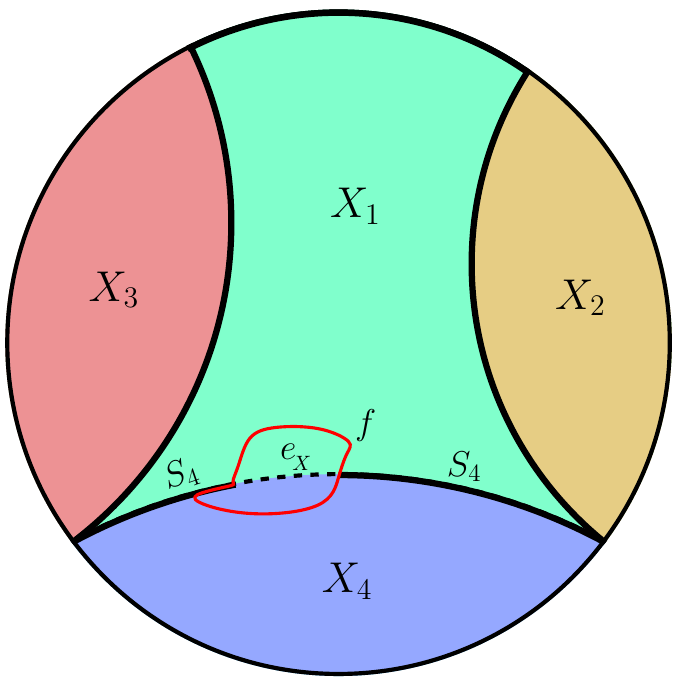} 
    \end{subfigure}
    \hspace{1 cm}
    \begin{subfigure}{0.42\textwidth}
    \includegraphics[width=1\linewidth, height=1\linewidth]{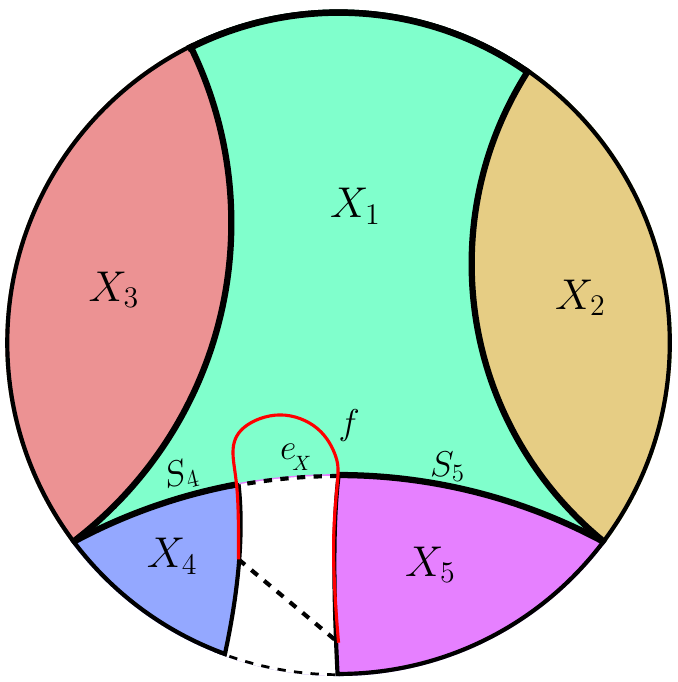}
    \end{subfigure}
    \caption{
    Two examples of a bag $X$. The bold cycle is $\SX$, and the edge $\eX$ is dashed. Distinct child bags of $X$ are of distinct colors. The child $X_1$ (green) is the interior of $\SX$. The external child bags $X_i$ are each defined by a single subpath $S_i$ of $\SX$. 
    One such subpath may contain $\eX$ ($S_4$ in the left image) if its external part (bag $X_4$ in the left image) is connected. Otherwise, the external part is broken into two (connected) bags ($X_4,X_5$ in the right image), each with its own $S_i$ that does not include $\eX$ (the additional dashed edges are some ancestor virtual edges).  In the left (resp. right) image, $X_1$ and $X_4$ (resp. $X_1,X_4,X_5$) contain a face-part of the critical face (face-part) $f$ (in red). 
    \label{fig: BDD_dteailed}}

\end{figure}

 The cycle $\SX$ defines the child bags of $X$ as follows (see \cref{fig: BDD_dteailed}): The interior of the cycle $\SX$ defines one child bag (denoted $X_1$) and the exterior may define several child bags. Each external child bag $X_i$ is identified with a unique (internally disjoint) subpath $S_i$ of $\SX$. That is, the only edges that belong to more than one $X_i$ are the edges of $\SX$. For each such edge, one of its darts $d$ belongs to $X_1$ and $\rev(d)$ belongs to some $X_i \neq X_1$ (note that $\rev(d)$ belongs to a hole in $X_1$ and that $d$ belongs to a hole in $X_i$). Note that every path between a vertex of $X_i$ and a vertex of $X_j \neq X_i$ must intersect a vertex of $S_i$ (Property~\ref{property: BDD_separator} of the BDD).

The critical face $f$ is partitioned (by $\eX$) into two face-parts. One part goes to $X_1$ and the other to an external child bag $X_i \neq X_1$. 
If $f$ is a face of $G$, then we need to count the two new face-parts that it generates. If $f$ is a face-part, then we don't need to count them (since we only care about faces of $G$ that get partitioned). However, in both cases we need to show that, except for $f$, no other faces can be partitioned between the child bags of $X$. For that we dive a bit more into the way the child bags are defined in~\cite{LP19}.

\medskip
\noindent
{\bf Case I: } When $\eX\in E(G)$.
In this case we claim that no face of $G$ gets partitioned among $X$'s children. Assume for the contrary that some face $g$ of $G$ gets split between some $X_i$ and $X_j$. 
The face $g$ must contain a dart $(a,b)$ in $X_i$ and a dart $(c,d)$ in $X_j$ such that neither $(a,b)$ nor $(c,d)$ is a dart of $S_i$. Since $g$ is a cycle of darts, it is composed of $(a,b), (c,d)$, and two dart-disjoint $b$-to-$c$ and $d$-to-$a$ paths. Each of these two paths must intersect the vertices of $S_i$, and so $g$ contains two vertices $x,y$ of $S_i$. 

\begin{description}
    \item[I.a.] If $g$ has some dart internal to $\SX$ ( w.l.o.g. $(a,b)$ in $X_j=X_1$) then, $g$ is a cycle that consists of two $x$-to-$y$ paths, one is internal to $\SX$ (in $X_1$) and the other is external to $\SX$, thus, $g$ must enclose the $x$-to-$y$ subpath of $S_i$ constituting a chord embedded inside the cycle $g$, contradicting that it is a face.

    \item[I.b.] If $g$ has no dart internal to $\SX$, then all of its darts are in external child bags  ($g$ has at least two darts each in  a distinct external child bag). If $g$ encloses a subpath of $S_i$ then we have our contradiction.  Otherwise, $g$ does not enclose anything and its reversal $\rev(g)$ encloses the entire graph. In that case, $g$ has to contain the first and last vertex in all the subpaths $S_i$ for all external bags $X_i$ (since from the construction of the BDD, these are the only vertices of $X_i$ that can be shared with another external child bag, and there is no edge connecting two different child bags). Hence, the only way to close a cycle $g$ that contains darts from different child bags is to contain all these vertices. There is only one face of $X$ that can do that, which by \cite{LP19} is a hole.
    In fact, the endpoints of all $S_i$ are detected in the algorithm of \cite{LP19} by looking for intersections between $g$ (referred to as the {\em boundary} in their paper) and $\SX$. $g$ is a hole because in \cite{LP19} the boundary is defined as the set of $\SH$ edges contained in $X$, where $H\neq X$ is any ancestor of $X$ in $\mathcal{T}$.
\end{description}

\medskip
\noindent
{\bf Case II: } When  $\eX\notin E(G)$.
In this case, there might be one child bag $X_i \neq X_1$ whose corresponding subpath $S_i$ contains $\eX$ (in which case, the above proof of Case I fails as the $x$-to-$y$ subpath might be missing an edge), meaning there is a face that gets partitioned. We argue that there is only one such face, the critical face, and that this can only happen (see \cref{fig: BDD_dteailed} left) if the endpoints of $\eX$ are connected in some $X_i\neq X_1$. 

\begin{description}
    \item[II.a.] When the endpoints of $\eX$ are connected in some $X_i\neq X_1$, then consider what happens if we add $\eX$ as a real edge of $E(G)$ and then remove it. By the proof of Case I, when $\eX$ is in $E(G)$, no face gets partitioned. Then, when we remove $\eX$, the only face that can be affected is the one enclosing both endpoints of $\eX$ (i.e. the critical face). Again there might be a hole that gets partitioned (exactly as in Case I) that we do not care about.

    \item[II.b.]
    When the endpoints of $\eX$ are not connected in any external bag, then no $S_i$ contains both endpoints of $\eX$. This can only happen if the critical face $f$ is a face-part (see \cref{fig: BDD_dteailed} right), for otherwise the endpoints of $\eX$ would be connected by $f$. In this case, the critical face-part $f$ is partitioned between three (and not two) child bags ($X_1$ and two external child bags). However, apart from $f$, no other face gets partitioned. As in Case I, there is a dart of $g$ in $X_1$, and then we can find a subpath of $\SX$ that crosses $g$, a contradiction. The proof follows the same proof as in Case I (all $S_i$ consist of $G$ edges); Otherwise, if no dart of $g$ is in $X_1$, then $g$ is not a cycle, a contradiction. That is, $g$ has to use the endpoints of all $S_i$ and somehow connect them to form a cycle, but the endpoints of $\eX$ are only connected in $X_1$. \qedhere

\end{description}

\end{proof}

The following completes the proof of Property~\ref{BDD_number_face_parts}.
\begin{restatable}{corollary}{corollaryNumberOfPartsInChildren}
\label{corollary_number_of_face_parts_in_children}
    The total number of face-parts resulting from partitioning faces and face-parts of a bag $X\in\mathcal{T}$ between its child bags is $O(D \log^2 n)$.
\end{restatable}
\begin{proof}
    By Lemma~\ref{lemma: number of new face-parts}, in a bag $X$, there are at most $O(\log n)$ face-parts and one critical face that get partitioned between child bags of $X$. We claim that any bag $X$ has at most $O(D\log n)$ child bags (and the corollary follows).
    To see why, recall that $\SX$ is a cycle separator of $X$ s.t its interior defines a child bag, and each external child bag is identified with a distinct (edge-disjoint) subpath of $\SX$. By Property~\ref{property: BDD_separator} of the BDD we have $|\SX|=O(D\log n)$.
\end{proof}

We prove the following, which will be useful in the labeling scheme and in proving properties \ref{BDD_Fx_cut} and \ref{BDD_dual_children} of the BDD. 
\begin{restatable}{lemma}{lemmaEachDartInOneBag}
\label{lemma_each_dart_in_one_bag}
    Each dart $d$ of $G$ belongs to exactly one bag $X$ in each level of $\mathcal{T}$. Moreover, if a dart $d$ is in $X$ and $\rev(d)$ is not in $X$ then $d$ belongs to $\SH$ for some ancestor $H$ of $X$.    
\end{restatable}
\begin{proof}
 In the root of $\mathcal{T}$, we have the graph $G$ and the claim trivially holds.
Assume the claim holds for level $\ell-1$ of the BDD, we prove for level $\ell$.
Recall that for a bag $X$, every edge $e$ of $X$ either belongs to one child bag of $X$ or to two child bags of $X$ (if $e \in \SX$). 
If $e$ is not an edge of $\SX$, then $e$ belongs to one child bag $X_i$ and its (one or two) darts in $X$  belong to $X_i$ (as both of $e$'s endpoints are in $X_i$).  
If $e$ is an edge of $\SX$ that is contained only in one child bag, then by the BDD algorithm  $e$ must be an edge of $\SH$ for some ancestor $H\neq X$ of $X$. Thus, $e$'s only dart is also contained in the same child bag that contains $e$. Finally, if $e$ is an edge of $\SX$ that is contained in two child bags of $\SX$, then $e$ is in $X$ and does not participate in $\SH$ of any ancestor $H$, then, $e$ has both of its darts in $X$. Thus, $e$ belongs to two child bags of $X$, in particular, one child bag is the internal (to $\SX$) and the other is external, each would contain one dart of $e$ as each dart of $e$ is in a face (not a hole), and in this case, $\SX$ separates those two faces (face-parts), each to a child bag. 
\end{proof}

\begin{restatable}{corollary}{FaceInOneChild}
\label{corollary_face_is_in_one_child_bag}
    Each face of $G$ in a bag $X\in \mathcal{T}$, except for the critical face (if exists), is entirely contained in one child bag of $X$. Face-parts of $X$ may get further partitioned between its child bags.
\end{restatable}

\subsubsection{Dual Bags and Their Properties}
\label{section_dual_bags}
We are finally prepared to prove the additional properties~\ref{BDD_dual_leaf_size},~\ref{BDD_Fx_cut}, and~\ref{BDD_dual_children} of the BDD,
starting with a proof for Property~\ref{BDD_dual_leaf_size}.
\begin{restatable}{lemma}{lemmaDualBagSize}
\label{lemma_dual_bag_size}
    The size of a dual leaf bag $X^*$ is $O(D\log n)$.
\end{restatable}
\begin{proof}
    By Property~\ref{property: BDD_leaf_size} of the BDD, we know that $X$ has $O(D\log n)$ vertices. Since $X$ is a simple planar graph, it therefore has at most $O(D\log n)$ edges (Euler formula).
    For each edge $e$ in $X$ there is at most one dual edge $e^*$ in $X^*$ as we define it. Then, $|E(X^*)|=O(D\log n)$, thus, $X^*$ has at most $O(D\log n)$ nodes as well (as each edge has two endpoints). Notice, there are no isolated nodes (i.e, we are not under-counting), as if there were, then they are not connected to any edge, and the only case that a primal edge has no dual is when it lies on a hole, thus, if there were isolated nodes, they would correspond to holes, contradicting the definition of $X^*$.
\end{proof}

Next, we prove Property~\ref{BDD_Fx_cut}.
\begin{lemma}[Dual separator]
        For a non-leaf bag $X^*$, let $\FX$ be the set of nodes whose incident edges are not contained in a single child bag of $X^*$. Then, $\FX$ is a node-cut (separator) of $X^*$ of size $|\FX|=O(D \log n)$. Concretely, $\FX$ is the set of (a) nodes incident to dual edges of $\SX$, and (b) nodes corresponding to faces or face-parts that are partitioned between child bags of $X^*$.
\end{lemma}
\begin{proof}
            First, we show that both definitions of $\FX$ are equivalent. I.e., a node (node-part) of $X^*$ does not have all of its edges in a single child bag if and only if it is contained in: (a) The set of $X^*$'s nodes (node-parts) which get partitioned further into node-parts between its child bags, or (b) The endpoints of edges dual to $\SX$ contained in $X^*$.

        {\em (if)} Note, the dual edges of $\SX$ contained in $X^*$ are not contained in any child bag of it as they lie on holes in them (\cref{lemma_each_dart_in_one_bag}), thus, their endpoints are in $\FX$. In addition, the nodes of $X^*$ that get split correspond to the critical face and a subset of face-parts in $X$ (\cref{corollary_face_is_in_one_child_bag}), thus, they have at least two distinct edges, each in a distinct child bag of $X^*$, hence, are in $\FX$.

        {\em (only if)} Consider a node $f\in \FX$, by definition, some of its incident edges either: (1) Not contained in any child bag of $X^*$, i.e, $f$ is incident to an $\SX$ edge. Or (2) there are at least two edges incident to $f$, each in a distinct child bag of $X^*$. Since a face of $G$ in $X$ is either fully contained in a child bag or is divided across child bags (\cref{corollary_face_is_in_one_child_bag}), we get that $f$ corresponds to the critical face or to one of the face-parts in $X$. 

        From the first definition of $\FX$, it is clear that it constitutes a separating set, as each path that crosses between child bags shall use at least two edges that share a node of $X^*$ which are not contained in a single child bag. I.e, that node is in $\FX$.

        Finally, since there are  at most $O(D\log n)$ edges of $\SX$ (Property~\ref{property: BDD_separator} of the BDD), there are at most twice the number of their endpoints in $X^*$. Moreover, there are at most $O(\log n)$ nodes (node-parts) in $X^*$ that correspond to the faces and face-parts that get partitioned between child bags of $X^*$ (Property~\ref{BDD_number_face_parts} of the BDD).
\end{proof}

Finally, we provide a proof of Property~\ref{BDD_dual_children}. 
\begin{lemma}[Assembling $X^*$ from child bags]
Let $X^*$ be a non-leaf bag with child bags $X^*_1,X^*_2,\ldots$; Then, $X^*$ is equivalent to $\cup_i X^*_i \cup X^*[\SX]$ after connecting all node-parts $f_1,f_2,\ldots$ in $X^*_1,X^*_2,\ldots$ (resp.) corresponding to the same $f\in \FX$ with a clique and contracting it.
\end{lemma}
\begin{proof}
            Let $K_f$ be the clique on the set of node-parts across child bags of $X^*$ corresponding to $f \in \FX$. 
        Denote $M:=\left( \cup_i X^*_i \right) \bigcup \left( X^*[\SX] \right) \bigcup \left( \cup_{f\in \FX} K_f \right)$.
        We start by showing that $M$ is very "close" to $X^*$, then show that the contraction suggested in the property indeed yields $X^*$.

        First, note that all edges of $X^*$ are contained in $M$, as any non-$\SX$ edge is contained in a child bag (\cref{lemma_each_dart_in_one_bag}) and $\SX$ edges that are contained in $X^*$ are added to $M$ by definition. 
        In addition, each node of $X^*$ is present in $M$, except for some nodes of $\FX$, as each node that is in some $X^*_i$ is contained in $M$ by definition, and any node of $X^*$ that is not in any $X^*_i$ must be an $\FX$ node (by Property~\ref{BDD_Fx_cut}). 

        Now, we show that the contraction produces $\FX$ nodes, connects them correctly and removes only non-$X^*$ edges from $M$ (the lemma follows).
        Consider the graph $M'$, given by $M$ after contracting $K_f$ for all $f\in \FX$. We claim that $M'=X^*$.
        \begin{enumerate}
            \item 
        The contraction gets rid of $K_f$ edges, and only them, as distinct node-parts across child bags of $X^*$ that correspond to the same $f\in \FX$ share no other edges. Thus, $M'$ has the exact same set of edges as $X^*$. 
        \item 
        Consider a node $f$ that was partitioned into node-parts $f_1,f_2,\ldots$, such that each $f_i$ is in a distinct child bag of $X^*$. 
        We want to show that contracting all $f_i$ into one node, results with one node in $M'$ that is incident to the same neighbors of $f$ in $X^*$, this node of $M'$ is identified with $f$. That follows from the following claims
        (1) Each edge of $f$ in $X$ goes to one face-part of $f$ in the child bag that contains it (by definition of face-parts). (2) The only edges without a dual in $X^*$ are edges of $\SH$ for some ancestor $H$ of $X$ (\cref{lemma_each_dart_in_one_bag}). Thus, if a primal edge does not have a dual in any of $X^*$'s child bags, it is either an edge of $\SH \neq \SX$, in which case it is not in $X^*$ either (the claim does not fail), or it is in $\SX$, thus, it is added to $M$ by definition.  
        Thus, $f$'s incident edges in $X^*$ are partitioned among $f_i$'s in $M$. Moreover, by the previous item, the contraction affects none of these edges and leaves only them incident to the resulting node,
        which we identify with $f$ in $X^*$. \qedhere
        \end{enumerate}

\end{proof}

\subsubsection{Distributed Knowledge Properties}
\label{section_distributed_knowledge}
To implement our algorithm, each vertex should know the IDs of all faces (face-parts) that contain it in every bag $X\in \mathcal{T}$ (i.e., the list of nodes $f\in X^*$ it participates in).
Here we show that the distributed knowledge properties~(\ref{BDD_know_faces}~and~\ref{BDD_know_edges}) of the BDD hold, which concludes the proof of \cref{theorem_additional_BDD_properties}.
As before, it will be convenient to think of an edge as having two darts. That is, every vertex incident to an edge $e=(u,v)$ considers itself incident to two darts $d^+,d^-$ of $e$.

Even though we think of the two darts of an edge as embedded one on top of the other, it would still be useful to maintain a local order for each vertex over its incident darts, extending its local ordering of edges:
In the local ordering of a vertex $v$ of its darts, a dart $d'$ of an edge $e'$ precedes a dart $d$ of an edge $e$ iff $e'$ precedes $e$ in the local ordering of $v$ for its edges.
To define well the local ordering of $v$ for its darts, it remains to define the order between each two darts that belong to the same edge. I.e., a vertex $v$ incident to an edge $e=(u,v)$ of darts $d^+,d^-$ orders $d^+,d^-$ in its local ordering, s.t. the dart that inters it precedes the dart leaving it. Note, the dart $d^+$ inters $v$ and leaves $u$. Symmetrically, the dart $d^-$ inters $u$ and leaves $v$. Meaning, this ordering is consistent (depends only on the direction of $e$).

We aim to assign each face (face-part) in any bag $X$ a unique $\tilde{O}(1)$ identifier and learn for each vertex $v\in X$, the set of faces (face-parts) containing it and its adjacent darts. 
This is done by keeping track of $G$'s darts along the decomposition. We start with (1) a local procedure that allows the endpoints of each dart $d$ to learn all the bags $X$ that contain $d$, then, (2) the endpoints of each dart $d$ learn the face of $G$ that $d$ participates in, and (3) we show how to extend this for the next level of $\mathcal{T}$. 
This allows us to (distributively) learn the nodes of $X^*$, after which, we have the necessary information to learn the edges of $X^*$.

\begin{restatable}{lemma}{lemmaKnowDarts}
\label{lemma_know_darts}
        In a single round, each vertex $v\in G$ learns for each incident dart, a list of the bags $X\in \mathcal{T}$ that contain it.          
\end{restatable}

\begin{proof}
    We wish to learn, for each dart, the ID of the single bag (by \cref{lemma_each_dart_in_one_bag}) in each level of $\mathcal{T}$ that contains it.
    For the root $G$ of $\mathcal{T}$, all vertices know that all their incident darts belong to $G$. 
    For a general bag $X$, we assume each vertex $v\in X$ knows which of its incident darts belong to $X$, and we show how to learn this for every child bag of $X$ that contains $v$. 
    If $v$ is in only one child bag $X_i$ of $X$ ($v$ knows it is in $X_i$), then $v$'s incident edges in $X_i$ are the same as in $X$. 
    Otherwise, $v$ must be a vertex of $\SX$ (and by the BDD properties, $v$ knows this). For each child bag $X_i$ that contains $v$, $v$ knows two incident edges $a,b$ such that: (1) $a$ precedes $b$ in $v$'s cw local ordering of incident edges, and (2) $v$'s incident edges in $X_i$ are exactly $a,b$ and all edges $I_v(X_i)$ strictly between $a$ and $b$ in $v$'s local ordering of incident edges. 
    Thus, each child bag that contains $v$ is identified with a pair of ordered edges incident to $v$. Observe that one of those edges might be the virtual edge $\eX$ (if $v$ is an endpoint of $\eX$). Though $\eX$ is not an actual edge of $X$, it (and its embedding) is known to its endpoints by the BDD construction. 
    To conclude, for any dart $d$ of an edge $e$ incident to $v$, $v$ can learn if that dart is in $X_i$ by checking if $d$ is the dart of $a$ leaving $v$, or $d$ is the dart of $b$ that inters $v$, or $e\in I_v(X_i)$.
    \end{proof}

We next show how, for all bags $X$, every vertex $v \in X$ learns the IDs of faces (face-parts) that contain it, and for each of its darts $d$ in $X$ the ID of the face (face-part) in $X$ containing $d$. We begin by showing this for $X=G$ and then for a general bag $X$.

\begin{restatable}{lemma}{lemmaKnowDartsofG}
\label{lemma_know_darts_G}
    There is a $\tilde{O}(D)$-round distributed algorithm that assigns unique IDs to all faces of $G$. By the end of the algorithm, each vertex $v\in G$ learns a list (of IDs) of faces that contain it. Moreover, $v$ learns for each incident dart $d$ of an incident edge $e$ in $G$ the face ID of the face of $G$ that contains $d$. 
\end{restatable}

\begin{proof}

    We first assign unique $O(\log n)$-bit IDs to the faces of the graph $G$ and let all vertices learn the IDs of all the faces of $G$ that contain them. We can do so in $\tilde{O}(D)$ rounds by using Property \ref{property: hat(G)_face_identification} of the face-disjoint graph (Section~\ref{sec: face_disjoint_graph}). 
    Once a vertex $v$ knows all faces containing it, it remains only to assign those faces to the darts incident to $v$ (which is done locally).

    Recall, each edge $e$ of $G$ has two darts $d^+,d^-$ in the set of edges $E_R$ of $\hat{G}$, and each vertex $v$ of $G$ has $\deg(v)$ many copies $v_i$ in $\hat{G}\setminus V_S$. Moreover, $\hat{G}[E_R]$ is a collection of disjoint cycles, each corresponding to a face of $G$.
    Let $e=(u,v)\in E(G)$ be the $i$'th and $j$'th edge in $u$'s and $v$'s local embedding (in $G$), respectively. Then, the copies $u_{i+1},v_j$ are connected by an edge in $E_R$ identified with the dart that inters $v$ and leaves $u$ of $e$ ($d^+$ since $e$ is directed from $u$ to $v$), and the copies $u_{i},v_{j+1}$ are connected an edge in $E_R$ identified with the dart that inters $u$ and leaves $v$ of $e$ ($d^-$ in this case).
    Note, this order of darts is the same order that we keep in other algorithms (lemmas) for consistency.
    We assign each of $d^+$ and $d^-$ the component (face) ID that contains them in $\hat{G}[E_R]$.
    Since $u,v$ in $G$ simulate all their darts in $\hat{G}$, each of them knows for each of $d^+,d^-$ the connected component ID in $\hat{G}[E_R]$ it participates in. This is the ID of the face that corresponds to that component. Thus, for each of $d^+,d^-$, both $u$ and $v$ learn the correspondence to faces of $G$.
    \end{proof}

The IDs assigned to faces and face parts are defined such that the ID of a face (face-part) $f$ of $X$ is contained in the ID of face-parts resulting from partitioning $f$ in child bags of $X$. Thus, we get the following lemma and corollary proving  property~\ref{BDD_know_faces} of the BDD:
\begin{restatable}[Distributed knowledge of faces]{lemma}{lemmaKnowFaces}
\label{lemma_know_faces}
    There is a $\tilde{O}(D)$-round distributed algorithm that assigns unique $\tilde{O}(1)$-bit IDs to all faces and face-parts in all bags $X\in \mathcal{T}$. By the end of the algorithm, each vertex $v\in G$ 
    learns a list (of IDs) of the faces and face-parts it lies on for each bag $X$ that contains it. Moreover, $v$ learns for each incident dart $d$ the (ID of) the face or face-part that $d$ participates in for each bag $X$ that contains $d$. The ID of a face (face-part) $f$ of $X$ is contained in the ID of face-parts $f'$ resulting from partitioning $f$ in child bags of $X$.
\end{restatable}

\begin{restatable}{corollary}{corollaryKnowFacesToPartition}
\label{corollary_know_face_to_partition}
    In $\tilde{O}(D)$ rounds, for each bag $X$, all vertices know for each of their incident darts whether they participate in the critical face (face-part) and whether they participate in any face-part of $X$.
\end{restatable}

\begin{algorithm}[htb]
    \caption{Distributed Knowledge of Faces}
    \label{algorithm_distributed_knowledge_of_faces}

    \KwIn{A BDD $\mathcal{T}$ of a $D$-diameter network $G$ and a BFS tree $T_{\!_X}$ for each $X\in\mathcal{T}$.}
    \KwResult{Each vertex $v$ of $G$ learns for each incident dart $d$ the list of faces and face-parts that contain $d$ in each bag $X\in\mathcal{T}$ that contains $d$.}

    Each vertex learns for each incident dart $d$ the list of bags that contains $d$ (\cref{lemma_know_darts})\;

    Each vertex learns for each incident dart $d$ the list of faces that contain $d$ in the bag $X=G$ (\cref{lemma_know_darts_G})\;

    \ForAll{bags $X$ of level $\ell\in [2,\Depth(\mathcal{T})]$, in parallel, }{

    \tcc*[f]{Global step}

     Detect the critical face (face-part) of $X$ if exists and broadcast a message on $T$ notifying all vertices that it shall be partitioned\;

     For each edge that participates in a face-part of $X$, broadcast the ID of the face-part and the child bag that contains it to the root of $T$\;

     The root of $T$ broadcasts to all vertices of $X$ the IDs of face(-parts) that shall be partitioned in between child bags of $X^*$\;

    \tcc*[f]{Local Step}

    Each vertex of $X$ learns locally for each incident dart $d$ the new ID of each face(-part) that contains $d$ in each child bag of $X^*$ that contains $d$ (if any)\;
    
    }
    
\end{algorithm}

\begin{proof}[Proof of \cref{lemma_know_faces}.]
    Throughout, we assume that for each $X\in\mathcal{T}$ we already computed a BFS tree $T_x$ of $X$ rooted at an arbitrary vertex of $X$. The subscript is dropped when $X$ is not ambiguous. This is done in $\tilde{O}(D)$ rounds using the properties of the BDD.
    The algorithm described in the proof is given in a high-level pseudocode by Algorithm~\ref{algorithm_distributed_knowledge_of_faces}. 

    First, we apply Lemma~\ref{lemma_know_darts} in order for each vertex to learn for each incident dart the bags that contain it in the decomposition. Then, for $X=G$ we apply Lemma~\ref{lemma_know_darts_G} assigning each dart a face in $G$.   
    For a general bag $X$, we assume we have the desired information and show how to maintain it for child bags of $X$.
    For a face or face-part $f$ that is contained in $X$ and is entirely contained in one of its child bags $X_i$, 
    endpoints of darts in $f$  inherit the ID of $f$ from $X$.
    However, in the case where $f$ gets partitioned between child bags of $X$, we need to identify each new face-part and assign it a unique ID.
   
    We work in two steps. First, a global broadcast over $X$ is performed in order to detect the faces and face-parts of $X$ that shall be partitioned. 
    Then, a local step of assigning IDs to the new resulting face-parts in the child bags of $X$.

   \medskip
    \noindent
    {\bf 1. Global step.}
    In this step we detect the face of $G$ and all face-parts $f$ in $X$ to be partitioned (if any). 
    We start with the unique (by Lemma~\ref{lemma: number of new face-parts}) critical face (or face-part) of $G$ in $X$. The endpoints of $\eX$ lie on $f$, are on $\SX$ and know if $\eX$ is virtual or not (if it is, then $f$ gets partitioned).
    Therefore, $f$ can be detected by the endpoints $u,v$ of $\eX$ by the BDD construction (i.e., using the algorithm of \cite{GP17} that \cite{LP19} use for computing a separator). 
    More concretely, the edge $\eX$ is embedded between two consecutive incident edges $e_1,e_2$ in  the clockwise embedding of $v$ ($e_1,\eX,e_2$ appear in that order in the embedding of $v$). Then, the dart of $e_1$ that leaves $v$ and the dart of $e_2$ that inters $v$ (in that order) define the face $f$. A similar argument holds for $u$. 
    Henceforth, an arbitrary endpoint of $\eX$ broadcasts the ID of the critical face (face-part) on $T$ saying it shall be partitioned as well (if it does).

    After identifying the critical face, we move on to identifying the other face-parts that shall be partitioned.
    We exploit the small (logarithmic) number of face-parts in $X$ (Lemma~\ref{lemma: number of new face-parts}) for performing the following broadcast.  
    Each vertex $v$ on a face-part $f$ of $X$ knows (its incident darts in) $f$. Then, $v$ upcasts $ID(f),ID(X_i)$ to the root of the BFS tree $T$ of $X$, where $X_i$ is a child bag that contains darts of $f$ incident to $v$ (the correspondence between darts and child bags is known from Lemma~\ref{lemma_know_darts}).
    We need to be careful in this upcast not to suffer unnecessary congestion, as a face-part might consist of many vertices and more than one connected component.
    So, if some vertex of $X$ receives the same message more than once, it passes it on only the first time. In particular, if a vertex of $X$  receives more than one child bag ID for the same face-part $f$, it stops passing messages regarding $f$, and upcasts only one special message to the root of $T_{\!_X}$ saying that $f$ shall be partitioned. 
    The upcast is performed in a pipelined method.
    Finally, after the root stops receiving messages, it broadcasts the IDs of all face-parts that shall be partitioned.
     
\medskip
    \noindent
    {\bf 2. Local step.}   
    For each face or a face-part $f$ that is not partitioned, the vertices lying on $f$ consider the ID of $f$ to be the same as from the parent level and remember that it also exists as a whole in the child bag it is contained in. 
    For a face or a face-part $f$ that is partitioned, the vertices $v$ of $f$ know for each incident dart $d$ that belongs to $f$ in which child bag $X_i$ $d$ is contained (Lemma~\ref{lemma_know_darts}). Thus, $v$ locally learns $(ID(f),ID(X_i))$ as the ID of the face-part that contains $d$ in $X_i$. We get that all vertices on a face-part of $f$ in some child bag of $X$ know the same ID for that face-part.    
    
    \medskip
    We claim that this procedure terminates within $\tilde{O}(D)$ rounds. 
    First, the IDs of face-parts are of size $\tilde{O}(1)$ since (1) by Lemma \ref{lemma_know_darts_G} the faces in $G$ are assigned IDs of size $O(\log n)$, (2) by the construction of the BDD in~\cite{LP19}, the ID of a child bag is of size $\tilde{O}(1)$ (and there are $O(\log n)$ levels of $\mathcal{T}$), and (3) a face-part in a level $\ell$ is assigned the ID of a face (or face-part) of level $\ell-1$ after appending $\tilde{O}(1)$ additional bits.
    Second, by Lemma~\ref{lemma: number of new face-parts} the number of face-parts in $X$ is at most $O(\log n)$ and there is at most one critical face (or face-part). Meaning, any vertex shall pass at most $O(\log n)$ messages of size $\tilde{O}(1)$ throughout the course of the procedure (one message for the critical face and one per  face-part of $X$). Finally, the BFS tree used for broadcast has a diameter of $O(D \log n)$ (Property~\ref{property: BDD_diameter} of the BDD), and all bags $X$ of the same level run the procedure in parallel with a constant factor overhead (Property~\ref{property: BDD_parallel_work_in_level} of the BDD).
\end{proof}

After each vertex $v$ knows the necessary information for each dart of an incident edge $e$, in an additional round, $v$ knows if $e$ exists in $X^*$, its weight in $X^*$ and its direction, constituting  property~\ref{BDD_know_edges} of the BDD.
\begin{restatable}{lemma}{lemmaKnowDualEdges}
\label{lemma_know_dual_edges}
    Let $e=(u,v)$ be an edge in a bag $X$. In one round after applying \cref{lemma_know_faces}, the endpoints of $e$ learn the corresponding dual edge $e^*$ in $X^*$ (if exists).
\end{restatable}
\begin{proof}
    Recall that by the definition of the dual bag, there is a dual edge $e^*$ between two faces or face-parts $f_1,f_2$ of $X$ (on both sides of $e$) iff the two darts of $e$ are contained in $X$ (i.e., none of them is on a hole).  By invoking Lemma~\ref{lemma_know_faces} and Lemma~\ref{lemma_know_darts}
    $v$  knows which of its incident edges $e$ has both darts in $X$, and the ID of the face (face-part) each dart participates in.
    The endpoints of such edges $e$ compute the correspondence locally as follows.
    Consider such an edge $e$ incident to a vertex $v\in X$ and let $f_1, f_2$ be the faces containing the darts $d^+, d^-$ of $e$, respectively. Then, the corresponding dual edge is directed from $f_1$ to $f_2$.
    Notice, this definition is consistent (independent from $v$) and relies only on the direction of $e$ and on faces it participate in. 
    If $G$ is undirected, then dual edges are undirected as well. 
    If $G$ is weighted, then, $e^*$ has the same weight as $e$.
   \qedhere

\end{proof}

\subsection{The Labeling Scheme}
\label{section_distance_labels}
Recall that, for every dual bag $X^*$, we wish to label the nodes of $X^*$, such that, from the labels alone of any two nodes we can deduce their distance in $X^*$.  
Recall, we refer to the dual node $f'$ in a child bag $X^*_i$ as a {\em node-part} of a node $f\in X^*$, if $f'$ corresponds to a face-part in $X_i$ resulting from partitioning the face (face-part) $f$ of $X$.

We already know that the set $\FX$ plays the role of a node-cut (separator) of $X^*$, and is crucial for the labeling scheme and algorithm. The main structural property that $\FX$ provides is: Any path in $X^*$ either intersects $\FX$ or is entirely contained in a child bag of $X^*$.

Note, a node $g\notin \FX$ of $X^*$ corresponds to real faces of $G$ (i.e. not face-parts) contained in $X$. The distance label $\Label_{X^*}(g)$ of $g\notin \FX$  is defined recursively.
If $X^*$ is a leaf-bag then $\Label_{X^*}(g)$ stores ID($g$), ID($X$) and the distances between $g$ and all other nodes $h\in X^*$. Otherwise, the label consists of the ID of $X$, the distances in $X^*$ between $g$ and all nodes of $\FX$, and (recursively) the label of $g$ in the unique (by Corollary~\ref{corollary_face_is_in_one_child_bag}) child bag  $X^*_i$ of $X^*$ that entirely contains $g$. Formally:
\[\Label_{X^*}(g)=(ID(X),ID(g),\{ID(f),\dist_{X^*}(f,g),\dist_{X^*}(g,f):{f\in \FX}\},\Label_{X^*_i}(g))\]
For the set of nodes $g\in \FX$, we remove the recursive part, and define   
\[\Label_{X^*}(g)=(ID(X),ID(g),\{ID(f),\dist_{X^*}(f,g),\dist_{X^*}(g,f):{f\in \FX}\})\]

As we showed in property~\ref{BDD_Fx_cut}, faces and face-parts that correspond to nodes in $\FX$ are the interface of the child bags of $X^*$ with each other. 

\begin{restatable}[Property~\ref{BDD_Fx_cut} of the BDD, rephrased]{lemma}{lemmaSufficentLabels}
\label{lemma_sufficent_labels}
    Let $g,h$ be two nodes in a non-leaf bag $X^*$, the (edges of the) shortest $g$-to-$h$ path in $X^*$ is either entirely contained in some $X^*_i$ or intersects $\FX$.
\end{restatable}

Thus, a path in $X^*$ from $g\in X^*_i$ to $h\in X^*_{j\neq i}$ must pass through some (part of a) node $f\in \FX$. I.e., a $g$-to-$h$ shortest path $P$ in $X^*$ is either: (1) Entirely contained in $X^*_i$, then $\dist_{X^*}(g,h)$ can be deduced from $\Label_{X^*_i}(g)$ and $\Label_{X^*_i}(h)$. Or (2), $P$ intersects $\FX$, then $\dist_{X^*}(g,h) = \min_{f\in \FX} \{\dist_{X^*}(g,f)+\dist_{X^*}(f,h)\}$.  
If one of $g$ or $h$ is a node of $\FX$ then distances are retrieved instantly, as each node in $X^*$ stores its distances to $\FX$. 
 
By this reasoning,
obtaining the distance between two nodes $g,h$ in a leaf bag is trivial, as all pairwise distances are saved in each label.
For a non-leaf bag $X^*$, by Lemma~\ref{lemma_sufficent_labels}, the information saved in the labels of $g,h$ in $X^*$ is sufficient in order to obtain the $g$-to-$h$ distances in $X^*$, constituting the following lemma. 
\begin{restatable}{lemma}{lemmaCorrectLabels}
\label{lemma_correct_labels}
    Let $g,h$ be two nodes in $X^*$, then we can decode their distance in $X^*$ from $\Label_{X^*}(g)$ and $ \Label_{X^*}(h)$ alone.
\end{restatable}

Note, we define the labels in $O(\log n)$ steps from the leaf to the root of $\mathcal{T}$, in each step, we append to the label at most $\tilde{O}(D)$ bits for distances from $\FX$ nodes. I.e., we have $|\FX|=\tilde{O}(D)$ (Property~\ref{BDD_Fx_cut} of the BDD), the weights are polynomial, the size of a leaf bag is $O(D\log n)$ (Property~\ref{BDD_dual_leaf_size} of the BDD) and IDs are of $\tilde{O}(1)$-bits (Property \ref{property: BDD_ID} and  Property~\ref{BDD_know_faces}). Hence, 
\begin{restatable}{lemma}{lemmaLabelSize} \label{lemma_label_size}
    The label of any node $g$ in a bag $X^*$ of $X\in\mathcal{T}$ is of size $\tilde{O}(D)$ bits.
\end{restatable}

\subsection{The Labeling Algorithm}
\label{section_dual_labeling_algorithm}
Here we show how to compute the labels of all nodes in all bags $X^*$ of $\mathcal{T}$. At the end of the algorithm, each vertex in a face $g$ of $X$ shall know the label of $g$ in $X^*$.
The planar network of communication $G$ is assumed to be directed and weighted, let $w(e)$ denote the weight of an edge $e$.
The BDD tree $\mathcal{T}$ of $G$ is computed in $\tilde{O}(D)$ rounds, by \cref{lem: BDD} and \cref{theorem_additional_BDD_properties}, such that, each (primal) vertex $v$ knows all bags $X$ that contain it, the IDs of all faces (face-parts) of $X$ that contain it, and the dual edges corresponding to each of its incident edges in $X$ (if any).

We compute the label of every dual node $g\in X^*$ for all $X^*\in \mathcal{T}$ in a bottom-up fashion, starting with the leaf-bags. For a non-leaf bag $X^*$, we have two main steps: (i) broadcasting labels of (parts of the dual separator) $\FX$ nodes contained in the child bags of $X^*$, broadcasting the edges dual to (the primal separator) $\SX$  and (ii) using the received information locally in each vertex of $X$ for computing labels in $X^*$.
For a high-level description, see Algorithm~\ref{algorithm_dual_labeling}.

\begin{algorithm}[htb]
    \caption{Dual Distance Labeling (high-level description) }
    \label{algorithm_dual_labeling}

    \KwIn{A BDD of a $D$-diameter network $G$ satisfying the  properties of \cref{theorem_additional_BDD_properties}.}
    \KwResult{Each vertex $v$ of $G$ learns $\Label(g)$ in $G^*$ for each face $g$ of $G$ that contains it. }

    \ForAll{bags $X$  of level $\ell\in [\Depth(\mathcal{T}),1]$, in parallel, }{
    \If{$X$ is a leaf-bag}{
    Collect the entire graph $X$ in each vertex $v\in X$\;
    
    }
    \Else(\tcc*[f]{Non-leaf bags}){
    
    \tcc*[f]{Broadcast step}

    Broadcast the dual edges of $\SX$ edges that exist in $X^*$ to the entire bag $X$\;

    Broadcast, in two steps, the labels computed in the previous level (in child bags of $X^*$) of faces (face-parts) that correspond to  $\FX$ to the entire bag $X$.

    \tcc*[f]{Local Step}

    Each vertex $v\in X$ uses the received information in the above broadcast in order to locally construct the Dense Distance Graph for each face (face-part) $g$ that contains it in $X$\;
}

    Each vertex $v\in X$ computes locally $\Label_{X^*}(g)$ for each face (face-part) $g$ of $X$ that contains it or reports a negative cycle aborting the algorithm\; 
    
    }
    
\end{algorithm}

\medskip
\noindent
{\bf Leaf bags.}  
For leaf-bags $X^*$, we show that we can broadcast the entire bag $X^*$ in $\tilde{O}(D)$ rounds. First, by Property~\ref{BDD_know_faces}, each vertex of $X$ knows all the IDs of faces of $X$ (both real faces of $G$ and face-parts) that contain it. These faces are exactly the nodes of the graph $X^*$, by Property~\ref{BDD_dual_leaf_size}, $|X^*|=\tilde{O}(D)$. That is, we can broadcast them to the whole bag in $\tilde{O}(D)$ rounds. Note that it may be the case that several different vertices broadcast the same ID, but since overall we have $\tilde{O}(D)$ distinct messages, we can use pipelined broadcast to broadcast all the IDs in $\tilde{O}(D)$ rounds. We can work simultaneously in all leaf bags, as each edge of $G$ is contained in at most two bags of the same level (Property \ref{property: BDD_parallel_work_in_level} of the BDD). Next,  we broadcast the edges of $X^*$. From Property~\ref{BDD_know_edges}, for each edge $e$ in $X$ the endpoints of the edge know if this edge exists in $X^*$, and if so they know the corresponding dual edge in the dual bag $X^*$.
We let the endpoint of $e$ with smaller ID broadcast a tuple ($ID(e^*)$, $w(e)$) to the bag, where $ID(e^*)=(ID(f),ID(g))$ if $e^*$ is directed towards $g$ in $X^*$. Notice, $X^*$ might be a multi-graph, thus, there might be multiple edges with the same ID of $e^*$. This does not require any special attention, we broadcast all of them as the total number of edges in $X^*$ is conveniently bounded.
Since for each edge we send $\tilde{O}(1)$-bits and in total we have $\tilde{O}(D)$ edges in $X^*$ (Property~\ref{BDD_dual_leaf_size}), the broadcast terminates within $\tilde{O}(D)$ rounds. Afterwards, all the vertices of $X$ know the complete structure of $X^*$.
Then, all vertices of $X$ can locally compute the labels of nodes of $X^*$ by locally computing the distances in the graph $X^*$. 
Finally, since vertices have full knowledge of $X^*$, they can check for negative cycles, if found, a special message is sent over a BFS tree of $G$, informing all vertices to abort.

\medskip
\noindent
{\bf Non-leaf bags. } 
We assume we have computed all labels of bags in level $\ell$ of $\mathcal{T}$, and show how to compute the labels of all bags in level $\ell-1$ simultaneously in $\tilde{O}(D^2)$ rounds. We describe the procedure for a single bag $X^*$ with child bags $X^*_1,X^*_2,\ldots$, by Property~\ref{property: BDD_parallel_work_in_level} of the BDD we can apply the same procedure on all bags of the same level without incurring more than a constant overhead in the total round complexity. The computation is done in two steps, a broadcast step and a local step.

\medskip
\noindent
{\bf  Broadcast step.}
We first broadcast the following information regarding $\FX$: (1) The edges of $\SX$ contained in $X^*$, (2) The labels of nodes (node-parts) of $\FX$ in the child bags of $X^*$.  I.e, if $f\in \FX$ is entirely contained in a child bag $X^*_i$ of $X^*$, we broadcast $\Label_{X^*_i}(f)$. Otherwise, $f$ is the critical face of $X$ or a face-part of $X$, $f$ may get partitioned to $f_1,f_2,\ldots,f_k$ in between the child bags of $X$. Thus, for each such $f\in \FX$, we broadcast the labels $\Label_{X^*_j}(f_i)$, where $X^*_j$ is the child bag that contains the node-part $f_i$.
Using this information, we shall compute the distances from any node $g$ to all nodes of $\FX$.
By induction, for each face (face-part) $f$ of a child bag $X_i$, every vertex $v$ that lies on $f$ knows the label $\Label_{X^*_i}(f)$ of $f$ in $X^*_i$.
\begin{enumerate}
\item 
Since each (primal) vertex $v$ knows its incident $\SX$ edges (by the BDD construction), $v$ also knows for each such incident edge whether its dual is in $X^*$ or not (By Property~\ref{BDD_know_edges}).
Thus, we can broadcast edges in $\SX$ contained in $X^*$ to the whole bag $X$. For each edge, one of its endpoints (say the one with smaller ID) broadcasts the edge (i.e., an $\tilde{O}(1)$-bit tuple $(ID(e^*),w(e))$).
The broadcast terminates in $\tilde{O}(D)$ rounds as $|\SX|=\tilde{O}(D)$ (Property~\ref{property: BDD_separator} of the BDD).
\end{enumerate}

 Next, we explain how to broadcast labels of nodes (node-parts) that correspond to $\FX$ in child bags of $X^*$. We do that in two steps, a step for nodes of $\FX$ that get partitioned further in child bags, and a step for the remaining nodes, which by definition of $\FX$, have to be endpoints of edges in $\SX$.
\begin{enumerate}
\setcounter{enumi}{1}

\item 
Consider a face or a face-part $f$ of $X$ that gets partitioned into parts $f_1,\ldots,f_k$ in between child bags of $X$. For every $f_i$, vertices $v$ lying on $f_i$ broadcast the label $\Label_{X^*_j}(f_i)$ to the entire graph $X$ (where $X^*_j$ is the child bag containing $f_i$). 
By Property~\ref{BDD_know_faces}, each vertex $v$ in $X$ on a face or a face-part $f$ knows $f$ as such, if it is partitioned, $v$ also knows the IDs of all parts $f_i$ that contain it. Thus, $v$ can broadcast the relevant labels.
The broadcast is pipelined, s.t. if a vertex receives the same message more than once, it passes that message once only.
A label is of $\tilde{O}(D)$-bits (Lemma~\ref{lemma_label_size}), the diameter of $X$ is  $\tilde{O}(D)$ (Property \ref{property: BDD_diameter} of the BDD) and by Property~\ref{BDD_number_face_parts} there are at most $\tilde{O}(D)$ face-parts in the child bags of $X$, i.e., at most $\tilde{O}(D)$ distinct labels get broadcast, thus, the broadcast terminates within $\tilde{O}(D^2)$ rounds.

\item 
By the first item, vertices incident to an edge $e\in \SX$ s.t. $e^*\in X^*$ know the IDs of the endpoints $f,g$ of $e^*$. Thus, if $f$ is not the critical face nor a face-part (i.e., $f$ is entirely contained in $X^*_i$), any vertex $v$ contained in $f$ knows this about $f$ (Property~\ref{BDD_know_faces}), then $v$ broadcasts the label $\Label_{X^*_i}(f)$ of $f$.
Notice that we send the labels of all (parts of) endpoints of $\SX$ in $X^*$ (if they are face-parts they get handled in the previous broadcast, if not, they get handled in this one).

Again, the broadcast is pipelined, and each vertex passes the same message only once.  There are at most $|\SX|=\tilde{O}(D)$ distinct labels being broadcast at any given moment.
The size of a label is $\tilde{O}(D)$, the diameter of $X$ is  $\tilde{O}(D)$. Thus, the broadcast terminates in $\tilde{O}(D^2)$ rounds.

\end{enumerate}

\medskip
\noindent
{\bf  Local step.}
For every face or face-part $g$ in $X$, the vertices lying on $g$ compute locally $\Label_{X^*}(g)$. 
Intuitively, this is done by decoding all pairwise distances of labels that were received in the previous step, then, constructing a graph known as the  {\em Dense Distance Graph (DDG)} and computing distances in it. The structure of this graph is described next, after that we shall prove that it indeed preserves distances in $X^*$.

\begin{figure}[htb]
    \centering
    \includegraphics[width=0.5\linewidth]{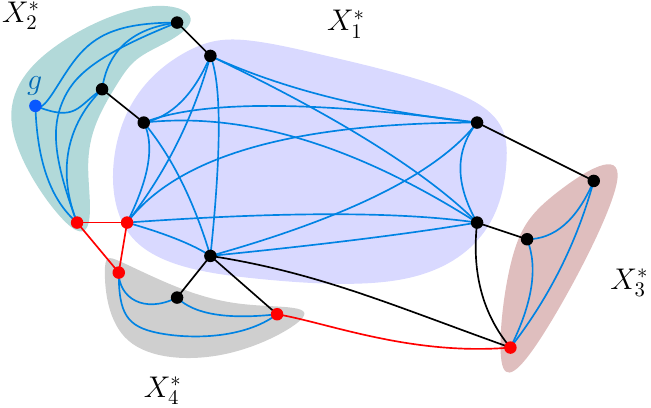}
    \caption{
    The graph $DDG(g)$. The blue node is $g$, incident red nodes correspond to face-parts of the same face  of $X^*$ across its different child bags, and black nodes correspond to endpoints of $\SX$ edges.
    The blue edges represent the clique on nodes that belong to the same child bag of $X^*$, red edges are bidirectional edges of  zero weight connecting face-parts of the same face in $X^*$, and black edges are dual edges to $\SX$ edges. For simplicity, the figure is undirected.
    }
    \label{fig: appendix_DDG}
\end{figure}

The DDG (see Figure~\ref{fig: appendix_DDG}) is a (non-planar) graph that preserves the distances in $X^*$ between a subset of nodes of $X^*$.
Traditionally, the DDG is a union of cliques on a (primal) separator, each clique representing distances inside a unique child bag, and the cliques overlap in nodes. 
In our setting, the DDG is slightly more complicated: (1) the nodes of the DDG are not exactly the separator nodes $\FX$, but nodes corresponding to faces and face-parts of $\FX$ in child bags of $X$. (2) different cliques do not overlap in nodes, instead, their nodes are connected by (dual) edges of $\SX$ that are also added to the DDG. (3) edges connecting face-parts of the same face but in different child bags are added to the DDG with weight zero.

Formally, the graph $DDG(g)$ is a weighted directed graph. The node set of $DDG(g)$ consists of (1) $g$, (2) the nodes in child bags of $X$ that correspond to nodes (node-parts) of $\FX$. 
Denote by $V_i$ the set $V(DDG(g))\cap V(X^*_i)$ for a child bag $X^*_i$ of $X^*$, then, the edge set of $DDG(g)$ is defined as:
\begin{itemize}
    \item An edge between every two nodes of $DDG(g)$ that belong to the same child bag of $X^*$. 
    The weight of such an edge $(f,h)\in V_i\times V_i$ is the $f$-to-$h$ distance in the graph $X^*_i$. By Lemma \ref{lemma_correct_labels}, each such distance can be obtained by pairwise decoding two labels of nodes (received in the broadcast step) that belong to the same child bag $X^*_i$.
    	
     \item The (dual) edges of $\SX$ which have both endpoints in $X^*$ (i.e., not edges incident to a hole). For each such edge $e=(f,h)$, it is either the case that $f$ exists in $X^*$ and in some $X^*_j$ or $f$ is divided into parts $f_1,f_2,\ldots$ in the child bags of $X^*$. In the first case, $e$ is incident to $f$ in the DDG. In the second case, $e$ is incident to the part $f_i$ of $f$ that contains $e$. 
     Notice, in case $f$ is not partitioned then it is in the DDG as it is in $\FX$ for being an endpoint of $e\in \SX$. If it does get partitioned, then all its parts are in the DDG as they are face-parts corresponding to a node of $\FX$.
     The weight and direction of $e$ is as in $X^*$.
     In either case, the label of those nodes ($f$ or its parts) got broadcast in the broadcasting step. In addition, the information about $e$ (ID, direction and weight) is known in vertices of $X$ also by the broadcast. Thus, vertices can know locally what nodes shall be incident to edges $e\in \SX$ that are contained in~$X^*$.

    \item A bidirectional zero weighted edge between each two nodes $f_i\in V_i$, $f_j\in V_j$ that correspond to node-parts in $X_i$, $X_j$ (respectively) of the same node $f$ in $\FX$.
    Again, this can be known locally in any vertex $v$ of $X$ since $v$ received the labels (which contain the IDs) of $f_i,f_j$ in the broadcast step, and since the correspondence of $f_i,f_j$ to $f$ can be known from the information received in the broadcast (specifically, the IDs of $f_i, f_j$ contain the ID of $f$ by \cref{lemma_know_faces}). 
\end{itemize}

Notice, the description of $DDG(g)$ does not directly address the case where $g$ is not contained in exactly one child bag of $X^*$ (i.e., $g$ gets partitioned). However, if that happens, we have a special case where (by definition) $g\in \FX$, and all of its parts are represented in the DDG. Meaning, the definition of $DDG(g)$ for a node $g\in X^*$ is general enough for both cases ($g\in \FX$ and $g\not \in \FX$). 

After each vertex $v$ on the face or face-part corresponding to $g\in X^*$ locally constructs $DDG(g)$ and computes distances in it, $v$ checks for a negative cycle. If found, $v$ broadcasts a special message informing all vertices of $G$ (via a global BFS tree of $G$ in $O(D)$ rounds) to abort their algorithm.
Otherwise, $v$ locally constructs $\Label_{X^*}(g)$ by appending distances computed in $DDG(g)$ to the recursive label of $g$ (if exists). That is, $v$ sets $\dist_{X^*}(g,f)$ (resp. $\dist_{X^*}(f,g)$), where $f\in \FX$, to be $\dist_{DDG(g)}(g,f')$ (resp. $\dist_{DDG(g)}(f',g)$) where $f'$ is $f$ (in case $f\in V(DDG(g)$) or is an arbitrary part of $f$ in some child bag of $X$. Due to the fact that all parts $f'$ of $f$ are connected by zero weight edges, then the distance to any of them is exactly the distance to all others.

\begin{restatable}{lemma}{lemmaAlgorithmCorrect}
\label{lemma_algorithm_computes_correct_labels}
    For all bags $X^*\in \mathcal{T}$, the above algorithm computes correct label as described in Section~\ref{section_distance_labels} for all nodes in $X^*$.
\end{restatable}

In order to see why does the above hold, we need to show that $DDG(g)$ of a node $g\in X^*$ is distance preserving w.r.t. the shortest paths metric of $X^*$.

\begin{proof}
    For leaf bags, there is no DDG and the labels of all nodes get computed in the first step of the algorithm. 
    For non-leaf bags, it remains to show that the distances computed between $g \in X^*$ and all $f\in \FX$ via $DDG(g)$ are correct in $X^*$.
    If $f\in \FX$ is contained in some child bag of $X^*$ then it has a node in $DDG(g)$. If not, then each of its parts has a node. These nodes are all connected by zero weight edges, thus, contracting this clique to a single node in the $DDG(g)$ does not affect distances in the DDG. 
    By  Property~\ref{BDD_dual_children} of the BDD, the resulting graph is a graph on a subset of nodes of $X^*$, which by construction  of $DDG(g)$ are connected with edges that represent paths in $X^*$ (the weights and direction were obtained by decoding labels in child bags of $X^*$, which are correct by Lemma~\ref{lemma_correct_labels}). So the distances in the DDG are at most their real distances in $X^*$. Clearly they are also at least the real distances, as all edges represent paths in $X^*$. \qedhere     
\end{proof}

Note, if there is a negative cycle in $X^*$ it gets detected in the leaf-most descendant of $X^*$. 
\begin{restatable}{lemma}{lemmaNegCyclesAlgorithmCorrect}
\label{lemma_algorithm_detects_neg_cycles}
    A negative cycle in $X^*$ gets detected if and only if it exists. 
\end{restatable}
\begin{proof}
If $X^*$ has no negative cycle then we never report a negative cycle because by Lemma~\ref{lemma_algorithm_computes_correct_labels} all the distances we consider are correct distances in $X^*$.  
    If $X^*$ does have a negative cycle, let $X^*$ be a leafmost such bag. 
    If $X^*$ is a leaf-bag, then we detect the negative cycle since each vertex of $X$ collects the whole graph $X^*$ and checks for it. For a non-leaf bag $X$, by Lemma~\ref{lemma_sufficent_labels}, all paths (and thus cycles) in $X^*$ are either contained in child bags of $X^*$ or cross the set $\FX$ of nodes.
    Since $X^*$ is a leafmost bag containing a negative cycle, the negative cycle must contain edges of at least two child bags. Therefore, it can be divided into subpaths that cross $\FX$ (\cref{lemma_sufficent_labels}). For any node $g$ of the negative cycle, the graph $DDG(g)$ represents all these subpaths correctly (\cref{lemma_algorithm_computes_correct_labels}), and so the cycle is detected by some vertex $v$ that lies on the face (face-part)~$g$. 
\end{proof}

To conclude, a BDD $\mathcal{T}$ is built in $\tilde{O}(D)$ rounds with each vertex $v\in G$ knowing all relevant information that is locally needed about $X^*$ (i.e., $X^*$ is distributively stored).
In each level of the $O(\log n)$ levels of $\mathcal{T}$ we have at most $\tilde{O}(D^2)$ rounds. Thus, by Lemmas~\ref{lemma_algorithm_computes_correct_labels} and~\ref{lemma_algorithm_detects_neg_cycles} the above gives a correct distance labeling algorithm of the dual network. Formally, 

\theoremDualLabeling*

\subsection{SSSP in the dual}
\label{section_dual_SSSP_tree}

After we compute the labeling of \cref{th: dual_distance_labeling}, we want to learn an SSSP tree in $G^*$ from a given source $s\in G^*$. That is, (1) all vertices of a face $f$ in $G$ shall know the distance from $s$ to $f$ in $G^*$, and 
(2) each vertex of $G$ shall know its incident edges whose dual counterparts are in the shortest paths tree.

We first choose the vertex $v$ of minimum ID in the face $s$, which can be done in $\tilde{O}(D)$ rounds by communication over a BFS tree. Then, $v$ broadcasts $\Label(s)$ to all vertices of the bag $G$, this can be done in $\tilde{O}(D)$ rounds, as we can broadcast the label in $\tilde{O}(D)$ messages of size $O(\log{n})$ over a BFS tree in $\tilde{O}(D)$ rounds.
Then, all vertices of a face $f$ can decode the distance from $s$ to $f$ in $G^*$ from $\Label(s), \Label(f)$ by Lemma \ref{lemma_correct_labels}.

To mark the SSSP tree edges, an aggregate operator is computed for each node $f$ of $G^*$ over its neighbors (using \cref{lem: PA_on_G^*}, we perform a single part-wise aggregation task). In particular, for a dual node, we mark its incident edge that minimizes the distance product with its neighbors, i.e, we take the  operator computed is $\min_{g\in N(f)}\{\dist(s,g)+w(g,f)\}$ (where $N(f)$ is $f$'s neighborhood). Finally, $G^*$ might be a multi-graph, so there might be many edges connecting $f$ with a neighbor. We mark the minimal weight edge connecting $f$ to the neighbor that minimizes the above (breaking ties by IDs for edges and for neighbors). Thus,

\lemmaDualSSSPTree*

The above theorem in addition to \cref{sec: dual_MA_model} imply distributed (exact and approximate) algorithms for Maximum $st$-Flow in $G$ (i.e., proves Theorems \ref{thmstflow} and \ref{thmstflowapprox}) by an  adaptation of the centralized algorithms of Miller-Naor~\cite{MN95} and of Hassin~\cite{Hassin}. The details are given in the next section.

\section{Primal Maximum {\em st}-Flow and Minimum {\em st}-Cut} 
\label{section_flow_cut}

In this section we show how to use the labeling algorithm of \cref{sec: dual SSSP} to compute the maximum $st$-flow in $G$, in addition, how to use the minor-aggregation model simulation on $G^*$ of \cref{sec: dual_MA_model} to approximate the maximum $st$-flow in $G$ in the special case when $G$ is undirected and both $s$ and $t$ lie on the same face in the given planar embedding. I.e, when $G$ is {\em $st$-planar}. The maximum $st$-flow results naturally extend to minimum $st$-cut algorithms.

\subsection{Maximum {\em st}-Flow}
\label{section_flow}
\thmstflow*

The proof mostly follows a centralized algorithm of Miller and Naor~\cite{MN95}, the general idea is to
perform a binary search on the value $\lambda$ of the maximum $st$-flow by a logarithmic number of SSSP computations in the dual graph, after which one can deduce the value of the max $st$-flow and a flow assignment. 

\begin{proof}
The algorithm follows the centralized algorithm of Miller and Naor~\cite{MN95}. It works by performing a binary search on the value $\lambda$ of the maximum $st$-flow. In each iteration, $\lambda$ units of flow are pushed along a (not necessarily shortest) $s$-to-$t$ path of darts $P$. If this does not violate the capacity of any dart in $P$ then it is a feasible flow and we conclude that the maximum $st$-flow is at least $\lambda$. Otherwise, a residual graph is defined by subtracting $\lambda$ from the capacity of every dart in $P$ and adding $\lambda$ to their  reverse dart (in the beginning every edge $e$ of capacity $c(e)$ corresponds to two darts - one  in the direction of $e$ with capacity $c(e)$, and its reverse with capacity zero). 
Notice that residual capacities may be negative for the darts of $P$. To fix this, an arbitrary face $f$ of $G$ is chosen and SSSP (with positive and negative edge lengths) in $G^*$ is computed from $f$. Miller and Naor~\cite{MN95} showed that if a negative cycle is detected then we can conclude that the maximum $st$-flow in $G$ is smaller than $\lambda$, and otherwise it is at least $\lambda$. Moreover, in the latter case, the following is a feasible $st$-flow assignment of value $\lambda$. For every dart $d\in G$ that maps to $d^*=(g,h)\in G^*$: the flow on $d$ is set to $\dist(f,h)-\dist(f,g)+\lambda$ if $d\in P$, to $\dist(f,h)-\dist(f,g)-\lambda$ if $\rev(d)\in P$, and to $\dist(f,h)-\dist(f,g)$ otherwise. 

The path $P$ can be found in $O(D)$ rounds using a standard breadth-first search (notice that $P$ is a directed path of darts but does not need to be a directed path of edges in $G$), and the dual SSSP computation is done in $\tilde O(D^2)$ rounds as we showed in \cref{sec: dual SSSP} (with a slight modification that we show next). Since $\lambda$ is assumed to be polynomial, all iterations of the binary search are done in $\tilde O(D^2)$ rounds. The flow assignment on the edges of $G$ is given by the assignment to its darts, that can be deduced locally from its endpoint vertices by Property~\ref{BDD_know_edges} of the BDD (since for a dart $d=(u,v)\in G$ with $d^*=(f,g)\in G^*$ the endpoints $u,v$ belong to both faces $f$ and $g$ in $G$). 

Notice that the SSSP algorithm on $G^*$ from Section~\ref{sec: dual SSSP} uses darts for obtaining the necessary distributed knowledge. However, the distance computations it performs considers edges and not darts. We next show that these computations can be easily modified to work with darts. I.e., to consider the two darts of an edge (with weights as described earlier) when computing the labeling and the shortest paths tree. To do that, we slightly modify the input network and use the algorithm as a black-box on it.
That is, we work with the network $G'$ obtained from $G$ by adding the reversed dart of each edge to the graph, such that, the two darts of an edge are present in $G'$ and its dual.
The reversed dart of an edge gets assigned a capacity of zero.
It is not hard to see that $G'$ can be constructed and embedded in the plane without any need for communication. This is because $G$ is embedded and the two darts $e,\rev(e)$ that map to the edge $e$ are embedded one over the other (the local cyclic ordering of a vertex for its incident edges extends naturally to darts).
In addition, each $\Congest$ round on $G'$ is simulated within two rounds on $G$.
Then, the dual $G'^*$ of $G'$ looks exactly like $G^*$ after augmenting it with an edge $\rev(e)=(g,f)$ of weight zero (the reversed dart) for each existing edge $e=(f,g)$ in $G^*$. Note, some of those edges $\rev(e)$ that we add are parallel edges, meaning, their endpoints may have an edge connecting them with the same direction of $\rev(e)$ (possibly with a different weight), however, this is fine as $G^*$ (before the the augmentation) may have parallel edges. I.e., our dual SSSP algorithm knows to handle that.
The graph $G'^*$ can be learnt distributively in the same manner that $G^*$ can be learnt, because each dart that we add in the primal is known to its endpoints. Furthermore, the (extended) BDD of $G'$ is the same as that of $G$ since $G'$ is $G$ with all of its darts present, and since the algorithm for constructing an extended BDD originally considers all darts of $G$.
Hence, the dual SSSP algorithm is simulated on $G'^*$ in the same $\tilde{O}(D^2)$ round complexity of running it on $G^*$.

Then, the maximum $st$-flow algorithm for $G$ runs as follows. First, $G'$ is constructed. Then, an $s$-to-$t$ path $P$ in $G'$ is found in $O(D)$ rounds via an undirected breadth-first search starting at $s$. Afterwards, $\lambda=1$ units of flow are pushed on $P$ (in the first iteration), and the capacities of $P$'s darts and their reversals are (locally) changed accordingly. Then, we compute an SSSP tree of $G'^*$ using \cref{lemma_dual_SSSP} with the source $f$ being the minimum ID face of $G'$ ($f$ is found in $O(D)$ rounds by aggregating on $G'$ the IDs of its faces, which are known by Property~\ref{BDD_know_faces} of the BDD). Following this, each node $v$ in $G'$ knows the distance from $f$ to each of the (dual nodes mapped to) faces it lies on. Hence, $v$ learns the flow value for each of its incident darts. Note, in each iteration, each vertex in $G'$ knows locally the current value of $\lambda$, depending on whether or not it has received a message reporting a dual negative cycle from the dual SSSP algorithm. When the algorithm termintes, each vertex $v$ of $G$ knows the $st$-flow value in $G$ and (locally) a corresponding flow assignment to its incident edges (the assignment induced by the darts of $G'$). \qedhere 

\end{proof}

Next we show how to use \cref{sec: dual_MA_model} for computing the {\em undirected} maximum $st$-flow when $s$ and $t$ lie on the same face of $G$.
The undirected maximum $st$-flow problem asks to find the maximum flow that can be pushed from $s$ to $t$ in an undirected network. Alternatively, where each two neighbors are connected by two opposite direction edges with the same capacity. Although the $s$-to-$t$ flow value equals $t$-to-$s$ flow value, the $s$-to-$t$ flow itself is directed (e.g. the flow entering $s$ is non-positive, the flow entering $t$ is non-negative).

\thmstflowapprox*

To prove this, we simulate a centralized algorithm of Hassin \cite{Hassin}, where he adds an edge $e=(t,s)$ and then computes SSSP in the dual of the resulting graph. $s$ and $t$ are required to be on the same face in order to ensure that $e$ does not violate planarity (it is embedded in the interior of the face containing $s$ and $t$). 
We use our minor-aggregation simulation on the dual (\cref{sec: dual_MA_model}) in order to apply a $(1+\epsilon)$-approximate SSSP algorithm of~\cite{GHSYZ22} on $(G\cup\{e\})^*$, the dual of $G\cup\{e\}$. We use an approximate algorithm since there is no exact one in the minor-aggregation model. 
This simulation instantly gives the value of the approximate maximum $st$-flow, however, in order to get an assignment we need to work a bit harder because, Hassin's algorithm is an exact algorithm whilst we use it approximately. Hassin defines the flow assignment on an edge $e'$ whose dual is $e'^*=(g,h)$ to be $\dist(f_1,h) - \dist(f_1,g)$, where $f_1$ is the source of the SSSP tree computed in $(G\cup\{e\})^*$ (note, we need the exact distances for the assignment). 
Thus, approximately, if we want the flow assignment to be capacity respecting (amongst other requirements), for each edge, we need to bound the difference of its endpoints' (approximate) distances, i.e, we need the approximate distances to obey the triangle inequality. Most (distributed) distance approximation algorithms do not have this feature. Meaning, the approximate flow assignment can be very far off the true answer. E.g, let $u,v$ be two neighbors of distance $\Omega(n)$ from the source, connected by a zero weight edge, we want the difference of their (approximate) distances from the source to be zero, however, the difference of the approximate distances might be at least as large as $\epsilon n$.
To overcome that, we show how to apply a recent algorithm of~\cite{RozhonHMGZ23} on the dual. This algorithm was not implemented in the minor-aggregation model before.

\begin{proof}
The algorithm follows the centralized algorithm of Hassin~\cite{Hassin}:
Let $f$ be a face of $G$ containing both $s$ and $t$.
First, we add an artificial edge $e=(t,s)$ to $G$ embedded inside the face $f$ (preserving planarity) with infinite capacity. This edge is assigned a weight of $nW$, where $W$ is the largest weight in the network, representing infinite capacity in $\tilde{O}(1)$-bits.

\begin{figure}[htb]
    \centering
    \includegraphics[width=.3\linewidth]{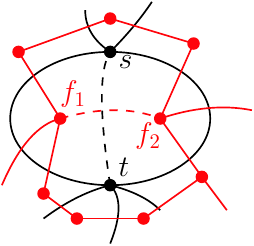}
    \caption{$G\cup\{e\}$ in black and its dual in red. $e$ ($e^*$) is the dashed edge. $f$ is the face in which $e$ is embedded,
    note, $f$ is split into two faces $f_1,f_2$. Each of $f_1$ and $f_2$ has a corresponding node in the dual.
    }
   \label{fig: approx_flow_graph}
\end{figure}

The edge $e$ divides $f$ into two faces, $f_1,f_2$, where $f_1$ is the face on the left
\footnote{Left and right are defined w.r.t. the direction of $e$ and can be deduced from the combinatorial embedding of $e$'s endpoints.}
of $e$ if $e$ were to be directed from $t$ to $s$ (this would be important for the flow direction later).  We compute SSSP from $f_1$ (this time all edge-lengths are non-negative) in $(G\cup \{e\})^*$. See \cref{fig: approx_flow_graph}.  Hassin showed that the value of the maximum $st$-flow in $G$ is equal to $\dist(f_1,f_2)$ in $(G\cup \{e\})^*$. 
We note that Hassin's algorithm works for directed and undirected graphs. For us however, it is crucial that the graph is undirected since the (distributed) SSSP algorithm that we use works only for undirected graphs. 

Distributively, all vertices need to know the IDs of $s,t$, and $f$. We detect the face $f$ by a simple part-wise aggregation in $\hat{G}$ (Properties \ref{property: hat(G)_diameter}, \ref{property: hat(G)_simulation} and \ref{property: hat(G)_face_identification} of $\hat{G}$), where the input of each copy of $s$ (resp. $t$) is $ID(s)$ ($ID(t)$) and an identity item for any other vertex (e.g. minus one, as no ID is negative). The aggregate operator simply counts the distinct values of input. Then, the face $f$ is the face that has the minimum ID of the (at most two) faces which got a count of two ($s,t$ know these IDs and can choose the minimum locally). 
Note, this aggregation terminates in $\tilde{O}(D)$ since the values aggregated are of size $\tilde{O}(1)$-bits and there are at most two distinct values aggregated on each face of $\hat{G}$ that maps to a face of $G$.
Then, both $s$ and $t$ add $e$ to the clockwise ordering of their incident edges in $G$, and embed it to be in between two consecutive edges of their incident edges in $f$.
We use the minor aggregation model implementation (Section~\ref{sec: dual_MA_model}) for the dual graph to simulate an SSSP minor-aggregation algorithm on $(G\cup \{e\})^*$. We use 
the algorithm of~\cite{GHSYZ22} that computes $(1+\epsilon)$-approximate SSSP in
 $2^{O(\log n \cdot \log \log n)^{3/4}} / \epsilon^2$
 minor-aggregation rounds, yielding a $(1-o(1))$ approximate $D\cdot n^{o(1)}$ round $\Congest$ algorithm in $G$ for the undirected  $st$ flow problem in $st$-planar networks after setting $\epsilon$ to $1/n^{o(1)}$.
 However, we want to simulate the algorithm on $(G\cup \{e\})^*$ rather than $G^*$, meaning the simulation result of Section~\ref{sec: dual_MA_model} cannot be used straightforwardly. Moreover,~\cite{GHSYZ22}'s algorithm assumes weights in $[1,n^{O(1)}]$ but not zero (not allowing zero capacities). 
So, we assume that we do not have zero weight edges and first dive into the details of the simulation, then we show how to get an assignment
and finally show a simple fix for dealing with zero weight edges.

\paragraph{Simulation.}

Here we desire to simulate \cite{GHSYZ22}'s approximate SSSP algorithm on $(G\cup\{e\})^*\neq G^*$ by applying \cref{sec: dual_MA_model} (Theorem~\ref{th: extended_MA_on_dual}) to simulate the extended minor aggregation model on $G^*$.
We exploit that the extended model can simulate any basic or extended minor aggregation algorithm on any virtual graph $G^*_{virt}$ that contains a small (poly-logarithmic) number of arbitrarily connected virtual nodes (see \cref{sec: dual_MA_model} for exact definitions). Thus, the only thing we need to do, is to use that (Theorem~\ref{th: extended_MA_on_dual}) in order to apply the approximate SSSP algorithm on $G^*_{virt}=(G\cup\{e\})^*$. To do so, we only need to distributively store $(G\cup\{e\})^*$ as needed (matching the input format of the algorithm stated in Theorem~\ref{th: extended_MA_on_dual}).
We explain; Recall, the face-disjoint graph $\hat{G}$ is used as the base graph for representing $G^*$, i.e., $\hat{G}$ is the distributed method for storing $G^*$; To store any virtual graph $G^*_{virt}$ that is obtained by replacing nodes or adding virtual nodes to $G^*$, we need that all vertices on a face $\hat{g}$ of $\hat{G}$  (maps to node $g\in G^*$) to know a list of (IDs of) $g$'s incident virtual edges and all virtual edges that connect any two virtual nodes (and their weight if exists) in $G^*_{virt}$. So, we need to (1) show how to delete $f$ from $G^*$, (2) add two nodes that play the roles of $f_1,f_2$, and (3), connect them to neighbors of $f$ and to themselves properly, represented in $\hat{G}$ as described; Such that, the obtained $G^*_{virt}$ has the exact same structure as $(G\cup\{e\})^*$. 
Note, changes need only to be made on $\hat{f}$'s vertices and vertices lying on (faces $\hat{g}$ in $\hat{G}$ corresponding to) neighbors $g$ of $f$ in $G^*$.

We start by: (1) Enumerating $\hat{f}$'s vertices in $\hat{G}$ in ascending order and deactivating them (deleting $f$). 
This enumeration would be useful to define $f_1,f_2$ (i.e, split $\hat{f}$).
(2) Learning (locally) in $\hat{G}$ for each edge of $E_C$ \footnote{Those edges map 1-1 to edges of $G^*$, see \cref{section: preliminaries} and \cref{appendix: preliminaries_hat{G}} for more details.} that is incident to $\hat{f}$,
whether its corresponding edge in $G^*$ shall be incident to $f_1$ or to $f_2$.
Then, (3) Each (face $\hat{g}$ in $\hat{G}$ that maps to a) neighbor $g$ of $f$ needs to know its minimal weight edge connecting it to $f_1$ and $f_2$, as we are interested in shortest paths and cannot afford to broadcast all edges of $f$, due to their possibly large quantity. To do so, we compute a minimum aggregate over weights of $\hat{g}$'s edges that are mapped to edges that shall be incident to $f_1$ ($f_2$),
so that all vertices in $\hat{g}$ 
know the edge that connects $g$ to $f_1$ ($f_2$).
Finally, (4) A global maximum operator is computed over edge weights of $G$, such that, all vertices of $\hat{G}$ learn that $f_1,f_2$ are connected by an edge whose weight is $nW$, where $W$ is the maximum weight found. We describe how each step is implemented; Assume $\hat{G}$ is already constructed and its faces that correspond to $G$ faces are identified (Properties \ref{property: hat(G)_construction_representation}, \ref{property: hat(G)_face_identification} of $\hat{G}$).
\begin{enumerate}
    \item For enumerating $\hat{f}$'s vertices in $\hat{G}$, we use a simple classic procedure of a subtree sum using part-wise aggregations on $\hat{G}$. Specifically, the input tree is $\hat{f}$ minus the $E_R$
    \footnote{For each edge $e$ of $G$, there are two edge copies in $E_R\subset E(\hat{G})$. Intuitively, there is an edge copy for each face of $G$ that contains $e$.}
    edge copy of the edge preceding $e$ in $s$'s local ordering of its incident edges (this edge is known to $s$ locally), the root is the copy of $s$ that is incident to the removed edge, and all vertices get an input value of one initially.
    The procedure we use is similar to those described in~\cite{GH16b, GP17}. We can do so in $\tilde{O}(D)$ rounds 
     by \cref{sec: dual_MA_model} (Lemma~\ref{cor: planar_PA}).
   
    Denote by $s'$, the copy of $s$ in $\hat{f}$ which is incident to the two consecutive edges $e_1', e_2'$ (of $E_R$ in $\hat{G}$), that map to the edges $e_1,e_2$ incident to $s$, s.t. $e=(t,s)$ is embedded in between $e_1,e_2$ (in $G$), we define $t'$ analogously. Let $e_a,e_b$ be the two edges incident to $s$ s.t. all edges between $e_a,e_b$ (in the local ordering of $s$) are in $f$, then all edges in between $e_a,e_1$ have to go to $f_1$ and all edges in between $e_2,e_b$ have to go to $f_2$, as $e$ is embedded right after $e_1$ and right before $e_2$. We define $s'$ to capture this in $\hat{G}$ (where edges have a lot of copies, sometimes in the same face
    \footnote{E.g, if an edge $e$ is internal to a face $f$, then it has its two $E_R$ copies in the face $\hat{f}$ of $\hat{G}$ that maps to $f$. More generally, if there is a tree attatched to a vertex in the interior of a face $f$, then each of those tree edges has two copies in $\hat{f}$, as $\hat{f}$ is basically an Euler tour on $f$. See \cref{section: preliminaries} and \cref{appendix: preliminaries_hat{G}} for more details.}),
    in order to define $f_1, f_2$ correctly. 
    
    Now, $s'$ (resp. $t'$) broadcasts its ID, assigned number, and $ID(f)$. The ID of $f_1,f_2$ is defined to be $(ID(f),0)$ and $(ID(f),1)$, respectively, all vertices now know these as well. The broadcast is pipelined and terminates in $\tilde{O}(D)$ rounds as there is a constant amount of distinct messages sent.
    From now on, all vertices in $\hat{f}$ consider themselves as deactivated - they might still be used for communication in the simulation, but they do not participate as an input.
    
    \item Each vertex on a face $\hat{g}$ of $\hat{G}$ already knows (by construction of $\hat{G}$) for each of its incident $E_C$ edge $e^*$, the (IDs of) faces it participates in and the ID of the edge $e$ of $G$ that $e^*$ corresponds to (recall, $E_C$ edges are meant to simulate $G^*$ edges in $\hat{G}$). Thus, in one round of communication, each vertex in $\hat{G}$ can know for each incident edge in $E_C$ whether it shall be in $f_1$ or $f_2$, by firstly receiving the numbering 
    of its neighbors in $\hat{f}$ and then, by deciding on each edge $e^*$ with both endpoints of $e$ having a number in between $s'$ and $t'$ to be in $f_1$, otherwise, to be in $f_2$. Note, this is consistent with $f_1$ being on the left of $e$ if directed from $t$ to $s$ due to the choice of $s'$ and $t'$.
    
    \item Now that edges incident to $f_1$ and $f_2$ are known to their endpoints in $\hat{G}$, two  minimum aggregations in $\tilde{O}(D)$ rounds are then computed and broadcast in each neighbor $\hat{g}$ of $\hat{f}$ in $\hat{G}$ over $E_C$ edges that connect $f_1$ to $g$ and $f_2$ to $g$, allowing all vertices on $\hat{g}$ to learn the two edges connecting $g$ and each of $f_1,f_2$ in $(G\cup\{e\})^*$, if any.
    
    \item Finally, an aggregate over $G$ edge weights is computed in $O(D)$ rounds for finding $W$, then, since each vertex in $\hat{G}$ knows the ID of $f_1,f_2$ and $W$, it learns that $f_1,f_2$ are connected by an edge
    with weight $nW$ in $(G\cup\{e\})^*$.
\end{enumerate}

Notice, all the above terminates within $\tilde{O}(D)$ rounds on $G$ by \cref{sec: dual_MA_model} (Lemma~\ref{cor: planar_PA}).
Finally, since $G^*_{virt}=(G\cup \{e\})^*$ now is properly distributively stored, then, we can simulate the extended minor-aggregation model on it (see \cref{th: extended_MA_on_dual} for details); So, we first use it in order to deactivate parallel edges (\cref{lem: deactivating_parallel_edges}), keeping only the minimal-weighted edge between its endpoints (we aim to compute SSSP). Then, since $G^*_{virt}$ is simple and we can simulate any minor aggregation algorithm on it, it satisfies the requirements for applying the mentioned theorem with \cite{GHSYZ22}'s SSSP algorithm, obtaining an approximate SSSP tree rooted at the source $f_1$.

\paragraph{Distances (flow value).}
Assuming we have computed an approximate SSSP tree $T$ (with zero weighted edges included) as promised, we can compute the distances of any node from the source by a simple procedure. That is, root the computed tree with the source, then compute the tree ancestor sum of every node $g\in T$, i.e., the weight of the $f_1$-to-$g$ $T$-path. For that, we use a procedure of \cite{GZ22} that runs in $\tilde{O}(1)$ minor aggregation rounds (Lemma 16 in their paper), thus, $\tilde{O}(D)$ rounds in $G$ (\cref{th: extended_MA_on_dual}).
After that process terminates, each vertex lying on $f_2$ (in $G$) knows the approximate distance from $f_1$ to $f_2$ in $G^*_{virt}$, thus, it can broadcast it to the whole graph. Note, that distance is a $(1+\epsilon)$ approximation of the maximum $st$-flow value (by Hassin~\cite{Hassin}). Since the algorithm guarantees to report $\alpha\cdot \dist(f_1,f_2)$ instead of $\dist(f_1,f_2)$, for $\alpha\in [1,1+\epsilon]$, multiplying by $(1-\epsilon)$, we get $(1-\epsilon)\alpha \in [1-\epsilon, 1-\epsilon^2]$, that is, a $(1-\epsilon)$ approximation of the maximum $st$-flow value.

\paragraph{Approximate flow assignment.} 
Note that in \cref{thmstflow} we obtained both the value of the maximum flow and a corresponding assignment of flow to the edges. Here however, we approximate the flow value but do not immediately obtain an assignment (or an approximate assignment) of the flow. The reason is as follows: 
Hassin's algorithm was designed as an exact algorithm (running an exact SSSP computation) while we use an approximate SSSP. For the flow value this is fine, instead of $\dist(f_1,f_2)$ we report a $(1-\epsilon)$-approximation of $\dist(f_1,f_2)$. However, for the corresponding flow assignment, recall   that the flow on edge $e'\in G$ with $e'^*=(g,h)\in (G\cup\{e\})^*$ is  $\dist(f_1,h)-\dist(f_1,g)$.
This means we need the approximation algorithm to bound $\dist(f_1,h)-\dist(f_1,g)$ by the edge's capacity (and not just approximate $\dist(f_1,h)$ and $\dist(f_1,g)$). I.e., we want the approximate distances to satisfy the triangle inequality. 
This indeed is possible using (a dual implementation of) a recent algorithm of Rozhon, Haeupler, Martinsson, Grunau and Zuzic \cite{RozhonHMGZ23}, however, their algorithm was not implemented in the minor-aggregation model.
Given access to a  $(1+\epsilon)$-approximate shortest paths oracle,~\cite{RozhonHMGZ23}'s algorithm produces {\em smooth} (defined next) approximate distances 
within a logarithmic number of calls to the oracle with a parameter $\epsilon'=O(\epsilon/ \log n)$. Before we dive into the simulation, we define what a smooth distance approximation is, suggest a flow assignment and prove it is a feasible $(1-\epsilon)$ approximate assignment using the smoothness property.

\medskip
\noindent
\underline{\em Smooth distances and a feasible flow assignment.}
In order to prove that the approximate SSSP computation gives a {\em feasible} $(1-\epsilon)$ approximate flow assignment, we need to show an assignment that is: (a) Capacity respecting: the flow being pushed on each edge is non-negative and is at most its capacity, (b) Obeys conservation: for all vertices except for $s$ and $t$, the flow pushed into a vertex equals the flow pushed by it. Finally, the assignment should be (c) $(1-\epsilon)$ approximate: the flow pushed from $s$ to $t$  is between $(1-\epsilon)$ and once the maximum $st$-flow. Note, even thought the graph of input is undericted, the flow itself is directed, so we address that as well.
We first define some notions and define the flow assignment, then proceed to prove that the above holds. 

In~\cite{RozhonHMGZ23}, a $(1+\epsilon)$ distance approximation $d(\cdot)$ from a given (single) source in a graph $\mathcal{G}$ is said to be {\em $(1+\epsilon)$-smooth} (definition 4.2 in their paper) if it satisfies the following:
\[\forall u,v \in V(\mathcal{G}): d(v) - d(u) \leq (1+\epsilon) \cdot \dist(u,v) \]
We assume we have computed a $(1+\epsilon)$ smooth approximate distances from $f_1$ in $(G\cup\{e\})^*$ by combining~\cite{GHSYZ22} and~\cite{RozhonHMGZ23}, denoted $d(\cdot)$, and we define $\delta(\cdot):=(1-\epsilon)d(\cdot)$. Then, for $e'=\{u,v\}\in G\cup\{e\}$ s.t. $e'^*=\{f,g\}$ in $(G\cup\{e\})^*$, we define the flow value pushed on $e'$ to be $|\delta(g)-\delta(f)|$. When $\delta(g) \geq \delta(f)$, the flow is being pushed on $e$ from $u$ to $v$, otherwise, it is pushed from $v$ to $u$. We denote this flow assignment by $\phi(\cdot)$, taking the induced assignment on $G$ gives the final assignment.

\begin{enumerate}[label=(\alph*)]
    \item {\em Capacity respecting.} Obviously, the flow pushed on each edge is non-negative (by definition).
    Given a $(1+\epsilon)$-smooth distance approximation from the source $f_1$ in $G^*_{virt}=(G\cup\{e\})^*$, we get a capacity respecting flow assignment as follows. Consider any edge $e'=(u,v)$ where $e'^*=(f,g)$ and the approximate distances $d(\cdot)$, s.t. $d(g) \geq d(f)$, then, from the smoothness property:
    \begin{equation}
    \label{equation: capacity_respecting_flow1}
    d(g) - d(f) \leq (1+\epsilon) \cdot \dist(f,g) \leq  (1+\epsilon) \cdot c(e') 
    \end{equation}
    multiplying by $(1-\epsilon)$, we get,
    \begin{equation}
    \label{equation: capacity_respecting_flow2}
    \delta(g) - \delta(f) \leq (1-\epsilon^2) \cdot \dist(f,g) \leq (1-\epsilon^2) \cdot c(e')
    \end{equation}
    Note, the left-hand side of \cref{equation: capacity_respecting_flow2} is exactly $\phi(e')$, the flow pushed on $e'$; $c(e')$ is its capacity and the property follows.
    The left inequality in \cref{equation: capacity_respecting_flow1}  follows simply from the smoothness property, whilst the right one follows since the the edge weights of $G^*_{virt}$ are the edge capacities of the corresponding primal edges and that the exact distances in $G^*_{virt}$ is a metric that obeys the triangle inequality, i.e, $\dist(f,g)$ is at most the weight of the edge connecting $f,g$.
    \cref{equation: capacity_respecting_flow2} follows from  \cref{equation: capacity_respecting_flow1} since, by definition, $\delta(\cdot)=(1-\epsilon)d(\cdot)$.

    \item {\em Conservation.} By~\cite{KM}, assuming a directed planar graph with edge capacities and an assignment of {\em potentials} (weights) $\rho(\cdot)$ on its faces, then, assigning for each edge a flow of $\rho(g)-\rho(f)$, where $g$ (resp. $f$) is the face on the right (left) of that edge,
    gives a flow assignment that obeys conservation in all vertices
    \footnote{This kind of a flow assignment is referred to as a {\em circulation} in the centralized literature on planar graphs.}.
    Considering $G\cup\{e\}$, the assignment we suggest, is deliberately defined so that the directions of the flow would be consistent with the above. I.e, directing an edge in its defined flow direction and pushing $|\delta(g)-\delta(f)|$ units of flow would be consistent with the definition above. 
    Thus, the flow assignment we suggest obeys conservation in every vertex of $G\cup\{e\}$. Note, $G$ is equivalent to $G\cup\{e\}$ after removing $e=\{s,t\}$, thus, the induced flow assignment on $G$ obeys conservation except in the endpoints of $e$, which are $s$ and $t$.  

    \item {\em Flow value of the assignment. } 
     $f_1$ was picked to be the source of the approximate SSSP computation ($\delta(f_1)=0$), thus, $\delta(f_2)\geq \delta(f_1)$, meaning, according to our definition of $\phi(\cdot)$, the flow on $e$ shall be directed from $t$ to $s$, as $f_1$ shall be the face on left of $e$ w.r.t. the flow direction on $e$. As a result, the flow pushed on $e$ from $t$ to $s$ is $\delta(f_2)-\delta(f_1)$. Note, $\phi(\cdot)$ obeys conservation in all vertices of $G\cup\{e\}$, thus, removing $e$ from $G\cup\{e\}$ would induce the flow assignment in $G$, directed from $s$ to $t$, of value:
    \begin{align}
    \label{equation: approximate_flow}
            \phi(e)=\delta(f_2)-\delta(f_1)=(1-\epsilon)\cdot d(f_2)-0 =(1-\epsilon)\cdot\alpha \cdot \dist(f_2,f_1)
    \end{align}

    where $\alpha\in[1,1+\epsilon]$.
    We already have shown that the right-hand side of \cref{equation: approximate_flow} is a $(1-\epsilon)$ approximation of the maximum $st$-flow value, thus, the induced flow assignment by $\phi(\cdot)$ on $G$ of value $\phi(e)$ is $(1-\epsilon)$ approximate.

\end{enumerate}

It remains only to show 
that we can apply~\cite{RozhonHMGZ23}'s algorithm (Theorem 5.1, Algorithms 3 and 4 in their paper), in order to obtain smooth $(1+\epsilon)$ approximate single source shortest paths from $f_1$ in $G^*_{virt}=(G\cup \{e\})^*$ by invoking the algorithm of \cite{GHSYZ22} that we use for approximating distances, which we refer to as the {\em oracle}. We note that \cite{RozhonHMGZ23}'s algorithm assumes real weights in $[1,n^{O(1)}]$, whereas we work with integral weights, however, they suggest a fix (by scaling) for this problem that appears in Appendix C of their paper. Note, scaling can be implemented efficiently in one minor-aggregation round, henceforth, we consider this issue resolved.

\medskip
\noindent
\underline{\em \cite{RozhonHMGZ23}'s algorithm in a high-level.}
We describe \cite{RozhonHMGZ23}'s algorithm for smooth distance approximation in a high-level, highlighting only the computational tasks necessary to implement it in the minor aggregation model. For the algorithm in full resolution, the reader is referred to Section 5 of~\cite{RozhonHMGZ23}.
Initially, nodes of the input graph shall know the source $s$, the approximation parameter $\epsilon$ and a trivial estimate $\Delta$ of the maximum distance in the graph ($\Delta:=2nW$ in our case), then, each node computes additional parameters that depend only on the former. Note, all of the former is implemented in only one minor-aggregation round. 

Next, the algorithm computes a distance approximation, based on a call to the approximation oracle (any SSSP approximation algorithm, without any additional special guarantees) with $s$ and $\epsilon'=O(\epsilon/ \log n)$ . 
Then, sequentially, $O(\log n)$ phases are performed, each defining a distance approximation, such that, the final output of the algorithm is the approximation computed in the last phase.
In each phase, two calls to the approximate oracle are performed with $s$ and $\epsilon'/100$.
Each call to the oracle is on a {\em level graph}. Afterwards, each level graph defines a distance approximation of the input graph, and the phase's distance approximation is computed by taking the element- (node-) wise minimum over: (1) the approximations computed on the level graphs, and (2) the approximation from the previous phase.
Again, taking the minimum requires only one minor-aggregation round.

\medskip
\noindent
\underline{\em Level graphs.}
The only vagueness in the implementation lies in the level graphs, which we now define, then, show that they can be constructed and simulated efficiently in the minor-aggregation model.
The level graph $H$ (Definition 5.7 in \cite{RozhonHMGZ23}) of an input graph $\mathcal{G}$ where each node of $\mathcal{G}$ knows its (approximate) distance from a source $s$, in a specific phase, is defined as follows. The node and edge sets of $H$ are the same as those of $\mathcal{G}$, then: (1) Each node assigns itself a level (a number), depending only on its distance from $s$ and some parameter $\omega$, (2) each edge that connects neighbors of different levels is removed, the weights of edges kept are altered depending only on previously computed parameters that all nodes of $\mathcal{G}$ know. Finally, (3) An edge between $s$ and each $u\in \mathcal{G}$ is added, with weight depending only on $\omega$ and the $s$-to-$u$ distance approximated in the previous phase.

\medskip
\noindent
\underline{\em Implementation of level graphs.}
First, we start by computing $\omega$ in one minor-aggregation round on $G^*_{virt}$ as it depends only on parameters known to all nodes of the input graph (as $\Delta, \epsilon$ and others). Then, we compute the level of each node, also in one round, as it only depends on $\omega$ and the $s$-to-$u$ distance approximated in the previous phase.
The remaining things to explain are: (1) How to "delete" edges that cross levels, and (2) How to connect $s$ to all nodes with the appropriate edge weights.
\begin{enumerate}
   \item Deactivating edges that cross levels is simple.
   Each edge $e=(u,v)$ in the input graph knows the level of both its endpoints, thus, it knows if it should participate in the level graph or not.
   In more detail, recall, $e$ in the minor aggregation model is simulated by its endpoints $u',v'$ in the base communication graph ($\hat{G}$ in our case). In one round on $\hat{G}$, $u'$ knows the level of the super-node that contains $v'$ (and vise-versa). Thus, $u',v'$ know that $e$ is prevented from participating in computations on $H$. In case $v'$ is virtual then all nodes of the graph simulate it (in particular $u'$), and know its information (in particular, its level).

   \item Luckily, we work in the extended minor-aggregation model, where adding virtual nodes and connecting them arbitrarily to any other virtual or non-virtual node, is a feature of the model. In the same manner that we described earlier (where we obtained $G^*_{virt}$), we alter $H$ and replace $s$ with a virtual node, all nodes in $H$ now know they are connected to the source. Notice, there are no problems regarding weights of edges, as the weight of an added edge $(s,u)$ depends only on $\omega$ and the approximated $s$-to-$u$ distance in the previous phase (already known to $u$). Thus, when nodes connect themselves to the replaced $s$, they connect themselves with the needed weights.
   Again, virtual nodes are simulated by all vertices, so for each virtual $v$, all real nodes know its info and locally consider it to be connected to $s$ with the right weight.
    
\end{enumerate}

Note, each of the $O(\log n)$ phases is of $\tilde{O}(1)+T_{SSSP(\epsilon')}$ minor-aggregation rounds, where $T_{SSSP(\epsilon')}$ denotes the minor-aggregation round complexity of a $(1+\epsilon')$-approximate SSSP  oracle. Thus, the overall  minor-aggregate round complexity of the $(1+\epsilon)$ smooth approximate distance computation is $O(\log n)\cdot \tilde{O}(T_{SSSP(\epsilon')})$ for $\epsilon'=(\epsilon/\log n)$. Implementing the approximate SSSP oracle with the algorithm of \cite{GHSYZ22}, results with a total minor-aggregate round complexity of $O(\log n)\cdot (\frac{1}{\epsilon'})^2\cdot 2^{O(\log n \cdot \log \log n)^{3/4}}$, i.e., $\frac{\log^3 n}{\epsilon^2} \cdot 2^{O(\log n \cdot \log \log n)^{3/4}}$, which by \cref{th: extended_MA_on_dual} translates to $\frac{D}{\epsilon^2}\cdot n^{o(1)}=D\cdot n^{o(1)}$ $\Congest$ rounds on the network of communication $G$, if we set $\epsilon=1/n^{o(1)}$.
For conclusion, since each vertex $v$ in $G$ simulates its copies in $\hat{G}$, which in turn simulate the faces $v$ participates in in $G^*_{virt}$, then, $v$ knows for each face it participates in, its distance from the source. Thus, after computing smooth distances in the relevant graph $G^*_{virt}$, in one round of communication, $v$ knows for each incident edge $e'$, such that $e'^*=(f,g)$, the values $\delta(f_1,g)$ and $\delta(f_1,f)$ and can locally compute $\phi(e')$.
Finally, we are ready to go through the last modification required to keep our promise, that is, we show next how to allow zero capacities (zero-weight edges in the dual shortest paths approximation).

\paragraph{Zero weighted edges.}
The algorithm of~\cite{GHSYZ22} that we use for the approximate SSSP and of~\cite{RozhonHMGZ23}  used to obtain the smoothness property
work for weights in $[1,n^{O(1)}]$, indeed all weights in $G^*_{virt}$, even that of the infinite capacity edge are in  $[1,n^{O(1)}]$ except the zero-weighted ones.
To deal with that issue,
we simulate the obtained smooth approximate SSSP algorithm above while considering zero weighted edges as follows: We run the algorithm on maximal connected components of $G^*_{virt}\setminus\{e:w(e)=0\}$, obtaining a partial SSSP tree of $G^*_{virt}$, then we augment it with some zero weighted edges without closing a cycle, obtaining an SSSP tree $T$ of $G^*_{virt}$, all done via the extended minor aggregation model.

Concretely, we work in two phases. In the first phase, we compute an approximate SSSP tree $T_1$ on the minor $G^*_1$ of $G^*_{virt}$ where all zero-weighted edges are contracted, where the source is the super-node that contains $f_1$.
Now, we need to choose what zero-weighted edges shall be added to complete the approximate SSSP tree $T_1$ of $G^*_1$ into an approximate SSSP tree $T$ of $G^*_{virt}$.
$T_1$ defines a set of connected components in $G^*_{virt}$, contracting those components
in $G^*_{virt}$ (after we undo the zero-weight edge contraction) results with another minor $G_2$, that contains only zero-weight edges. 
I.e., we are looking for some spanning tree in $G_2$. So, we compute a minimum spanning tree $T_2$ of $G^*_2$ by a minor aggregation procedure of \cite{GHSYZ22} (Example 4.4 in their paper).
Hence, the tree $T=T_1\cup T_2$ (considering the endpoints of edges of $T_1,T_2$ in $G^*_{virt}$ rather than $G^*_1,G^*_2$ when taking the union) constitutes an approximate SSSP tree in $G^*_{virt}$. 

Note, $T$ obeys the triangle-inequality (which is crucial for a flow assignment) for the following reasoning. Distances in $T_1$ represent distances in $G^*_{virt}$, by construction any pair of nodes in $T_1$ satisfies the triangle inequality. Thus, map each node $f\in G^*_{virt}$ to the super-node $a$ of $T_1$ that contains it, the property follows since $f$ and $a$ have the same distances from the source because $T_1$ was obtained by contracting zero-weight edges only. 
I.e, $T=T_1\cup T_2$ satisfies the triangle inequality because each pair of (super-)nodes in $T_1$ does.
\qedhere 
\end{proof}

\subsection{Minimum {\em st}-Cut}
\label{section_st_cut}
By the well known Max-Flow Min-Cut Theorem \cite{ff56}, having the exact (approximate) value of the maximum $st$-flow in $G$ immediately gives the exact (approximate) value of the minimum $st$-cut in $G$. So, here we focus on finding the cut edges and bisection. We use our flow algorithms to do so.

\begin{restatable}[Distributed exact planar directed minimum $st$-cut]{theorem}{theoremExactstCut}
   \label{th: exact_min_cut}
   Given a directed planar graph $G$ with non-negative edge weights and two vertices $s,t$, there is a $\tilde{O}(D^2)$ round randomized algorithm that finds the minimum $st$-cut w.h.p..
\end{restatable}

To prove this, we use \cref{thmstflow} to compute a maximum $st$-flow $f$ in $G$ in $\tilde{O}(D^2)$ rounds.
By \cite{ff56}, a corresponding minimum $st$-cut is a cut defined by a set of saturated edges (an edge is saturated if the flow it pushes equals its capacity) which are usually found by solving a reachability problem in the {\em residual graph}, which for technical reasons we reduce to an SSSP problem that is solved with 
an exact SSSP algorithm that runs in  $\tilde{O}(D^2)$ rounds for planar graphs by~\cite{LP19}.
The cut edges and bisection are then deduced from the distance computation.

\begin{proof}
   At first, we use \cref{thmstflow} to compute a maximum $st$-flow $f$ in $G$ in $\tilde{O}(D^2)$ rounds. 
   A minimum $st$-cut is a cut determined by the reachable vertices from $s$ in the {\em residual graph} $R$ w.r.t. the flow $f$. This cut is denoted by $\overrightarrow{\delta}(S,V\setminus S)$, where $S$ (respectively $V\setminus S$) is the side of the cut containing $s$ (respectively $t$).
    The residual graph $R$ is defined as follows. For each edge $(u,v)$ in $G$ with capacity $c(\cdot)$, $R$ has two edges $(u,v)$ and $(v,u)$ with weight $c(u,v)-f(u,v)$ and $f(u,v)$ respectively. Edges of $R$ with weight zero are then removed. 

    Thus, our goal is to construct $R$ and compute reachability on it. Notice that the removal of zero-weight edges may increase the diameter of the graph. Hence, we instead construct a network $R'$ that is similar to $R$ except that in $R'$ zero-weight edge (instead of being removed) are assigned a weight of $\infty$ and all other edges are assigned a weight of zero. 
    Then, a vertex $u$ is reachable in $R$ from $s$ iff $\dist(s,u)=0$ in $R'$. Note, since the weights of the edges in $R'$ are either zero or $\infty$, we can represent $\infty$ by $1$.

    We construct $R'$ locally (i.e., without communication): since each vertex $v\in G$ knows its neighbors, incident edges, their capacity and amount of flow pushed on them, then it learns locally its incident edges in $R'$. $R'$ is embedded in the plane similarly to the embedding of $G$. E.g., the new copy $(v,u)$ of the existing edge $(u,v)$ succeeds it in the local ordering of the higher ID endpoint (say $v$) and precedes it in the local ordering of the other endpoint (say $u$).
    Note that $R'$ might be a multi-graph, however, one round of communication on $R'$ can be simulated in two rounds on $G$.

   To compute a directed SSSP tree from $s$, we use the exact SSSP algorithm of Li and Parter~\cite{LP19} that runs in $\tilde{O}(D^2)$ rounds (Theorem 1.6, Lemma 5.3 in their paper). Although it is stated in their paper for the undirected case, it can be easily adapted to the directed case as noted by Parter in~\cite{ParterReachability20a}. In addition, we note that the same algorithm works for planar multi-graphs as well.

    Finally, each $u\in G$ knows whether it is in $S$ or $V\setminus S$ according to its distance from $s$ in $R'$. So, every edge $(u,v)$ of $G$ s.t. $\dist(s,u)=0$ and $\dist(s,v)\neq 0$ in $R'$ is marked by its endpoints as a cut edge. This is computed in one round.

\end{proof}

Next, we show an almost optimal algorithm for $(1+o(1))$ approximation of the undirected minimum $st$-cut, where $s$ and $t$ lie on the same face.
\begin{theorem}
   [Distributed approximate $st$-planar undirected minimum $st$-cut]
   \label{th: approx_min_cut}
  Given an undirected planar graph $G$ with non-negative edge weights and two vertices $s,t$ that lie on the same face of $G$, there is a randomized algorithm that finds an approximate $(1+o(1))$ minimum $st$-cut and runs in $D\cdot n^{o(1)}$ rounds w.h.p..
\end{theorem}

\begin{proof}
   Note, the method used in \cref{th: exact_min_cut} does not apply instantly here, for two reasons, (1) the round complexity of the primal SSSP computation is very high compared to our promise, and (2) we cannot compute the residual graph, as we only have an approximation of the flow, meaning, we may not distinguish between saturated edges and non-saturated edges, a crucial step in the previous method. 
   Eventually, the cut edges are going to be the path $P$ found by the SSSP computation we perform when we approximate the max $st$-flow (see the proof of \cref{thmstflowapprox} for more details).
   
   Concretely, we take a primal-dual detour, where we consider faces in the dual graph that map to the vertices $s$ and $t$ in the primal\footnote{The dual of the dual is the primal, i.e, primal vertices map to faces of the dual graph. Moreover, the dual face that is mapped to any vertex $v$ is the (only) face that consists of all darts entering $v$. See~\cite{KM} for more details.}. See \cref{fig: approx_cut_graph}.
   We refer to faces of the dual as {\em dual faces}.  A special case of the cycle-cut duality (\cref{fac: cycle_cut_duality}) is given by Reif~\cite{Reif83}, stating: An $st$-separating cycle in the dual graph is the dual of an $st$-cut in the primal, where a cycle in the dual graph is said to be $st$-separating, iff it contains the the dual face that is mapped to $s$  in its interior, and the dual face mapped to $t$ in its exterior (or vice versa). Thus, we find a $(1+\epsilon)$-approximation of the minimum $st$-separating cycle in $G^*$ in order to find its dual, the desired $(1+\epsilon)$-approximate minimum $st$-cut in $G$ (then, setting $\epsilon=1/n^{o(1)}$ gives our promised $(1+o(1))$-approximation).

    \begin{figure}[htb]
    \centering
    \includegraphics[width=.4\linewidth]{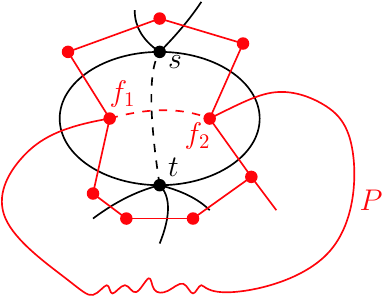}
    \caption{$G\cup\{e\}$ in black and its dual in red. $e$ ($e^*$) is the dashed edge. Each of $s$ and $t$ correspond to a face in the dual. Both of those dual faces contain $e$. 
    Note, the cycle $C$ is demonstrated by the $f_1$-to$f_2$ path $P$ and $e^*$, where the dual face $t$ is in its interior and the dual face $s$ in its exterior.  
    }
   \label{fig: approx_cut_graph}
    \end{figure}

    To do so, we compute a $(1+\epsilon)$-approximate max $st$-flow by \cref{thmstflowapprox}, not $(1-\epsilon)$, see the proof of that theorem where we first obtain a $(1+\epsilon)$ approximation (and then transform it to a $(1-\epsilon)$ one). Then we consider $G^*_{virt}=(G\cup \{e\})^*$, where $e=(s,t)$ splits the face $f$ in $G$ on which both $s,t$ lie into two faces, $f_1$ and $f_2$. Recall, the algorithm from \cref{thmstflowapprox}, computes a shortest path $P$ from $f_1$ to $f_2$ in $G^*_{virt}$ (see the proof of that theorem, its correctness is used in this proof). We claim that $P$ corresponds exactly to the approximate minimum $st$-cut in $G$. See \cref{fig: approx_cut_graph}.

    Note, any $st$-cut in $G\cup\{e\}$ shall contain $e$, and removing it gives $G$. Thus, to have the min $st$-cut in $G$, we can look for the min $st$-cut in $G\cup\{e\}$. In particular, there is a mapping between $st$-cuts of $G$ and $G\cup\{e\}$, where the value of the cut in $G\cup\{e\}$ is the same value of the corresponding cut in $G$ plus the weight ($nW$) of $e$.
    Thus, when applying the duality of Reif we consider $G^*_{virt}$. That is, we look for an $st$-separating cycle in $G^*_{virt}$. By duality, this cycle should contain $e^*=(f_1,f_2)$.
    
    Note, the edge $e^*=(f_1,f_2)$ closes a cycle with $P$, denoted $C=P\cup\{e^*\}$, that would define the desired $st$-separating cycle. 
    $C$ is separating as any cycle in a planar graph is separating (Jordan curve theorem~\cite{Jordan-Curve}), in particular it is $st$-separating, for: (1) $s$ and $t$ are faces in the dual, each has to be either in the interior or the exterior of $C$, (2) one dart of $e$ is in $C$'s interior and the other in its exterior, each such dart corresponds to a distinct dual face, one of which maps to $s$ and the other maps to $t$ because $e$ connects $s$ and $t$ in the primal.

    Since $C=P\cup\{e\}$, then the length of $C$ is that of $P$ plus $nW$. 
    Contracting $e$ subtracts $nW$ from the weight of $C$ and results with a $(1+\epsilon)$-approximate $st$-separating cycle in $G^*$, for otherwise, 
    $P$ is not a $(1+\epsilon)$ approximation to the shortest $f_1$-to-$f_2$ path, a contradiction. I.e., any $f_1$-to-$f_2$ path $P$ in $(G\cup\{e\})^*$ maps to a  $f$-to-$f$ path $P$ (cycle) in $G^*$ of the same weight.

    Next, we aim to mark those $st$-cut edges in $G$. By the above, it is sufficient to mark edges in $G$ who are the duals of edges $P$ in $G^*_{virt}$. Assuming we already applied \cref{thmstflowapprox}, in particular, we have computed an approximate SSSP tree $T$ from $f_1$, then, we want to mark $P$. We already know that we can simulate any minor-aggregation algorithm on $G^*_{virt}$ (for details, one is referred to the proof of \cref{thmstflowapprox}), thus, we can simulate any minor-aggregation algorithm on any subgraph of $G^*_{virt}$ (Corollary 11 of \cite{GZ22}). I.e., we can invoke any minor-aggregation algorithm on $T$.
   To mark $P$ edges, a minor-aggregation tree-rooting and subtree-sum procedures of~\cite{GZ22} are invoked, where we root $T$ with $f_1$ and compute subtree-sum on the rooted tree,  such that, the input value of all nodes is zero, except for $f_2$ that would have an input of one. The edges whose both endpoints have subtree-sum value of one are marked, those exactly are the $P$-edges.
   Those procedures are of $\tilde{O}(1)$ minor-aggregation rounds, which are simulated within $\tilde{O}(D)$ rounds in $G$ by applying \cref{th: extended_MA_on_dual}, the exact details of this simulation on $G^*_{virt}$ are provided in the proof of \cref{thmstflowapprox}.

   Next, we show how to find the cut vertices, in particular, we show how to find a set $S$ such that $P= \delta (S,V\setminus S)$. To do that, we take an approach similar to that described in the proof of \cref{th: exact_min_cut}, where we define a graph $R'$, in which, each $P$-edge is assigned a weight of one, and all other edges a weight of zero, we then again use the minor-aggregation SSSP approximation algorithm of \cite{GHSYZ22} with an arbitrary constant approximation parameter $\epsilon$ in order to compute a $(1+\epsilon)$-approximate SSSP tree $T$ from $s$ in $R'$. In particular, we modify the algorithm to work with edges with weights in $[0,n^{O(1)}]$ rather in $[1,n^{O(1)}]$, as shown in the proof of \cref{thmstflowapprox}.
   Finally, vertices whose distance is zero from $s$ are in $S$ (other vertices are in $V\setminus S$).
   
   The round complexity of simulating the SSSP approximation algorithm is $D\cdot n^{o(1)}$, since the part-wise aggregation problem can be solved on $G$ in  $\tilde{O}(D)$ $\Congest$ rounds (\cref{cor: planar_PA}), and the approximate SSSP algorithm of \cite{GHSYZ22} with $\epsilon=1/n^{o(1)}$ runs in $n^{o(1)}$ minor-aggregation rounds. Thus, by \cref{lem: MA_via_PA}, simulating the algorithm incurs a total of $D\cdot n^{o(1)}$ $\Congest$ rounds in $G$. Note, after applying the distance approximation algorithm, vertices of $G$ know locally to what side of the cut do they belong.
\end{proof}

\section{Directed Global Minimum Cut}
\label{section_directed_global_mincut}
In this section, we show that our labeling algorithm from \cref{sec: dual SSSP} for the dual graph $G^*$ can be used to compute the directed global minimum cut in the primal graph $G$.
The problem asks to find a bisection $(S,V\setminus S)$ of the vertex set, such that, the set of edges $C$ leaving $S$ to $V\setminus S$ is of minimal total weight.

\theoremDirectedGlobalCut*

In order to prove \cref{th: directed_global_min_cut}, we work with {\em darts}: each directed edge is represented by two darts in opposite directions.
For a definition, see \cref{def: darts}.
Specifically, we rely on a dart version of the cycle-cut duality (Theorem 4.6.2 in~\cite{KM}).

\begin{fact}[Dart Cycle-Cut Duality]
    \label{fact: directed_cycle_cut_duality}
    Let $G$ be a connected embedded planar graph (in the undirected sense). 
    A set of darts $C$ is a {\em simple dart cycle} in $G^*$ if and only if $C$ is a simple cut in $G$.
    Where a cycle is a simple dart cycle iff it is a simple cycle that uses no dart and its reversal, and,
    a cut is simple iff when removed it disconnects the graph into two connected components (in the undirected sense).
\end{fact}

\begin{figure}[htb]
    \centering
    \includegraphics[width=.27\linewidth]{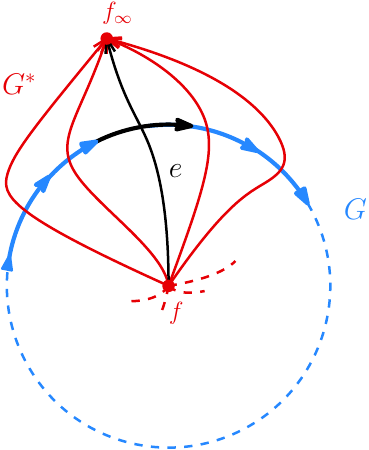}
    \caption{ In blue, the primal graph $G$ (a clockwise directed cycle). In red, the dual $G^*$ (two nodes with parallel edges in only one direction).
    The black edge $e$ is the minimal weight edge in $G$, constituting the directed global minimum cut in $G$, whilst its dual is the (single) black edge in $G^*$. $e$ is not a cycle in $G^*$.
    }
   \label{fig: cycle_cut_fail}
    \end{figure}

The reason we chose to work with darts is simply that the duality does {\em not} hold for general directed planar graphs, i.e., when an edge may be present without its reversal dart.
For example, the minimum cut in $G$, when $G$ is a directed cycle, is simply a single directed edge (i.e., there are two edges in the cut but only one of them is directed from $S$ to $V\setminus S$). Hence, in the dual it corresponds to the edge $e=(f,f_\infty)$, where $f$ (resp. $f_\infty$) is the dual node that corresponds to the interior (exterior) of the cycle, since $e$ is a single edge that connects two distinct nodes, it is not a cycle. 
See \cref{fig: cycle_cut_fail}.
\\
Thus, we augment $G$ with the reversal dart of all of its edges. Since the problem of interest is the minimum cut, all of those darts are assigned a weight of zero. This guarantees that the value of the minimum cut remains the same.
Henceforth, by~\cref{fact: directed_cycle_cut_duality}, 
it is enough to compute the directed minimum weight simple dart cycle in $G^*$.

\begin{proof}
    By \cref{fact: directed_cycle_cut_duality}, assuming $G$ is connected in the undirected sense, the minimum weight directed global cut $C$ of $G$ is given by the minimum weight simple cycle of darts in $G^*$.  
    We compute this cycle in $G^*$ in a bottom-up fashion on the extended BDD of $G$. Some terms are borrowed from the proof of \cref{{th: dual_distance_labeling}} (the distance labeling algorithm of $G^*$), e.g. the dual separator $\FX$ and the dense distance graph $DDG(g)$. Thus, the reader is referred to \cref{section_dual_labeling_algorithm} to recall the details of \cref{{th: dual_distance_labeling}}'s proof.

    The general idea is simple, given an extended BDD of $G$, by~\cref{lemma_sufficent_labels} , any shortest path (or cycle) of $G^*$ either intersects a dual node in $F_{\!_G}$ (the separator of $G^*$) or is entirely contained in one of $G^*$'s child bags. Thus, in order to find $C$, we take the minimum over: (i) directed cycles that intersect $F_{\!_G}$, and (ii) minimum directed cycles in $G^*$'s child bags (found recursively).
    Note, we need to ensure that all cycles found are simple dart cycles, which introduces a further complication to the general approach.
    The following describes in detail how to find the desired cycle of $G^*$ and then infer the cut in $G$.\\

\noindent {\bf Preprocessing:}
    We work with the network obtained from $G$ by adding the reversal dart of each edge to the graph as described in the proof of \cref{th: dual_distance_labeling}.  
    We abuse notation and denote this network by $G$.
    The reversal dart gets assigned a weight of zero.
    It is not hard to see that the new network can be constructed and embedded in the plane without any need for communication (assuming an embedding of the original network). 
    Note that any $\Congest$ round on the augmented network is simulated within two rounds on the communication graph.

    Next, in $\tilde{O}(D^2)$-rounds, by \cref{th: dual_distance_labeling}, we construct an extended BDD $\mathcal{T}$ and compute distance labels for each node in each bag $X^*\in \mathcal{T}$.
    The algorithm of \cref{th: dual_distance_labeling} succeeds to compute distance labels in $G^*$ and does not abort since all weights are non-negative (i.e., there are no negative cycles in $G^*$).\\

\noindent {\bf Leaf bag case:}
    Each leaf bag $X^*$ of $\mathcal{T}$ is gathered in all vertices of $X$ in $\tilde{O}(D)$ rounds (Property~\ref{BDD_dual_leaf_size} of $\mathcal{T}$). Thus, the minimum weight simple dart cycle of each such $X^*$ is computed locally and is known to all vertices in $X$, this cycle is denoted by $C_X$ and is assigned a unique ID, that of the minimal ID node on it concatenated with its weight.\\

\noindent {\bf Non-leaf bag case:}
    Let $X^*$ be a non-leaf bag of $\mathcal{T}$, we first find the minimum weight cycle that crosses between $X^*$'s child bags, i.e., a cycle that did not appear in any child bag of $X^*$, (hence, crosses $\FX$), by distance computations in $DDG(f)$ for an arbitrary $f\in \FX$. Then, we take the minimum between this cycle and cycles found in the previous recursion on the child bags of $X^*$ (cycles entirely contained in a child bag of $X^*$). Note that: (1) $DDG(f)$ is the same graph for all $f\in \FX$, and (2) all (primal) vertices on each $\FX$ node know the distance labels of all nodes in $\FX$.
    \begin{enumerate}
    \item $ID(C_{X_i})$ for each child bag $X_i$ of $X$ is aggregated on $X$ in $\tilde{O}(D)$ rounds so that the ID
    of the minimum weight cycle $C_X$ (that does not cross $\FX$) is known to all vertices in $X$.
    Note, 
    vertices in $X_i$s know $ID(C_{X_i})$, thus they can broadcast it (this is obviously true for leaf bags, for non-leaf bags, this is recursively maintained in the next two steps).
    Since $X$ is of  $\tilde{O}(D)$ diameter, this aggregation terminates in $\tilde{O}(D)$ rounds.
    Next, we consider cycles that cross $\FX$.

    \item The $f\in \FX$ with maximum ID is found first. Due to the fact that $X$ is of diameter  $\tilde{O}(D)$ and since all vertices of $X$  know the IDs of faces and face-parts they participate in (Properties~\ref{property: BDD_diameter} and~\ref{BDD_know_faces} of $\mathcal{T}$), this aggregation of IDs terminates in $\tilde{O}(D)$ rounds.

    \item Finally, all vertices of $X$ locally construct $DDG(f)$ and compute the minimum weight simple dart cycle $C$ by investigating the following two options related to $\SX$ (the edges of the primal separator, see \cref{section_BDD} for exact definitions).

    \begin{enumerate}
        \item {\em $C$ uses an $\SX$ dart:}
        In this case, one can deduce $C$ by iterating over $\SX$ darts present in $DDG(f)$. For each such dart $d=(g,h)$, we remove its reversal from $DDG(f)$ and compute SSSP from $h$.
        \begin{figure}[htb]
        \centering
        \includegraphics[width=.45\linewidth]{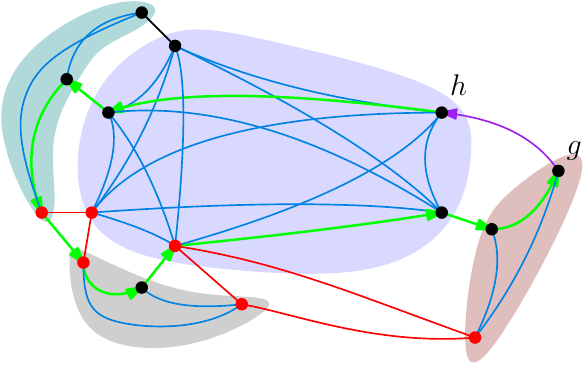}
        \captionsetup{margin={1.8cm, 0cm}}
        \caption{$DDG(f)$ in the iteration of the dart $(g,h)$ (purple), where each shadow color demonstrates a distinct child bag of $X^*$. Black vertices are endpoints of $\SX$ edges and red vertices are node parts in $X^*$'s child bags. The reversal dart of $(g,h)$ is removed from the DDG. All other edges have both their darts present (for simplicity, some are undirected in the picture). The green directed path demonstrates a shortest $h$-to-$g$ path that closes a cycle with the dart $(g,h)$.}
       \label{fig: dual_girth_cycle_uses_Sx}
        \end{figure}
        Note, the shortest $h$-to-$g$ path found maps to a shortest simple $h$-to-$g$ path $P$ in $X^*$ that does not use $\rev(d)$. $P$ is simple because there are no negative cycles in $G^*$.
        The following claim is a corollary of \cref{lemma_algorithm_computes_correct_labels}.
        \begin{claim*}
            $DDG(f)\setminus\{D_X\}$ preserves distances in $X^*\setminus\{D_X\}$, where $D_X$ is a subset of $\SX$ darts that are present in $DDG(f)$.
        \end{claim*}
        Due to the above claim, we get that $P$ combined with $d$ gives the minimum weight simple dart cycle that uses $d$, denoted $C_d$.
        We define $C$ to be the minimum over $C_d$ for all $\SX$ darts $d$.
        If $C$ is of weight strictly less than $C_X$, then $C_X$ is set to $C$ and is identified with the ID of an endpoint of $d$ ($d$ is the dart s.t. $C=C_d)$) concatenated with the weight of $C$.

        \item {\em $C$ does not use $\SX$:}
        In this case, we iterate over node-parts $g_i$ in $\FX$, in each iteration we remove $\SX$ darts from $DDG(f)$ and all clique edges between all node-parts $g_1,g_2,\ldots$ that correspond to the same face $g$ of $G$ as $g_i$.
        \begin{figure}[htb]
        \centering
        \includegraphics[width=.45\linewidth]{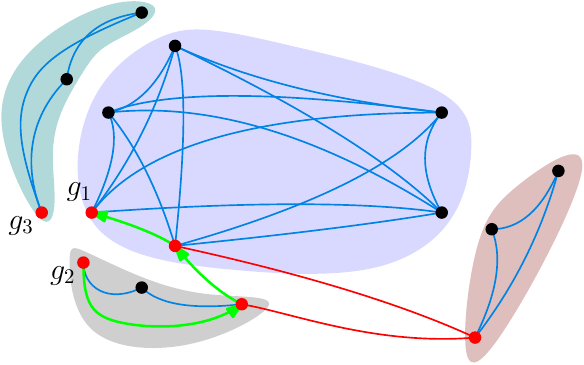}
        \captionsetup{margin={1.8cm, 0cm}}
        \caption{$DDG(f)$ in the iteration of node-part $g_2$. $g_1,g_2,g_3$ correspond to the face $g$ of $G$, where each shadow color demonstrates a distinct child bag of $X^*$. Black vertices are endpoints of $\SX$ edges and red vertices are node parts in $X^*$'s child bags. All $\SX$ darts are removed from the DDG, as well as clique edges on $g_1,g_2,g_3$. All other edges have both their darts present (for simplicity, some are undirected in the picture). The green directed path demonstrates a shortest $g_2$-to-$g_1$ path,
        representing a cycle that contains $g$ in $G^*$.}
       \label{fig: dual_girth_cycle_doesnt_use_Sx}
        \end{figure}
        Then, we compute the shortest path in $DDG(f)$ from $g_i$. We define $C_g$ to be the shortest path found from any $g_i$ to any $g_j\neq g_i$.
        If $C_g$ is of weight strictly less than $C_X$, then $C_X$ is set to $C_g$ and is identified with the ID of $g$ concatenated with the cycle's weight.

        $C_g$ corresponds to a simple dart cycle in $G^*$. A cycle since $g_i,g_j$ are parts of the same node $g$ of $G^*$; And simple since $C_g$ uses a dart incident to $g_i$ and a dart incident to $g_j$, those darts are not the reversal of each other because each of $g_i,g_j$ is in a distinct child-bag of $X^*$. I.e., the cycle $C_g$ in $G^*$ leaves $g$ with a dart incident to $g_i$ and ends entering $g$ with a dart incident to $g_j$ and those darts are not the reversal of each other. The removal of the dart incident to $g_j$ from $C_g$ in $G^*$ gives a path $P$, $P$ is simple due to the fact that it is a shortest path, hence $P$ with a dart that its reversal does not appear in $P$ constitutes a simple dart cycle, $C_g$. Note, we removed the clique edges in the DDG to have that the shortest $g_i$-to-$g_j$ path in the DDG is not both darts of a clique edge (which is also not an edge of $G^*$).
        
        The correctness of the cycle's weight is again due to the claim above, where $D_X$ is set to contain all $\SX$ darts. Note, removal of clique edges may only increase distances, however, all distances computed still map to paths in $G^*$.
        
    \end{enumerate}
    \end{enumerate}

\noindent {\bf Cut edges and bisection:}
    After the recursive process terminates, each $v\in G$ knows the weight of $C_G$  and an ID of a node $f$ of $G^*$ that participates in $C_G$. We aim to extend this information to find the edges and a corresponding bisection of the cut in $G$.
    Note, even if the node identified with the cycle was a face-part $f'$ in the bag $X$ where the cycle was found, we can locally identify the face $f$ of $G$ that $f'$ is a part of, as the ID of $f$ is contained in the ID of $f'$ (Property \ref{BDD_know_faces} of the extended BDD).

    \begin{claim*}
        The minimal weight simple dart cycle in $G^*$ can be decomposed into a shortest $f$-to-$g$ path $P$ and a dart $d=(g,f)$, where $g$ is the node that minimizes the total weight of $P \cup \{d\}$ s.t. $P\neq \rev(d)$. 
    \end{claim*}
    To prove the above claim, note, the desired cycle contains $f$, thus, $f$ has two incident darts in the cycle, specifically, there is a dart $d$ that enters it, denote $d$'s other endpoint by $g$, we get that the cycle minus $d$ is a shortest $f$-to-$g$ path $P$ (otherwise, a shorter cycle exists). Since the cycle is a simple dart cycle, we have that $P$ must not be the reversal of $d$. $g$ obviously minimizes the weight of $P \cup \{d\}$ (otherwise, a shorter cycle exists).
    
    Henceforth, to find the edges of the cycle in $G^*$, we compute an SSSP tree $T$ rooted at $f$ in $G^*$. Then, each non-$T$ dart entering $f$ determines a single cycle. We need to find such a simple dart cycle $P \cup \{d\}$ of total minimal weight where $P\neq \rev(d)$. 
    To do that, we first need to discard all reversals $\rev(d)$ of the darts $d$ that leave $f$ in $T$,
    because all such $\rev(d)$s close non-simple dart cycles with $T$.
    Then, we can indeed look for the dart $d=(g,f)$ entering $f$ that minimizes the (only) cycle of $T\cup \{d\}$.
    We do that by aggregations on $G^*$ using the minor-aggregation model (\cref{sec: dual_MA_model}). 

    \begin{enumerate}
    \item 
    By the proof of \cref{lemma_dual_SSSP}, within additional $\tilde{O}(D)$ rounds after computing distance labels, using aggregations on $G^*$, we can obtain the tree $T$.

Note that 
    $T$ is represented by $\hat{G}$, our tool for aggregations on $G^*$. Thus, we next work with it and with minor-aggregations on $G^*$ and $T$, which by \cref{sec: dual_MA_model} is possible since (1) we can do that on $G^*$ or any minor of it, and (2) $T$ is a subgraph (minor) of $G^*$.
    For an extended discussion of $\hat{G}$ and minor-aggregations on $G^*$, see Sections~\ref{section: preliminaries},~\ref{sec: dual_MA_model}, and Appendix~\ref{appendix: preliminaries_hat{G}}.

    \item $f$ and each node $g$ of its children in $T$ both discard the dart $\rev(d)$, where $d$ connects $f$ and $g$ in $T$, from participating in the algorithm, ensuring that no dart closes a non-simple dart cycle with $T$. 
    This is done locally in $\hat{G}$: the endpoints $u,v$ of the edge simulating $d$ in $\hat{G}$, know that $d$ connects $f$ and a child of it $g$ in $T$, thus, $u,v$ stop considering their other common edge that simulates $\rev(d)$ as an input for the aggregations on $\hat{G}$ that simulate aggregations on $G^*$ (and~$T$).

    \item Now we work with minor-aggregations on $G^*$. Since each node $g$ of $G^*$ knows its distance from $f$ (the weight of the $f$-to-$g$ $T$-path $P$), for the next aggregation, all active darts $(g,f)$ choose an input value of their weight plus the weight of $P$, other darts (that $f$ is not their head) chose an identity value. 
    Thus, within one minor-aggregation round, $f$ computes the minimum over the input values of its incident active darts. This allows $f$ to learn the dart $d$ that closes the minimal weight simple dart cycle with $T$, determining $C_G$ as the only cycle in $T\cup\{d\}$.
    $d$ marks itself as a dart of $C_G$.

    \item Next we want to mark the rest of the cycle (i.e. the path $P$).
    In order to do so, $f$ broadcasts the ID of $d$ to the entire graph $G^*$ in one minor-aggregation round. Allowing the other endpoint $g$ of $d$ to learn that $C_G\setminus\{d\}$ is the $T$-path $P$ from $f$ to it. 
    Then, the node $g$ chooses an input value of one, all other nodes of $G^*$ chose zero, and a subtree sum is computed over $T$ in $\tilde{O}(1)$ minor-aggregation rounds as shown in~\cite{GZ22}. The only path of $T$ whose nodes got a sum of one is $P$. In one minor-aggregation round, darts on $P$ mark themselves (those darts are the only darts with sum of one for each of their endpoints). 
    
    Turning back to the primal graph $G$, we have that vertices incident to darts of $C_G$ know those darts. I.e., the cut edges are known.

    \item 
    To obtain a bisection, by \cref{fact: directed_cycle_cut_duality}, it is sufficient to compute connected components in the underlying undirected graph after the removal of $C_G$ darts and their reversal $\rev(C_G)$ from $G$. I.e., each component is a side of the cut, and $S$ is the component which the darts $C_G$ leave.

    We do that by a Boruvka-like algorithm using aggregations (low-congestion shortcuts) on $G$ (e.g. the connectivity algorithm of~\cite{GH16b}), where we discard $C_G$ and $\rev(C_G)$ corresponding edges 
    \footnote{Edges and not darts, as we talk about an undirected graph.} 
    from participating as an input.
    Thus, each vertex of $G$ learns the ID of the connected component containing it out of two.

    Finally, we identify the component $S$: any vertex $v$ incident to a dart of $C_G$ leaving it, broadcasts its component ID to the entire graph $G$ over a BFS tree. Since there are only two distinct IDs passed, the broadcast terminates in $O(D)$ rounds. 
    Hence, vertices with the same component ID learn they are in $S$, other vertices learn they are in $V\setminus S$.
    
 \end{enumerate}   
    Note, this phase requires $\tilde{O}(D)$ $\Congest$ rounds, as each step either performs $\tilde{O}(D)$ $\Congest$ rounds on $G$, or $\tilde{O}(1)$ minor aggregation rounds on $G^*$, which also boil down to $\tilde{O}(D)$ $\Congest$ rounds by \cref{sec: dual_MA_model}.\qedhere

\end{proof}

\bibliographystyle{plainurl}
\bibliography{main}

\appendix

\section{Deferred Proofs from \cref{section: preliminaries} (Properties of \boldmath$\hat{G}$)}
\label{appendix: preliminaries_hat{G}}

We prove the following properties of the face-disjoint graph. All of those properties (except for Property~\ref{property: hat(G)_mapping_to_E(G^*)}) were proven in \cite{GP17}. However, for completeness a proof is provided for all properties. 
\begin{enumerate}
    \item    
    \label{property: appendix_hat(G)_construction_representation}
    $\hat{G}$ is planar and can be constructed in $O(1)$ rounds (after which every vertex in $G$ knows the information of all its copies in $\hat{G}$ and their adjacent edges). 
    \item
    \label{property: appendix_hat(G)_diameter}
    $\hat{G}$ has diameter at most $3D$.
    \item
    \label{property: appendix_hat(G)_simulation}
    Any $r$-round algorithm on $\hat{G}$ can be simulated by a  $2r$-round  algorithm on  $G$.
    \item
    \label{property: appendix_hat(G)_face_identification}
    There is an $\tilde{O}(D)$-round algorithm that identifies $G$'s faces by detecting their corresponding faces in $\hat{G}$. 
    When the algorithm terminates, every such face of $\hat{G}$ is assigned a face leader that knows the face's ID. Finally, the vertices of $G$ know the IDs of all faces that contain them, and for each of their incident edges the two IDs of the faces that contain them.
    Thus, each vertex of $G$ knows for each pair of consecutive edges adjacent to it (using its clockwise ordering of edges) the ID of the face that contains them.
    \item
    \label{property: appendix_hat(G)_mapping_to_E(G^*)}
    There is a 1-1 mapping between edges of $G^*$ and $E_C$. Both endpoints of an edge in $E_C$ know the weight and direction of its corresponding edge in $G^*$ (if $G^*$ is directed and/or weighted). 
\end{enumerate}

\begin{proof}
$\;$

\begin{enumerate}
    \item    
    [(\ref{property: appendix_hat(G)_construction_representation})]
      $\hat{G}$ can be constructed in $O(1)$ rounds, as the only information transmitted on an edge is its local clockwise numbering in its endpoints' combinatorial embedding.
      
       For proving that $\hat{G}$ is planar, we show an explicit  planar embedding, which can be deduced from the planar embedding of $G$. 
    Consider first the star-center copy of $v$ in $\hat{G}$. Recall that $v$ has $\deg(v)$ copies $v_1,v_2,\ldots$ that are incident to the star-center, the star center local clockwise embedding is $(v,v_1),(v,v_2),\ldots$. 
    Now consider a non star-center copy $v_i$ of $v$, such that $u$ and $w$ are neighbors of $v$ in $G$ and the edge $(u,v)$ precedes the edge $(w,v)$ in $v$'s local clockwise embedding. $v_i$ is possibly incident to the following edges $(v_i,u_j),(v_i,v_{i+1}),(v_i,v),(v_i,v_{i-1}),(v_i,w_k)$ which we embed in that order. 
 
    \item
    [(\ref{property: appendix_hat(G)_diameter})]
    $\hat{G}$ has diameter at most $3D$ as each edge in $G$ becomes a path $(u-u_i-v_j-v)$ in $\hat{G}\setminus E_C$, and adding $E_C$ can only make the diameter smaller.
    \item
    [(\ref{property: appendix_hat(G)_simulation})]
    Any $r$-round $\Congest$ algorithm on $\hat{G}$ can be simulated in $2r$ $\Congest$ rounds in $G$, because each edge in $G$ has exactly two copies in $E_R$. These are the only edges that require simulation as edges $(v_i,v_j),(v,v_i)$ in $E_C \cup E_S$ are simulated locally by $v$.  
    \item
    [(\ref{property: appendix_hat(G)_face_identification})]
    There is an $\tilde{O}(D)$ round algorithm that identifies $G$'s faces by detecting their corresponding (vertex and edge disjoint) connected components of $\hat{G}[E_R]$ while considering $\hat{G}$ as the communication network. 
    The algorithm is a variant of \cite{GH16a}'s connectivity algorithm, working in a Boruvka-like fashion performing $O(\log n)$ phases of merges.\footnote{The mentioned algorithm is randomized, however, the only randomized step in it is performing star-shaped merges, a primitive which was derandomized by \cite{GZ22} and can be used as a black box in \cite{GH16a}'s algorithm to derandomize it.}
    When the algorithm terminates, every non star-center vertex of $\hat{G}$ knows its face ID and each face is assigned a leader. 
    Hence, the vertices of $G$ know the IDs of all faces that contain them, and for each of their incident edges the  IDs of the two faces that contain them.
    \item
    [(\ref{property: appendix_hat(G)_mapping_to_E(G^*)})]
    There is a 1-1 mapping between edges of $G^*$ and $E_C$. 
    Every edge $e^*$ of $G^*$ corresponds to exactly one edge of $e$ of $G$ so we show the 1-1 mapping between edges of $G$ and $E_C$.  
    Each edge $e$ in $G$ maps to one edge $\hat{e}$ of $E_C$. Each edge $\hat{e}$ in $E_C$ is of the form  $(v_i,v_{i+1})$. $v_{i+1}$ is connected to some $u_j$  by an edge and $v_i$ is connected to $u_{j+1}$ by an edge. These two edge belong to $E_R$ and correspond to  one edge $e=(u,v)$ in $G$. We map the edge $\hat{e}$ to $e$. 
        
    Consider $e^*\in G^*$ that maps to $\hat{e}=(v_i,v_{i+1})$ in $E_C$. Since $\hat{e}$ connects copies of $v\in G$ then it is mapped to an incident edge $e$ of $v$, meaning, $v$ knows $e$'s weight and direction. Thus, $\hat{e}$ is assigned the same weight as $e$, and the direction of $\hat{e}$ is defined to be $(v_i,v_{i+1})$  (resp. $(v_{i+1},v_i)$) if and only if $e=(v,u)$ (resp. $e=(u,v)$). The directions are defined this way for the following reason. Consider an edge $e=(v,u)\in G$, consider the clockwise local numbering of $v$ of $e$, denote the face to the left of $e$ by $v$'s numbering by $f$ and the face to the right of $e$ by $v$'s numbering by $g$, then, $e^*:=(f,g)$. And indeed, when $e=(v,u)$, then, $v_i$ is in the face that represents $f$ in $\hat{G}$ (the face to the left of $e$) and $v_{i+1}$ is in the face that represents $g$ (the face to the right of $e$), and vise versa in case $e=(u,v)$. Meaning, when $e^*$ is directed from $f$ to $g$, $\hat{e}$ is directed from the face that  represents $f$ to the face that represents $g$.

    The above can be achieved locally after construction, as the only information needed to one endpoint of an edge is the other endpoint's local numbering of that edge.\qedhere
 \end{enumerate}     
\end{proof}

\end{document}